\documentclass[longauth]{aa}

\usepackage{graphicx}
\usepackage{txfonts}
\usepackage{multirow}
\usepackage{multicol}
\usepackage{lscape}
\usepackage{booktabs}
\usepackage{threeparttablex}
\usepackage{wasysym}
\usepackage{arydshln}
\usepackage{placeins}
\usepackage[bookmarks=false,colorlinks=true,linkcolor=black,citecolor=cyan,filecolor=black,urlcolor=cyan]{hyperref}
\newcommand{\SRG}{{\rm SRG}\xspace}
\newcommand{\xmm}{{\it XMM-Newton}\xspace}
\newcommand{\ROSAT}{{\rm ROSAT}\xspace}
\newcommand{\cxo}{{\it Chandra}\xspace}
\newcommand{\einstein}{{\it Einstein}\xspace}
\newcommand{\uhuru}{{\it Uhuru}\xspace}
\newcommand{\swift}{{\it Swift}\xspace}

\newcommand{\EXT}{{\texttt{EXT}}\xspace}
\newcommand{\EXTLIKE}{{\texttt{EXT\_LIKE}}\xspace}
\newcommand{\DETLIKE}[1]{{\texttt{DET\_LIKE\_#1}}\xspace}

\begin{document}

   \title{The SRG/eROSITA all-sky survey}
\subtitle{First X-ray catalogues and data release of the western Galactic hemisphere\thanks{The catalogue is available at the CDS via anonymous ftp to \url{cdsarc.cds.unistra.fr} (130.79.128.5) or via \url{https:// cdsarc.cds.unistra.fr/viz- bin/cat/J/A+A/682/A34}}}   
\titlerunning{The eRASS1 all-sky survey}

  \author{A.~Merloni\inst{1}\thanks{\email{am@mpe.mpg.de}},
G.~Lamer\inst{2},
T.~Liu\inst{1,25,26},
M.~E.~Ramos-Ceja\inst{1},
H.~Brunner\inst{1},
E.~Bulbul\inst{1},
K.~Dennerl\inst{1},
V.~Doroshenko\inst{3},
M.~J.~Freyberg\inst{1},
S.~Friedrich\inst{1}
% ...
}

\institute{
   Max-Planck-Institut f\"ur Extraterrestrische Physik, Gie{\ss}enbachstra{\ss}e, D-85748 Garching, Germany
\and 
    Leibniz Institut f\"ur Astrophysik Potsdam, An der Sternwarte 16, D-14482 Potsdam, Germany
\and 
    Institut f\"ur Astronomie und Astrophysik, Universit\"at T\"ubingen, Sand 1, D-72076 T\"ubingen, Germany
\and 
    Institute for Astronomy and Astrophysics, National Observatory of Athens, V. Paulou and I. Metaxa, 11532, Greece
\and
Exzellenzcluster ORIGINS, Boltzmannstr. 2, 85748, Garching, Germany
\and
Dr. Karl Remeis-Sternwarte and Erlangen Centre for Astroparticle Physics, Friedrich-Alexander Universität Erlangen-Nürnberg, Sternwartstra{\ss}e 7, 96049 Bamberg, Germany
\and
Hamburger Sternwarte, Gojenbergsweg 112, 21029 Hamburg, Germany
\and
MIT Kavli Institute for Astrophysics and Space Research, Massachusetts Institute of Technology, Cambridge, MA 02139, USA
\and
Argelander Institute for Astronomy, University of Bonn, Auf dem H{\"u}gel 71, 53121 Bonn, Germany
\and
Max-Planck-Institut f{\"u}r Radioastronomie, Auf dem H{\"u}gel 69,53121 Bonn, Germany
\and
Dipartimento di Fisica e Astronomia "Augusto Righi", Universit{\`a} di Bologna, via Gobetti 93/2, 40129 Bologna, Italy
\and
INAF - Osservatorio di Astrofisica e Scienza dello Spazio di Bologna, via Gobetti 93/3, 40129 Bologna, Italy
\and
IRAP, Universit{\'e} de Toulouse, CNRS, UPS, CNES, Toulouse, France 
\and
Tautenburg Landessternwarte, Sternwarte 5, 07778 Tautenburg, Germany
\and
University Observatory Munich, Faculty of Physics, Ludwig-Maximilians-Universit{\"a}t, Scheinerstr. 1, 81679 Munich, Germany
\and
Department of Physics \& McDonnell Center for the Space Sciences, Washington University in St. Louis, One Brookings Drive, St. Louis, MO 63130, USA
\and
Universit\"at Innsbruck, Institut f\"ur Astro- und Teilchenphysik, Technikerstra{\ss}e 25/8, 6020 Innsbruck, Austria
\and
TU Dresden, Institute of Nuclear and Particle Physics, 01062 Dresden, Germany
\and
DESY, Notkestraße 85, 22607 Hamburg, Germany
\and
Department of Astronomy and Astrophysics, University of Chicago, Chicago, IL 60637, USA
\and
Leiden Observatory, Leiden University, Niels Bohrweg 2, 2333 CA Leiden, The Netherlands
\and
Istituto di Radioastronomia IRA-INAF, via Gobetti 101, 40129 Bologna, Italy
\and
INAF, Osservatorio Astronomico di Brera, Via E. Bianchi 46, 23807 Merate, (LC), Italy
\and
Max-Planck Institut f{\"u}r Astronomie, Königstuhl 17, 69117 Heidelberg
\and
CAS Key Laboratory for Research in Galaxies and Cosmology, Department of Astronomy, University of Science and Technology of China, Hefei 230026, China
\and
School of Astronomy and Space Science, University of Science and Technology of China, Hefei 230026, China
             }

   \date{Received 12 June 2023; accepted 26 November 2023}

\authorrunning{Merloni et al.}

% \abstract{}{}{}{}{} 
% 5 {} token are mandatory
 
  \abstract
  % context heading (optional)
  % {} leave it empty if necessary  
   {The eROSITA telescope array aboard the {\it Spektrum Roentgen Gamma} (\SRG) satellite began surveying the sky in December 2019, with the aim of producing all-sky X-ray source lists and sky maps of an unprecedented depth. Here we present catalogues of both point-like and extended sources using the data acquired in the first six months of survey operations (eRASS1; completed June 2020) over the half sky whose proprietary data rights lie with the German eROSITA Consortium. We describe the observation process, the data analysis pipelines, and the characteristics of the X-ray sources. With nearly 930\,000 entries detected in the most sensitive 0.2--2.3 keV energy range, the eRASS1 main catalogue presented here increases the number of known X-ray sources in the published literature by more than  60\%, and provides a comprehensive inventory of all classes of X-ray celestial objects, covering a wide range of physical processes. A smaller catalogue of 5466 sources detected in the less sensitive but harder 2.3--5 keV band is the result of the first true imaging survey of the entire sky above 2 keV. We present methods to identify and flag potential spurious sources in the catalogues, which we applied for this work, and we tested and validated the astrometric accuracy via cross-comparison with other X-ray and multi-wavelength catalogues. We show that the number counts of X-ray sources in eRASS1 are consistent with those derived over narrower fields by past X-ray surveys of a similar depth, and we explore the number counts variation as a function of the location in the sky. Adopting a uniform all-sky flux limit (at 50\% completeness) of $F_{\rm 0.5-2~keV} > 5 \times 10^{-14}$ erg\,s$^{-1}$\,cm$^{-2}$, we estimate that the eROSITA all-sky survey resolves into individual sources about 20\% of the cosmic X-ray background in the 1--2 keV range. The catalogues presented here form part of the first data release (DR1) of the \SRG/eROSITA all-sky survey. Beyond the X-ray catalogues, DR1 contains all detected and calibrated event files, source products (light curves and spectra), and all-sky maps. Illustrative examples of these are provided.  
}
   \keywords{catalogues --
                surveys --
                X-ray: general
               }

   \maketitle
%
%-------------------------------------------------------------------

\section{Introduction}
\label{sec:intro}
Wide-area multi-wavelength (and multi-messenger) sky surveys  play a key role in the development of astrophysics. The design of these surveys is driven both by the desire to open up new discovery spaces, and by the realisation that charting the structure of the Universe in detail may help solve long-standing open questions of fundamental physics \citep[see e.g.][]{Peebles1980,Weinberg2013}.

Sky surveys also occupy a central position in the relatively short history of X-ray astronomy \citep[see e.g.][for a succinct recap]{Elvis2020}. These surveys were initiated by the SAS-A \uhuru satellite \citep[1970--1973,][]{1971giacc}, a mission designed to conduct a survey of the X-ray sky in the 2--20~keV energy range, resulting in the identification of 339 X-ray sources \citep{1978form}. Likewise, an X-ray survey was carried out by the cosmic X-ray experiment aboard the HEAO-1 observatory \citep[1977--1979,][]{1979roths} covering the 0.25--25~keV energy range, enabling the discovery of 842 X-ray sources \citep{1984wood}. The \einstein (HEAO-2) X-ray Observatory \citep[1978--1981,][]{1979giacc} performed a medium-sensitive survey in the 0.3--3.5~keV energy range, covering roughly one-third of the sky and providing the detection of about 4000 X-ray sources \citep{Harris1990}. 
However, the first comprehensive all-sky X-ray survey was performed by the \ROSAT X-ray observatory in the 0.1--2.4~keV energy range \citep[1990--1999,][]{1982truem}.

The \ROSAT All-Sky Survey (RASS)  
and its associated catalogues of X-ray sources, The Bright Source Catalogue (BSC) \citep{Voges1999}, Faint Source Catalogue (FSC) \citep{Voges2000}, and the more comprehensive second catalogue release 2RXS \citep{Boller2016}, marked a new milestone in the quantity and quality of X-ray detections.
With more than $10^5$ unique sources detected in only six months of survey observations, RASS outnumbered all previous all-sky catalogues by more than two orders of magnitude.

Since the turn of the century, \xmm and \cxo, with their large collecting area and high spatial resolution, respectively, have increased the number of known X-ray sources significantly. \cxo and \xmm serendipitous source catalogues, however, only cover a small fraction of the sky, due to their relatively small fields of view and their pointed observation strategy.

These and other X-ray surveys provide a unique view of many celestial phenomena. X-ray emission is a universal signature of accretion of matter onto compact objects. In binary systems, these mark the graveyards of stellar evolution \citep[see e.g.][]{Shapiro1983,Warner1995,BEcker1997,Fender2004, Remillard2006,Haberl2007,Ozel2016}, where black holes, neutron stars, and white dwarfs accrete matter from a companion. The brightest X-ray sources in the sky, including Sco X--1, the first extrasolar one discovered \citep{Giacconi1962}, fall into this category. 
At cosmological distances, X-rays signpost the growth of the supermassive black holes (SMBHs) that sit at the centres of galaxies and which may strongly influence their formation and subsequent evolution \citep{Brandt2005,Hopkins2008,Hickox2009,Fabian2012,Alexander2012,Kormendy2013,Brandt2015}. 

X-ray emission is seen from gas heated by the shocks generated by rapidly expanding supernova remnants, which are in addition likely responsible for a large fraction of the accelerated particles that diffuse through interstellar space \citep[see e.g.][and references therein]{Vink2012,Blasi2013}. Strong X-ray emission is also seen from the coronae of stars \citep{Schmitt1997,Pizzolato2003,Wright2011}, which plays an important role in the determination of the physical characteristics of the orbiting planets' atmospheres, with implications for the potential habitability of such planets \citep{Lammer2003,Sanz-Forcada2011,Poppenhaeger2021,Foster2022}.

A further source of X-ray emission of great wider importance is from hot gas associated with large-scale structures. Most of the baryons in the Universe are indeed predicted to be locked into a warm-hot (X-ray emitting) tenuous phase \citep{Cen1999,Dave2001,Nicastro2018,Tanimura2020}. Direct or indirect detections of these baryons often require sensitive mapping of large volumes in X-rays. Moreover, in the hierarchical distribution of matter characteristic of our observed Universe, the densest knots of the large-scale structure are signposted by the hottest and most massive concentration of diffuse baryons. The clusters of galaxies that identify them are thus extremely sensitive tracers of the underlying geometry and growth history of the cosmos, and therefore prime cosmological tools \citep{Bahcall1977,Cavaliere1978,Sarazin1986,Rosati2002,Voit2005,Arnaud2005,Norman2005,Borgani2006,Vikhlinin2009,Borgani2009,Allen2011,Reiprich2013}. It is the potential for sensitive cosmological measurements with galaxy clusters that provided the main motivation for the development of eROSITA in  the early 2000s, when it became apparent that a significantly larger number of clusters of galaxies compared to those provided by narrow-field instruments would be required to effectively  constrain the fundamental parameters of cosmological models \citep{2005haim, Merloni2012, Pillepich2012}.
 In particular, these authors argued that sample sizes of order $10^5$ clusters were required to provide constraints competitive with other prime cosmological measurement tools. 

%\subsection{The eROSITA all-sky survey}

eROSITA \citep[extended ROentgen Survey with an Imaging Telescope Array; ][]{Predehl2021}, on board the {\it Spektrum Roentgen Gamma} (\SRG) orbital observatory \citep{Sunyaev2021}, was developed in the period 2007--2019. It is a sensitive, wide-field focusing X-ray telescope array, optimised to deliver large effective area and field-of view (hence also grasp and survey speed) in the soft X-ray band. The angular resolution is sufficient to distinguish between the two largest extragalactic source populations, that is, AGN and clusters of galaxies, via measurement of their X-ray extent. 

The observing strategy was designed to achieve the needed sensitivity in the soft X-ray band (below 2 keV) to detect at least the requisite 10$^5$ clusters of galaxies by scanning the entire sky eight times over a period of four years (the eROSITA All-Sky Surveys, hereafter: eRASS). 

\SRG was launched on July 13, 2019 from Baikonur, Kazakhstan, using a Proton-M rocket and a BLOK DM-03 upper stage. On its three months cruise to the second Lagrangian point (L2) of the Earth-Sun system the spacecraft and instruments underwent commissioning and checkout. Since mid-October 2019, \SRG was placed in a six-month-periodic halo orbit around L2, with a semi-major axis of about 750\,000~km within the ecliptic plane and semi-minor axis of about 400\,000~km perpendicular to it \citep{Sunyaev2021}; periodic orbit correction manoeuvres over the intervening years have slightly reduced the size of the orbit in order to satisfy ground segment visibility constraints. 

Following First Light \citep{Maitra2022, Haberl2022}, a two-months long Calibration and Performance Verification (CalPV) programme was executed between October and December 2019. 
The large body of publications based on  CalPV observations \citep[see][and the associated articles of the A\&A special issue]{Campana2021} have demonstrated the capabilities of eROSITA, and confirmed the main design characteristics. In particular, the eROSITA Final Equatorial Depth Survey \citep[eFEDS;][]{Brunner2022}, designed to provide uniform exposure over a sufficiently large field (140 deg$^2$) about 50\% deeper than what is expected for eRASS at the end of the 4 year all-sky survey programme, has shown that large samples of X-ray sources of different classes can be detected, identified, and characterised making use of the synergy with existing multi-wavelength surveys \citep[see, e.g.][Nandra et al., submitted]{Ghirardini2021,LiuA2022,LiuT2022b,Salvato2022,Klein2022,Schneider2022,Bulbul2022,Pasini2022,Ramos-Ceja2022}.

%\subsection{Data distribution rights}
The eROSITA data are shared equally between German and Russian scientists, following an inter-agency agreement signed in 2009. Two hemispheres of the sky have been defined, over which each team has unique scientific data exploitation rights. These data rights are split by Galactic longitude ($l$) and latitude ($b$), with a division marked by the great circle passing through the Galactic poles $(l,b)=(0\degr,+90\degr);(0\degr,-90\degr)$ and the Galactic Center Sgr\,A* $(l,b)=(359.9442\degr,-0.04616\degr)$: data with $-0.05576\degr < l < 179.9442\degr$ (eastern Galactic hemisphere) belong to the Russian consortium, while data with $359.9442\degr > l > 179.9442\degr$ (western Galactic hemisphere) belong to the German eROSITA consortium (eROSITA-DE). Here we only describe and release the data collected in the western Galactic hemisphere.

%\subsection{Structure of the paper}
After a brief introduction to the salient technical aspects of eROSITA and its calibration (Sect.\,\ref{sec:tech}), in Sect.\,\ref{sec:observations} we describe in detail the eROSITA observing procedures in all-sky survey mode. Section\,\ref{sec:data_processing} is then devoted to a summary of the main data processing stages. Most of the details of the software system used to process eROSITA data have already been presented in \citet{Brunner2022}, and we refer the interested reader to that work for more information. The catalogues generated by our standard processing pipeline analysing the first eROSITA all-sky survey (eRASS1) are presented in Sect.\,\ref{sec:catalog}. There we present the preliminary astrometric correction applied to the detected sources as well as our general attempt to isolate and flag potential spurious sources and other artefacts. Section\,\ref{sec:cons_checks} describes further consistency checks on the X-ray catalogue performed by comparing the eRASS1 sources with those of the \xmm serendipitous catalogue (photometry) and other multi-wavelength Quasi-Stellar Object (QSO) catalogues (astrometry). Section\,\ref{sec:dr1} then gives an overview of all the available products, including all-sky maps, and source-specific light curves and X-ray spectra. To conclude, we summarise our work in Sect.\,\ref{sec:summary}.

\section{eROSITA technical facts}
\label{sec:tech}
In this section, we provide a compact summary of the main technical characteristics of eROSITA and its calibration; more details, including a summary of the on-ground calibration, can be found in \citet{Predehl2021}. Specific technical descriptions of the instrument subsystems can be found in \citet{Meidinger2021} (camera system) and \citet{Friedrich2012} (mirror system), while a thorough description of the on-ground calibration campaign and its results can be found in \citet{Dennerl2020}. Indeed, most of the data analysis and pipeline settings adopted for the reduction of the eRASS1 data presented here rely on this extensive on-ground calibration of the instrument. As for all other X-ray space observatories, in-flight calibration is a long-term endeavour; here we present some preliminary results that demonstrate the fidelity of the calibration and the reliability of the data released. An in-depth analysis of the in-flight calibration will be presented elsewhere, and on the website of the first data release (DR1).

\subsection{Instrument characteristics}
eROSITA consists of seven identical and co-aligned X-ray telescopes arranged in a common optical bench. A system of carbon fibre honeycomb panels connects the seven mirror assemblies on the front side with the associated seven camera assemblies on the focal plane side.  

Each of the mirrors comprises 54 mirror shells in a Wolter-I geometry, with an outer diameter of 360 mm and a common focal length of 1\,600 mm \citep{Friedrich2008,Arcangeli2017}. 
The average on-axis resolution of the seven mirrors, as measured during the on-ground calibration,  
is $16.1$\arcsec\  Half-Energy Width (HEW) at 1.5\,keV. 
The unavoidable off-axis blurring typical of Wolter-I optics is compensated by a 0.4~mm shift of the 
cameras towards the mirrors. This puts each telescope slightly out of focus, leading to a slight 
degradation of the on-axis performance ($18''$ HEW), but improved angular resolution averaged over the field of view. Indeed, in the scanning observational mode adopted for the all-sky survey (see section~\ref{sec:scanning_strategy}), it is the field-of-view average HEW that matters. This is discussed in Section~\ref{sec:psf_cal} and Appendix~\ref{appendix:psf}.

Each Mirror Assembly has a CCD camera in its focus \citep{Meidinger2014}. The eROSITA CCDs are advanced versions of the EPIC-pn CCDs on \xmm  \citep{Strueder2001}. They have $384\times384$ pixels in an image area of $28.8\,\mathrm{mm} \times 28.8\,\mathrm{mm}$, yielding a square field of view of $1\fdg{}03 \times 1\fdg{}03$. 
Each pixel corresponds to a sky area of $9\farcs{}6 \times 9\farcs{}6$. The nominal integration time for all eROSITA CCDs is 50\,msec. The additional presence of a frame-store area in the CCD  reduces substantially the amount of so-called out-of-time events, which are recorded during the CCD read-out, a significant improvement with respect to the EPIC-pn camera on \xmm. For optimal performance during operations, the CCDs are cooled down to about $-85^\circ$ by means of passive elements \citep{Fuermetz2008}. During survey operation (i.e. in scanning mode of observations), the angle between the scanning direction projected onto the sky and the CCD read-out direction in the focal plane is not the same for the seven cameras\footnote{The angles between the scanning direction projected onto the sky and the CCD read-out direction in the focal plane are $\theta_C = (\frac{3}{2}\pi,\frac{5}{3}\pi, \frac{\pi}{2}, \frac{\pi}{2}, \frac{\pi}{3}, \frac{\pi}{3}, \frac{5}{3}\pi)$ for (TM1, TM2, TM3, TM4, TM5, TM6, TM7), respecively.}. This contributes to averaging out any possible non-circular symmetry of the PSF as well as camera-induced defects. 
For calibration purposes, each camera has its own filter wheel with a radioactive $^{55}$Fe source and an aluminium and titanium target providing three spectral lines at 5.9\,keV (Mn K$\alpha$), 4.5\,keV (Ti K$\alpha$) and 1.5\,keV (Al K$\alpha$). 

The electronics for onboard-processing of the camera data is provided by seven sets of Camera Electronics (CE), each one mounted and interfacing to the Cameras. Each of the CEs provide the proper voltage control and readout timing of the associated camera, and performs the on-board data processing within the time constraints of the camera integration time.

\subsection{eROSITA calibration}

The quantities derived from the eROSITA calibration are stored in a calibration database (CalDB), which is accessed by various processing tasks. Information about the content of the CalDB and how important entries were derived can be found in \citet{Brunner2022}, Appendix B. A `live' online repository can be found on the DR1 website\footnote{\url{https://erosita.mpe.mpg.de/dr1/eSASS4DR1/eSASS4DR1_CALDB/}}. Here we present a brief assessment of the current status of the in-flight calibration program.

\subsubsection{PSF calibration}
\label{sec:psf_cal}
 Given the scientific objectives of eROSITA, i.e. imaging the whole sky at soft-to-hard X-ray energies with good sensitivity to low surface brightness diffuse and extended emission, clusters of galaxies in particular, an accurate calibration of the telescopes' Point-Spread Function (PSF) is critical. As we describe in greater detail below (\S ~\ref{sec:data_processing}), the catalogues presented here have  been constructed using a single photon mode, in which the PSF of each telescope module at the location of each detected event on the CCD is accounted for, using a shapelet-based model calibrated on the extensive dataset accumulated on ground before the launch \citep{Dennerl2020}. A description of the shapelet-based PSF modelling, and its usage in scanning mode can be found in Appendix B.1 of \citet{Brunner2022}.

While work is ongoing to accurately characterize eROSITA's PSF based on the survey data themselves, here we report on a preliminary analysis that confirms the reliability of the on-ground calibration adopted for the DR1 datasets. Appendix~\ref{appendix:psf} shows a direct comparison of the average PSF shape (obtained by combining all seven Telescope Modules) from stacking point sources detected in the all-sky survey with the PSF model from the PANTER on-ground calibration and its shapelet representation. 
The differences between the PSF models are within the $\sim20$\% level in the inner 1\arcmin, although the shapelet PSF drops below the measured PSFs beyond this radius, where it is not used in the source detection process, except for normalisation.
The measured HEWs from the source stacking method applied to survey data are 30.0\arcsec\ in the 0.2--2.3~keV band and 34.4\arcsec\ in the 2.3--5.0~keV band, very close to the pre-flights estimates of 28.3\arcsec\ and 36.2\arcsec, respectively, for the shapelet representation, and 32.0\arcsec\ and 38.0\arcsec\ for the PANTER ground-based values. The PSF does not appear to be varying across the sky, at least within the limited statistics of our preliminary analysis (see Tab.~\ref{tab:psf_ecl}). In Appendix~\ref{appendix:psf} we also show an estimate of the PSF azimutal symmetry, and compare positional offsets of eRASS1 point sources against Gaia QSOs along equatorial and ecliptic coordinates, demonstrating the goodness of our circular symmetric approximation for the positional uncertainty of the detected sources.

\subsubsection{Energy calibration}

After launch, the energy calibration obtained on ground was checked by using extensive measurements with the internal $^{55}$Fe calibration sources. These measurements showed that the in-flight energy resolution of the detectors was essentially the same as on ground \citep[Tables 1 and 2 in][]{Dennerl2020}. The $^{55}$Fe calibration measurements were also used to derive updated values of the Charge Transfer Inefficiency (CTI) and Gain, in order to minimise their impact on the absolute energy scale. Additional tests of the energy calibration were made with dedicated observations of astrophysical calibration targets, especially of the isolated neutron star RX\,J1856--3754 and of the supernova remnant 1E\,0102.2--7219. These demonstrated that the energy calibration is sufficiently accurate for the analysis of survey data with their limited photon statistics\footnote{For detailed spectral studies of sources in long pointed observations with high photon statistics further improvements of the RMF and ARF appear to be possible; this work is ongoing at present.}. Appendix~\ref{appendix:energy_calibration} provides more details on these energy calibration experiments. The currently available energy calibration has already been used successfully for spectroscopic studies \citep[e.g.][]{Camilloni2023,Mayer2023,Ponti2023,Yeung2023}.

\subsubsection{Flux calibration: Effective area and vignetting}

The challenge of accurately flux-calibrating space-based X-ray telescopes is as old as X-ray astronomy itself \citep[see e.g.][and references therein for recent discussions]{Marshall2021, Madsen2021}. While the Quantum Efficiency (QE) of the CCD detectors can be accurately measured on ground, but is subject to degradation in the harsh space environment, the main difficulty rests with the paucity of  suitable stable standard candles and the impossibility to accurately reproduce in the laboratory the observing conditions in space needed to calibrate a telescope effective area and vignetting function (i.e. variation of effective area across the focal plane). Clusters of galaxies, possibly the brightest intrinsically non-varying X-ray sources in the sky, are usually adopted as cross-calibrators among different X-ray observatories, but even after decades, uncertainties remain \citep{Nevalainen2023}. For eROSITA, we can take advantage of the all-sky survey nature of the observations to build large samples that can be used to validate the flux calibration in a statistical sense. For point sources, we report here (Section~\ref{sec:flux_compare}) a comparison with \xmm in the 0.5-2 keV band, resulting in a possible residual flux systematic uncertainty of just about 6\%, similarly to what was found by \citet{Maitra2022} in the early phases of the mission. We have also tested the effect of  possible PSF calibration uncertainties on the reconstructed flux, by running an alternative source detection pipeline which makes use of the PSF image preliminary stacks, and found only a 3\% source counts systematic offset compared to the catalogue described here.

The situation for clusters of galaxies in the survey is still under examination \citep[but see e.g.][for pointed Calibration observations]{Whelan2022,Sanders2022}, with Bulbul et al. (\citeyear{Bulbul2023}) reporting a systematic flux deficit with respect to \cxo of about 15\%, while \citet{LiuA2023} and Migkas et al.\ (submitted)\ found a temperature offset with respect to both \xmm and \cxo (with eROSITA measuring lower tempertures than both) that increases with the cluster temperature itself. Work is ongoing to determine to what extent this is induced by calibration uncertainties or by cluster-related astrophysical effects (such as multi-temperature ICM distributions).

\section{Observations}
\label{sec:observations}

\subsection{Scanning strategy}\label{sec:scanning_strategy}

In order to complete an all-sky survey, the \SRG observatory rotates continuously with a period of four hours around an axis pointed near the direction of the Sun. This gives a scan rate of 0.025~deg~s$^{-1}$. The rotation axis slowly shifts by approximately one degree per day following the motion of the Earth (and of the L2 point) around the Sun. Following this scanning pattern, eROSITA observes the entire sky in about six months, and observes each point in the sky typically six times (`visits') for up to 40 seconds over a day at the ecliptic equator; towards the ecliptic poles, the sources remain observable for more than 24 hours, and are therefore scanned more than six times. Indeed, all great circles (individual scans) intersect in the north and south ecliptic poles in the sky, creating regions of deep exposure and longer visibility periods. 
In addition, a slight inhomogeneity in the sky coverage is introduced by the elongated halo orbit around L2 and the nonuniform angular movement of the spacecraft rotation axis, which compensate the separations between Sun and Earth, as seen from the spacecraft, in order to maintain the Earth in the cone of the downlink antenna \citep[further details in][]{Predehl2021,Sunyaev2021}.

\begin{table*}
    \caption{Timeline of the main eROSITA operations and major events during eRASS1.}
    \label{tab:erass1_operations}
    \centering
    %\begin{tabular}{p{3cm}|p{13cm}}
    \begin{tabular}{ll}
        \hline\hline
        Date & Description \\
        \hline
         12.12.2019 & Start of eRASS1 \\
         30.01.2020 & Routine orbit correction (for station keeping)\\  % for station keeping
         10.02.2020 & Major ITC anomaly\\
         01.04.2020 & Routine orbit correction (for station keeping)\\  % for station keeping
         02.04.2020 & Observation of 1RXS\,J185635.1-375433 for contamination monitoring\\
         04.04-24.05.2020 & ESTRACK support due to low visibility from Russian antennas\\
         11.06.2020 & End of eRASS1. \\
        \hline
    \end{tabular}
\end{table*}

\subsection{All-sky survey operations}\label{sec:operations}

After the commissioning of the instrument and the CalPV phase, the first all-sky survey started on December 12, 2019, and was completed on June 11, 2020, for a total survey duration of 184 days. During this period, ground contacts with \SRG and eROSITA took place every day, without interruption. A timeline of the most significant operations milestones during eRASS1 is presented in Table~\ref{tab:erass1_operations}.

\begin{table}[htbp]
    \caption{Timeline of extended diagnostic and engineering exposures during eRASS1.}
    \label{tab:erass1_engineering}
    \centering
    \begin{tabular}{lll}
        \hline\hline
        Date & Time [UTC] & Description \\
        \hline
         30.01.2020 & 13:40 -- 18:15 & Filter wheel closed \\
                    & 18:15 -- 18:30 & Raw frames\\
                    & 18:30 -- 19:18 & Filter wheel closed \\
                    & 19:18 -- 19:33 & Raw frames\\
                    & 19:33 -- 22:30 & Filter wheel closed \\
         01.04.2020 & 13:40 -- 18:15 & Filter wheel closed \\
                    & 18:15 -- 18:30 & Raw frames \\
                    & 18:30 -- 19:18 & Filter wheel closed \\
                    & 19:18 -- 19:33 & Raw frames \\
                    & 19:33 -- 23:08 & Filter wheel closed \\
        \hline
    \end{tabular}
    \tablefoot{``Raw Frames'' are read-out cycles where each pixel is transmitted (by-passing the event processor), 10 such frames were commanded for each TM consecutively (due to the large amount of telemetry there is some time gap in between). After the seven TMs a second round of ``raw frames'' was commanded, giving in total $10\times 7 \times 2 = 140$ such frames per epoch.}
\end{table}

As described in \citet{Coutinho2022} and \citet{Predehl2021}, the Mission Control Center, located in Moscow at NPO Lavochkin (NPOL), has mainly used two deep-space antennas for the science downlink (Ussuriysk and Bear Lakes)\footnote{Due to a low spacecraft visibility window during the spring period of 2020 \citep{Sunyaev2021}, two deep-space antennas of the ESA tracking stations network (ESTRACK) were also used on 15 occasions for the science data downlink of about 3.5 GB.}. In total, approximately 75 GB of telemetry data from eROSITA were dumped during eRASS1, with an average of $\sim$407~MB per day.

\begin{figure}
    \centering
    \includegraphics[scale=0.29]{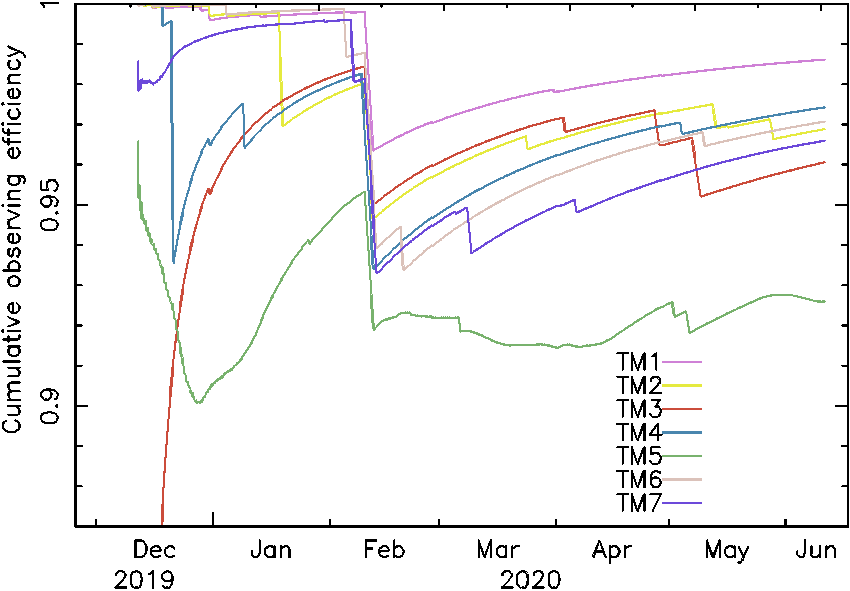}
    \caption{Cumulative observing efficiency (as a percentage of the elapsed survey time) as a function of time for each eROSITA camera (labeled according to the associated Telescope Module, TM) during eRASS1. With the exception of TM5, badly affected by light leak, all cameras achieved an efficiency of more than 95\% by the end of eRASS1, with TM1 reaching close to 99\%. The largest loss of net observing time was caused by an ITC malfunction on February 10, 2020.}
    \label{fig:erass1_obseff}
\end{figure}

The overall observing efficiency of eROSITA during eRASS1 was $\sim$96.5\%. This efficiency has been calculated by taking the average of the Good Time Intervals (GTI) with respect to the total observing time for each camera, as they are operating independently. The main observation disruptions responsible for the loss of efficiency come from Camera Electronics (CE) and Interface and Thermal Controller (ITC) anomalies: the CEs and ITC are susceptible to Single Event Upsets (SEU), which can interrupt the functioning of the instrument. Moreover, Telescope Modules (TM) 5 and 7 suffer from a time-varying light leak, which can lead to loss of data and consequently to gaps in GTI \citep[see Sect.\,\ref{subsection:proc_issues} and][for further details]{Predehl2021,Coutinho2022}. Figure~\ref{fig:erass1_obseff} shows the cumulative observing efficiency as a function of time for each TM during eRASS1. 

Regular `Filter Wheel Closed' (FWC) observations were carried out during the survey to monitor the instrumental background, starting in February 2020. The filter wheel of one camera per day was set to the closed position for $30$~minutes. To reduce the impact on the exposure of the survey observations, the same camera had an FWC observation every seven days. 
In total, around 18 short FWC observations for each camera were performed in eRASS1 (exposure fraction 0.3\%), providing a precious data set to model and monitor the background \citep[see][for more details about the instrumental background model]{Yeung2023}. In addition, more extended FWC observations were performed during periods of orbit corrections (see Table~\ref{tab:erass1_engineering}), along with other diagnostic and engineering exposures.
The viewing direction of the telescope was reset after the orbit corrections (and monitoring pointed observations) to ensure survey coverage without gaps.

With the exception of the short-lived SEU-induced malfunctions of the camera electronics and of the ITC (on Feb. 10, 2020), all eROSITA sub-systems were fully functional during eRASS1, and they have not suffered any permanent damage, apart from the expected degradation caused by external environmental effects along \SRG's L2 orbit. In \citet{Coutinho2022} the interested reader can find more details about the technical performance of eROSITA during the first two years of operations (from eRASS1 to eRASS4).

\subsection{Pre-processing and archiving}
The eROSITA data received at a ground station are forwarded in
real time via a socket connection to the \SRG operations centre and to IKI (the Space Research Institute of the Russian Academy of Sciences), where they are stored in binary files with a size of approximately 7.5\,MB each. These files are transferred to MPE via a
data exchange server. The files
are then picked up by the pre-processing pipeline, which converts the telemetry into FITS files and forward them to two separate pipelines:
the preproc-archiver and the Near Real Time Analysis (NRTA). 

The archive is organised into `erodays', i.e. fixed intervals of 4 hours (also corresponding to the duration of one revolution of \SRG in all-sky survey mode).  
Once an eroday is completed, the data are moved to the regular archive and the post-ingest pipeline is triggered. 

The NRTA pipeline has been developed in order to (i) monitor interactively the instrument parameters and check the health of all sub-systems on the shortest possible timescales, and (ii) alert the team and the community of possible interesting time-domain phenomena observed by eROSITA. The NRTA and its functionality is described in Appendix~\ref{sec:NRTA}. At this point, the archived data are ready to be processed by the standard analysis pipeline.              

\section{Standard data processing}
\label{sec:data_processing}
The eRASS1 data were processed with the eROSITA standard data processing pipeline, operated at MPE.
The pipeline consists of modules for event processing (TEL processing chain), event file and image creation (EXP processing chain), exposure and background map creation and source detection (DET processing chain), as well as for the 
creation of source-specific products such as spectra and light curves (SOU processing chain). Each processing chain executes a number of
software tasks, which are part of the eROSITA Science Analysis Software System (eSASS). 

In comparison with the data processing version (001) from the Early Data Release (EDR)\footnote{\url{https://erosita.mpe.mpg.de/edr/}}, the event calibration in the processing version (010) of eRASS1 has a stronger telescope module (TM) specific noise suppression of doubles, triples and quadruples\footnote{Single, double, triple or quadruple events (known as `patterns') refer to the possible number of pixels within which the charge cloud generated by an X-ray photon can be distributed. See \url{erosita.mpe.mpg.de/edr/eROSITATechnical/patternfractions.html} for details on how this information can be used to reconstruct the energy and position of the incoming photons.}, a better computation of the subpixel position, a corrected flagging of pixels next to bad pixels, and improved accuracy of projection. The list of eSASS task and calibration database
versions used for this work is provided in Appendix~\ref{appendix:esass}. A list of standard data products
included in the eROSITA DR1 data release is provided in Sect.~\ref{sec:dr1}. A detailed description of the main
eSASS tasks, associated eROSITA calibration database, and calibrated data products is available in \citet{Brunner2022} (Appendix A--C). In the rest of this section we briefly summarize the organisation of the pipeline and the function of each pipeline chain.

\subsection{TEL chain}\label{sec:tel_chain}

The TEL chains perform the functions of event file preparation (tasks {\tt evprep}, {\tt ftfindhotpix}),
pattern recombination (task {\tt pattern}), energy calibration (task {\tt energy}), and attitude calculation (tasks
{\tt attprep}, {\tt telatt}, and {\tt evatt}). The TEL chains are executed separately for each telescope module
for data intervals of one eroday. 

\begin{figure}
    \centering
    \includegraphics[scale=0.39]{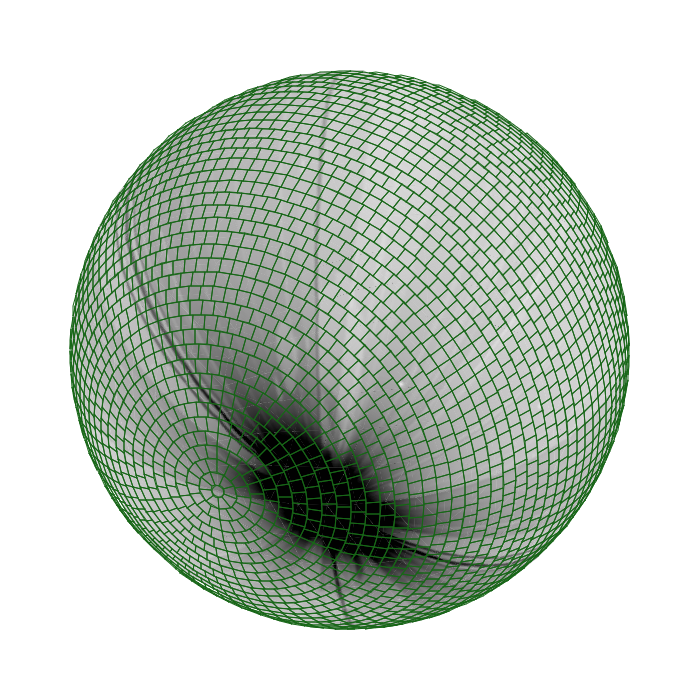}
    \caption{Spherical projection of the eRASS1 1B exposure map centred on the western Galactic hemisphere (i.e. the eROSITA-DE area) with over-imposed in green the tiling of the sky into equal-area overlapping $3.6\degr \times 3.6\degr$ tiles into which data are organised and analysed by the pipeline.}
    \label{fig:erass1_tiles}
\end{figure}

The sky is divided into 4700 non-overlapping, unique areas ("sky tiles"), organized into 61 equatorial declination zones (see Fig.~\ref{fig:erass1_tiles} for a visual representation of the adopted sky tiling). The size of these sky tiles is exactly three degrees in declination and approximately three degrees in right ascension (on the side facing the equator), resulting in an average area of about 8.78 square degrees per tile. Each unique sky tile is embedded in an overlapping, square sky field of size $3.6\degr \times 3.6\degr$ centred on the sky tile. The minimum overlap between neighbouring sky fields, introduced in order to avoid edge effects in the source detection, ranges between 15\arcmin\ for polar fields and 18\arcmin\ for equatorial fields\footnote{See also \url{https://erosita.mpe.mpg.de/dr1/eSASS4DR1/eSASS4DR1_ProductsDescription/skytile.html}.}. After the completion of the
event calibration, the event data of each TEL chain are sorted into these overlapping sky fields, based
on the right ascension and declination of each photon (tasks {\tt telgti}, {\tt telselect}, {\tt telstage}). Source detection and further source-level analysis are then performed in each sky field.

\subsection{EXP chain}\label{sec:exp_chain}

The event data collected in each sky tile are merged on a `per TM' basis (task {\tt expmerge}) and sky pixel
event coordinates with pixel size $0\farcs05$ centred on the sky tile centre are computed (task {\tt radec2xy}).
Combining these TM-specific files, filtered event files and images that include all TMs (pixel size $4\arcsec$,
$3240 \times 3240$ pixels) are created in a variety of energy bands suitable for source detection  (task {\tt evtool}). 
The event filtering excludes invalid patterns (i.e. pixel patterns with a low probability of having been caused by a single X--ray photon),
events on or close to bad pixels, as well as events outside a circular
detection mask of radius 0.516\degr. In addition, good time intervals
free of background flares are created, if necessary (task {\tt flaregti})\footnote{For eRASS1, due to the very low frequency of background flares, the {\tt flaregti} were not applied to the images used for source detection.}.

\subsection{DET chain}
\label{sec:det_chain}

\subsubsection{Exposure maps}

Exposure maps are created on a tile-by-tile basis by the task {\tt expmap}. For each energy band and for each TM, a map of the active pixels of the CCD (that is, excluding flagged bad pixels, their orthogonal neighbours, as well as out-of-FOV pixels) is first divided by a vignetting map. The latter is generated as a weighted mean of the energy-dependent vignetting function across the respective energy band, the weighting being a power law of spectral index\footnote{As customary, the spectral index $\Gamma$ is here defined such that for the photon spectral density at energy $E$, $n(E)dE \propto  E^{-\Gamma}$.} $\Gamma=1.7$. The vignetted CCD image is then projected onto the sky (using the same projection as the template image supplied to the task) at a series of time samples (nominally one per second) of the TM attitude, using only samples during GTIs for that TM. The resulting TM-specific sky exposure maps are then combined in a weighted mean for each band, the weights (corresponding to the fraction of area supplied by each TM with respect to the nominal all-TM area) being read from the calibration database. The results are $N$ maps in the $N$ specified energy bands, each giving the spatially varying mean (vignetted) exposure in that given band, in seconds.

\begin{figure*}
    \centering
    \includegraphics[scale=0.41]{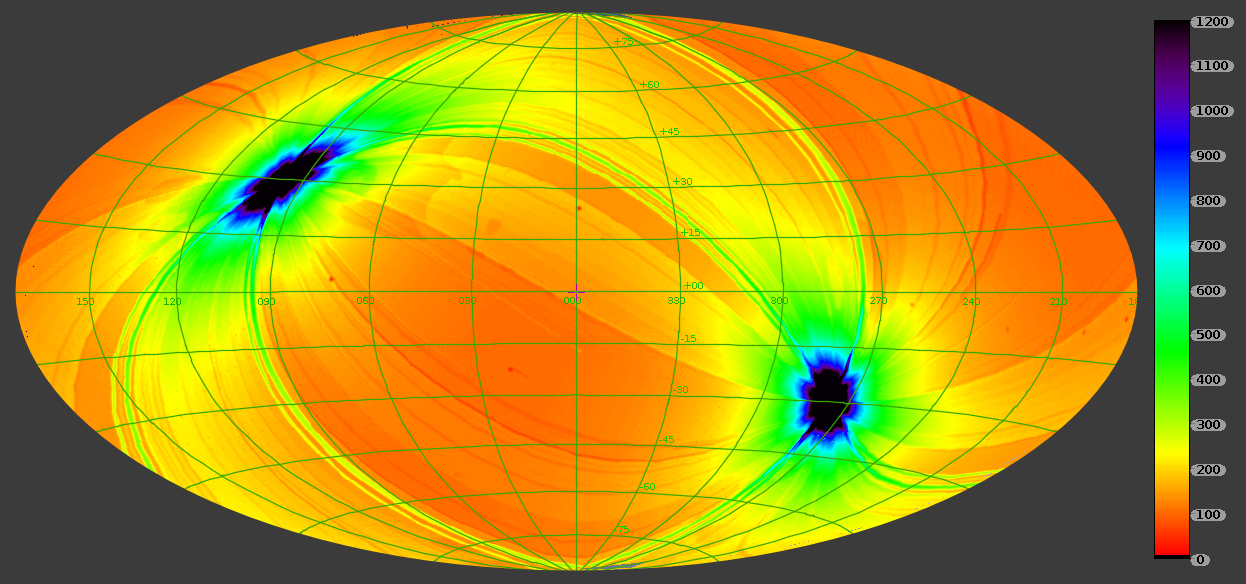}
    \caption{Effective (vignetted) eRASS1 exposure map (Galactic coordinates, Aitoff projection). The values in the map show the exposure time multiplied by the average of the ratio of the (vignetted) effective area to the on-axis effective area in the energy band 0.6--2.3\,keV. 
Effective exposure values range from $\sim$100\,s at the ecliptic equator to more than 10\,000\,s close to the ecliptic poles (not visible on the colour scale of this image). The eROSITA-DE (western) Galactic hemisphere is on the right of the central meridian in this map.}
    \label{fig:erass1_aitoff_projection}
\end{figure*}

As an example, Fig.~\ref{fig:erass1_aitoff_projection} shows the combined all-sky exposure map (where darker regions mark longer exposures) for the 0.6--2.3 keV band. The main qualitative features of this map are determined by the scanning law and the associated event milestones discussed above (i.e. scan interruptions, overlaps, etc.); these features are common to all exposure maps. Quantitative measures, however, such as the overall normalisation of the exposure, may differ for different energy bands, mainly due to the variations of the vignetting function of the telescope at different energies.

\subsubsection{Source detection}
The eSASS source detection pipeline was applied to each of the $3.6\degr \times 3.6\degr$ tiles with the following steps: 

 \begin{enumerate}
     \item Creation of masks to define image regions with valid data (task {\tt ermask});
     \item Initial source finding (task {\tt erbox});
     \item Background determination (task {\tt erbackmap});
     \item Search for candidate sources (task {\tt erbox});
     \item Creation of PSF fitting catalogue (task {\tt ermldet});
     \item Forced PSF photometry in additional energy bands (task {\tt ermldet});
     \item Aperture photometry in all energy bands and the corresponding sensitivity maps (task {\tt apetool}); the aperture is chosen as a fixed one at 75\% encircled energy fraction\footnote{Note that the catalogue only lists raw photometry data products, i.e. total number of photons within the aperture or expected background level within the aperture. Users can use the raw photometry data products to reconstruct the fluxes or count-rates.};
     \item Creation of sensitivity maps (task {\tt ersensmap}).  
 \end{enumerate}
Steps 3. and 4. were iterated twice for an improved separation of sources and background.
The source detection setup is very similar to the configuration used for the creation of the eFEDS catalogues \citep{Brunner2022}, where the individual detection tasks are described in more detail. 

Reliable detection and characterisation of X-ray extended sources is crucial for any subsequent application and analysis of galaxy cluster samples.
It should be noted that for both point sources and extended sources the rate, count and flux values reported in the eRASS1 catalogues are based
on the scaling of the best fit PSF or PSF-folded extent model.  By definition,  these quantities correspond to values integrated to infinite radii while the model fits are performed on circular
sub-images of 1\arcmin\ radius. Therefore, the flux related source parameters may be subject to significant
systematic uncertainties, in particular for large extended sources. In Bulbul et al. (\citeyear{Bulbul2023}) the interested reader will find a detailed description of the procedure used to define clean clusters samples starting from the extended sources catalogues, and also to derive robust physical quantities from the X-ray data alone.

In each sky tile two catalogues were created: the single-band catalogue with sources detected in the broad 0.2--2.3 keV band (hereafter `1B') for maximal sensitivity,
and a 3-bands catalogue (0.2--0.6 keV, 0.6--2.3 keV, 2.3--5.0 keV, hereafter `3B'), for which detection is carried simultaneously and single-band and combined likelihoods for each source are computed.

\subsubsection{Sensitivity maps: {\tt ersensmap}}
\label{sec:methodsensmaps}

For both the 1B  catalogue and the 3B catalogue
sensitivity maps were calculated  with the task
{\tt ersensmap}. As described in \citet{Brunner2022}, the task {\tt ersensmap} uses the eSASS exposure maps and background maps to estimate the detection limits in eROSITA observations. For eRASS1, the maps for the master catalogue contain the 0.2--2.3 keV source flux required to reach a typical detection likelihood of 5.0 at the respective pixel position for consistency with the threshold of the PSF fitting catalog created by \texttt{ermldet}.
The map values for the 3-band catalogue correspond
to the respective 0.2--5 keV flux. For the  conversion between the map fluxes and count rates in the detection images the energy conversion factors (ECFs) listed in Table~\ref{tab:sens_ecf}  were used. 
Based on spectral analysis of eFEDS AGN \citep{LiuT2022}, typical eROSITA-detected AGN have a median power-law slope of 2.0. So we adopted this slope and an absorbing column density $N_\mathrm{H} = 3 \times 10^{20}$\,cm$^{-2}$ (typical value of the Galactic absorption) to calculate the ECFs \citep{Brunner2022}. The impact of the assumed power-law slope is negligible here.

\begin{table}[htbp]
    \caption{ECFs used to calculate the 1-band and 3-band sensitivity maps.}
    \label{tab:sens_ecf}
    \centering
    \begin{tabular}{cc}
       \hline\hline
         Image energy range  &  ECF  \\
         $\mathrm{[keV]}$    & $\mathrm{[cm^2/erg]}$ \\
       \hline
         \multicolumn{2}{c}{1B: 0.2--2.3 keV} \\         
         0.2--2.3  &  $1.074\times 10^{12} $   \\    
       \hline
         \multicolumn{2}{c}{3B: 0.2--5.0 keV} \\
         0.2--0.6 &  $2.300\times 10^{11} $   \\
         0.6--2.3 &  $5.253\times 10^{11} $  \\
         2.3--5.0 &  $3.356\times 10^{10} $   \\  
       \hline
    \end{tabular}
    \tablefoot{For the 3B maps the ECFs refer to the conversion between the individual bands count rates and the total 0.2--5 keV flux. These ECFs are for conversion between an observed count rate and an observed flux, not corrected for absorption, under the spectral model assumption.}
\end{table}

\subsubsection{Sensitivity maps: {\tt apetool}}
\label{sec:methodsensmaps_apetool}
The {\tt apetool} sensitivity maps use aperture photometry to determine the selection function of a sample of X-ray point sources detected in a given spectral band. The selection function is defined as the probability of detecting an X-ray point source with a given count rate or flux in the band of interest across the eROSITA field of view, across all observations of the source throughout the duration of
eRASS1. In generating these sensitivity maps, the stacked PSF at a given sky position is the exposure-time weighted superposition of the individual PSFs from all the pixels (detector coordinates) that contribute to that sky position as eROSITA slews across the sky. The {\tt apetool} sensitivity maps can only be used in combination with the aperture photometry of individual X-ray sources provided in the eROSITA catalogues. We refer to the appendices of \cite{Brunner2022} for a full description of the {\tt apetool} functionality and \cite{Georgakakis2008} for details on the calculation of X-ray sensitivity maps based on aperture photometry. 

There are two key parameters that are relevant for the {\tt apetool} sensitivity maps: the radius of the circular aperture adopted for photometry and the Poisson probability that the observed counts within an aperture are produced by random fluctuations of the background (Poisson false detection probability, or False Alarm Probability, FAP). The lower the latter, the less likely it is that the counts within an aperture are produced by the background, thereby suggesting the presence of an astrophysical source. For a given background level (as specified in the background maps), we use Poisson statistics to estimate the minimum number of photons within the aperture, $N_\mathrm{min}$ so that the  corresponding FAP is below an adopted threshold, thus yielding X-ray source detections to a given confidence level. 
The {\tt apetool} sensitivity map is an image of $N_{\rm min}$  across the field of view of the eROSITA observations. The sensitivity map can be further combined with the eROSITA exposure and background maps to determine the probability of detecting a source with FAP $\le P_{\rm thresh}$ integrated over the eROSITA field of view. This probability as a function of count rate or flux (once an ECF is adopted) is referred to as the sensitivity or area curve and is also provided by {\tt apetool}. This allows us to study the flux limit at each position, based on the local background level and exposure depth. Two applications of the {\tt apetool} sensitivity maps (computation of flux limits and number counts) are described in Sects.\,\ref{subsec:flux_limit} and \ref{subsec:lognlogs}.

\subsection{SOU chain}
Source products (spectra, background spectra, response matrices, ancillary response files and light curves) were created using
the \texttt{srctool} task,  for the subset of bright sources with a detection likelihood greater than 20 in the single-band 1B catalogue. In terms of net counts, this corresponds to a sample with a median number of $\sim$21, and 90$^{\rm th}$ (10$^{\rm th}$) percentile of $\sim$79 ($\sim$12) counts.
Further details about the source products are given in Sect.~\ref{sect:srctool}.

\subsection{Bright sources: pile-up and other losses}

Source parameters derived from the detection pipeline, such as total flux or temporal or spectral variability, carry systematic uncertainties, in particular for very bright X-ray sources.
This is mainly due to the pile-up effect \citep{Ballet1999,Davis2001,Tamba2022}, which occurs if two or more photons hit the same CCD pixel in the same (50 ms) read-out cycle and the sum of the charges created will enter the event analyser (``energy pile-up''). Pile up also occurs if these photons are recorded in two adjacent pixels where, after recombination of the individual charges, a higher energy value is reconstructed (``pattern pile-up''). 
It is important to note that for pile-up the total energy band is relevant and also events below the lower event trigger threshold, including optical photons.
The probability for these effects mainly depends on the source brightness in photons/cycle/pixel and on the actual shape of the PSF, for example, it is reduced for a deteriorated PSF far off-axis. This can lead to apparent spectral and temporal variability within a scan through the FOV, but also between one (e.g., more central) scan and another (more off-axis). All this complexity in general requires detailed simulations for an accurate estimate of pile-up effects \citep[see e.g.][]{Koenig2022}. Preliminary studies using \texttt{SIXTE} \citep{Dauser2019}, indicate that for eRASS1 pile-up starts to become important (more than a few per-cent effect) for point sources brighter than $\approx 10^{-11}$ erg\,s$^{-1}$\,cm$^{-2}$ in the 0.2--2.3 keV band.

Photon counts can also be lost in eROSITA due to effects in the event analyser and telemetry limitations between camera electronics and ITC.
The `event quota' mechanism for each camera ensures a reasonable telemetry rate, even in the case of, e.g., a CCD column becoming bright between two ground contacts.
In the current implementation, the event quota is triggered if in one TM there are more than 50 events for four consecutive read-out cycles after onboard rejection of minimum ionising particles (MIP) and bad pixels. In that case, for one minute only frames containing less than 50 events are telemetered. After one minute this is reset and all frames are transmitted
until the trigger criteria are fulfilled again. Due to this event quota, complete read-out frames are lost, and all sources within the FOV are affected, differently from pile-up, which only affects the piled-up source itself. This mechanism was designed for instrumental reasons, but very bright point sources, such as Sco X--1 (or even bright optical stars) and extended sources (for example Puppis A) are also (celestial) triggers.
Missing read-out cycles are properly handled in the exposure computation, but there remains a bias towards fainter read-out cycles during active event quota triggers.

Finally, the source parameters of some of the brightest sources in the catalogues  (\texttt{ML\_CTS\_1} $> 1000$) suffer from
poor convergence of the PSF fits due to the large number of individual events to be included in the modelling. This may result in  larger than  expected deviations in both photometry and astrometry.

\subsection{Other known data processing issues}
\label{subsection:proc_issues}

During the commissioning phase of eROSITA, it was noticed that TM5 and TM7 (the CCDs not equipped with an on-chip Al optical blocking filter) were contaminated by optical light: a small fraction of sunlight reaches the CCD, by-passing the filter wheel. The intensity of this optical contamination depends on the orientation of the telescope with respect to the Sun.
This `light leak', mostly restricted to very low photon energies (typically $<0.3$ keV), generates a non-negligible amount of telemetry data and decreases the low energy coverage and spectroscopic capabilities for these two cameras. In order to reduce the amount of transmitted data from TM5 and TM7, their primary thresholds are higher (about 125--145 eV) than the ones in the other TMs (about 65--95 eV). The fact that the amount of contamination by optical light is spatially and temporally variable makes it very difficult to derive an accurate energy calibration for these TMs. More details about the light leak in eRASS1 can be found online on the DR1 web portal\footnote{\url{https://erosita.mpe.mpg.de/dr1/eROSITA\_issues\_dr1/}}.

The effect of optical loading, i.e. the appearance of fake X-ray sources, or the distortion of the measured X-ray properties of sources associated to (very) bright optical stars, is discussed in \S~\ref{subsec:optical_loading}.

In some cases, the source detection algorithm failed to converge during its error estimate procedure, leaving some sources without
reliable uncertainty estimates for the position, counts, or extent. Many of the affected sources have low detection likelihood or are related to spurious detections in areas of extended emission.
In some cases, however, also significant detections can be affected by this issue.
We flag these sources with the labels \texttt{FLAG\_NO\_RADEC\_ERR}, \texttt{FLAG\_NO\_CTS\_ERR} and \texttt{FLAG\_NO\_EXT\_ERR}, respectively (see Tables~\ref{tab:flags} and \ref{tab:source_flag_summary}). 

Finally, in calculating the number of seconds elapsed between the time reference datum of the mission (00:00 hrs Moscow time on January 1$^{\rm st}$, 2000) and a given later date, leap seconds should be added. Five leap seconds occurred between the reference datum and the start of the mission. However, the pipeline software used to create the DR1 data omitted to add these seconds.
Therefore, when converting UTC times in the {\tt badcamt} (bad time intervals) and {\tt timeoff} (instrumental one second time shifts) calibration components to spacecraft clock a five second shift is introduced, leading to the exclusion of five seconds of good data. In both cases, changes in status (i.e. date entries in the component table) tend to occur when the camera is in an anomalous state, not receiving data \citep[see][Appendix B]{Brunner2022}.  As the cameras are not yet registering photons at these times, not considering leap seconds does not have any other negative consequences (such as inclusion of bad time interval in the data) in this case. These five leap seconds will be properly corrected in the next data release. 

\section{The eRASS1 X-ray catalogues}
\label{sec:catalog}

Following the approach devised for the Performance Verification eFEDS survey \citep{Brunner2022}, we present here two distinct X-ray catalogues: 
a catalogue of sources detected in the 0.2--2.3 keV band (selected from the 1B detection process; {\it Main} catalogue) and a catalogue of sources detected in the 2.3--5 keV band (selected from the 3B detection process; {\it Hard} catalogue). 

We generate the single-band 1B catalogue including sources down to a low detection likelihood (\DETLIKE{0} $\geqslant 5$), to maximise completeness. We then make use of the eRASS1 {\it digital twin} simulations \citep{Comparat2019, Comparat2020, Seppi2022} to estimate the amount of spurious detections as a function of the detection likelihood threshold \citep[see also][]{LiuT2022}. Based on these all-sky survey simulations, we define our {\it Main} sample as the one comprising all extended sources and all point sources with  \DETLIKE{0} $\geqslant 6$. Table 3 of \citet{Seppi2022} indicates that the Main catalogue should contain $\approx 14$\% spurious detections. This reduces to about 1\% for \DETLIKE{0} $>10$. Point-like sources with $5\leqslant$ \DETLIKE{0} $<6$ are released as a (highly contaminated) {\it Supplementary} catalogue. 

To extract the hard sample from the 3B catalogue, we apply a threshold for the detection likelihood in the 2.3--5 keV band of \DETLIKE{3} $\geqslant 12$. The threshold is higher than the one applied for the Main catalogue as the lower sensitivity and higher background in that energy range significantly increases the number of spurious detection at a given detection likelihood \citep{LiuT2022}. Based on the same simulations described in \citet{Seppi2022}, we estimate for this threshold a spurious sources fraction of about 8-10\% in the hard-band selected sample. We note here that, 
as shown in \citet{Seppi2022}, the expected fraction of spurious sources depends on the exposure, with lower fraction of spurious detections predicted for higher exposures. Here, for simplicity, we have adopted all-sky average estimates.

\begin{table}
    \caption{Basic eRASS1 catalogue properties.}
    \centering
    \begin{tabular}{lccr}
    \hline\hline
    \multicolumn{4}{c}{1B detection [0.2--2.3 keV]} \\
    \hline
            Catalogue & \DETLIKE{0} & \EXTLIKE & \# of Sources\\
    \hline
    All & $\geqslant$5 & $\geqslant$0 & 1277477 \\
        Main, PS  & $\geqslant$6 & $=$0 & 903521 \\
        Main, Ext.  & $\geqslant$5  & $>$0 & 26682 \\
        Supplementary & $<$6 & $=$0  & 347274 \\
        
        \hline\hline
                \multicolumn{4}{c}{3B detection [2.3--5 keV only]} \\
                \hline
        Catalogue & \DETLIKE{3} & \EXTLIKE & \# of Sources\\
        \hline
        Hard  & $\geqslant$12 &  $\geqslant$0 & 5466 \\
        Hard, PS & $\geqslant$12 &  $=$0 & 5087 \\
        Hard, Ext. & $\geqslant$12 &  $>$0 & 379 \\
        \hline
    \end{tabular}
    \tablefoot{`PS' indicates point sources (i.e. those with \EXTLIKE $=0$), and `Ext.' indicates extent-selected sources with \EXTLIKE $>0$. We note here that there is a slight difference in \EXTLIKE parameter in the 1B and 3B detections, as for the latter all photons in the 0.2--5 keV band are used to evaluate the source extent.}
    \label{tab:selection}
\end{table}

\begin{figure*}[hptb]
\begin{center}
\begin{tabular}{cc}
\includegraphics[width=\columnwidth]{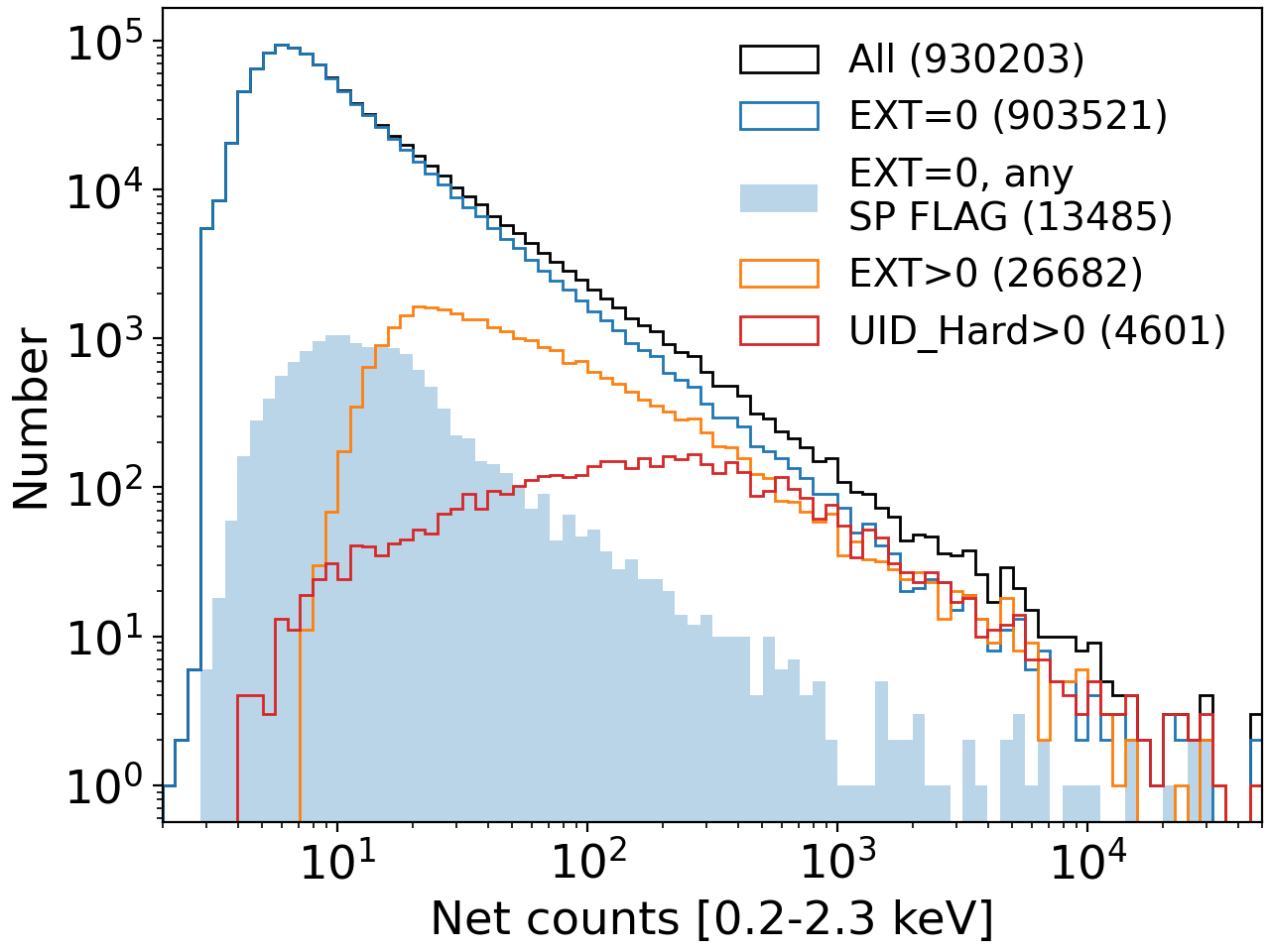}
\includegraphics[width=\columnwidth]{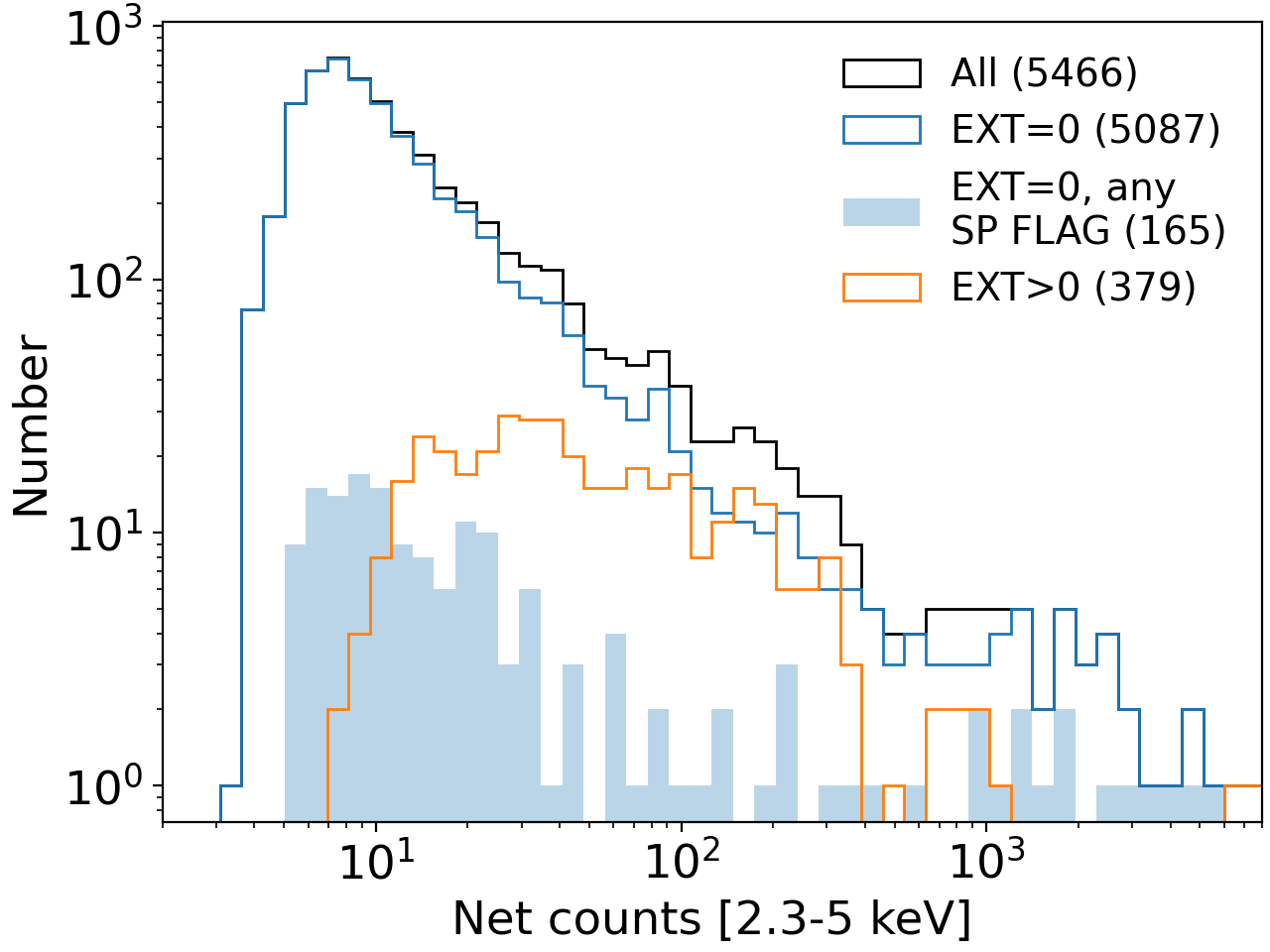} \\
\end{tabular}
\caption{Distributions of net counts for the Main (0.2--2.3 keV; {\it Left panel}) and Hard (2.3--5 keV; {\it Right panel}) catalogues. Point sources (\EXT = 0) and extended sources (\EXT > 0) are plotted in blue and orange, respectively. Point sources with any SP flag (see Table~\ref{tab:source_flag_summary}) are displayed as blue shaded histograms. The sources that appear in both the Main and the Hard catalogues are plotted in red in the left panel.}
\label{fig:hist}
\end{center}
\end{figure*}

Table~\ref{tab:selection} presents a summary of the catalogues selection criteria and properties. In Fig.~\ref{fig:hist} we show the distributions of  net counts for the {\it Main} and the {\it Hard} samples, respectively.

A summary of the content of each catalogue is presented in the Appendix~\ref{sec:column_description} (column descriptions, units). Below we describe in greater detail the catalogue creation procedure, the astrometric verification steps and our attempt to flag potential spurious sources.

\subsection{Catalogue creation and preliminary astrometric correction}
\label{subsection:prelim_astrom}

The catalogues resulting from the PSF fitting with task {\tt ermldet} were re-formatted
using the eSASS task {\tt catprep} and then merged into hemisphere catalogues. 
Since the survey sky tiles overlap each other, the sources outside the nominal, non-overlapping area were removed from each sky tile catalogue. In order to avoid the loss of significant sources
detected by chance just outside the nominal areas in two adjacent tiles,
detections in a $\pm 30\arcsec$ strip near the nominal borders were matched, using a matching radius of $r=15\arcsec$  for point sources and $r=30\arcsec$ for extended sources, and only one detection for each match was kept in the merged catalogue. These matching radii approximately correspond to  half of the HEW and one HEW  of the survey averaged PSF (see Fig. \ref{fig:psf}).

As we discussed above, the eROSITA field of view and survey scanning strategy imply that a source near the ecliptic equator is visited about six times  within a time span of one day in each of the eROSITA six-months surveys. The time span and the number of visits increase with the ecliptic latitude of the source.
In the catalogues, the epoch of survey coverage was estimated for each source by using the attitude time series for camera TM1 to calculate the times of the observation closest to the optical axis as well as the first and last appearance of the source in the camera field of view.
These epochs are listed in the columns \texttt{MJD}, \texttt{MJD\_MIN}, and \texttt{MJD\_MAX}.

The positional uncertainty of X-ray sources is an important parameter for their association with multi-wavelength counterparts, especially given the relatively large PSF of eROSITA. The \texttt{ermldet} task provides a statistical estimate of this quantity for individual X-ray sources based on the spatial distribution of their X-ray photons and a PSF model \citep[see][Appendix A.5, for more details]{Brunner2022}. 
These measurements may underestimate the true positional errors because of, e.g., calibration uncertainties and other systematic effects \citep[e.g.][]{Webb2020}. In particular, the astrometric accuracy of the eROSITA all-sky catalogues can be affected by the following systematic factors:
\begin{itemize}
    \item Errors in the timing between attitude measurements and event arrival;
    \item Boresight calibration errors;
    \item Systematic errors introduced by the PSF fitting.  
\end{itemize}
As for the latter point, eRASS1 data, just like those of the CalPV phase, have been analysed using the PSF model derived from on-ground calibration. The analysis presented in Appendix~\ref{appendix:psf} demonstrate that this effect is negligible, so we discuss only the former two below.

Due to the \SRG scanning geometry, a timing mismatch between the attitude measurements 
and event timing would result in an offset along the scanning direction, i.e., ecliptic latitude $\beta$. 
An offset in the boresight between the \SRG attitude solution and the eROSITA cameras would result in a  constant astrometric offset in ecliptic coordinates during a 180\degr\ scan between the two ecliptic poles.
Assuming that any timing or boresight offsets vary only slowly with time, in order to correct for this effect we
divided the merged catalogue into stripes of 1\degr\ width in ecliptic longitude $\lambda$, corresponding to $\sim$1 day of survey scanning. For each stripe the  X-ray positions of point-like detections at ecliptic latitudes ($-60$\degr\ to $+60$\degr)  were  matched with mid-infrared counterparts from the AllWISE catalogue \citep{Cutri2014}.
After applying a colour cut ($0.3\,\mathrm{mag}<W1-W2<1.7\,\mathrm{mag}$ and
$2.2\,\mathrm{mag}<W2-W3<3.6\,\mathrm{mag}$) to select for likely QSO matches, median values 
of the offsets $\beta_X-\beta_{\rm IR}$ and $(\lambda_{\rm X}-\lambda_{\rm IR})\times \mathrm{cos}(\beta)$ were 
calculated. All X-ray positions within each latitude stripe were then corrected using the median offsets in ecliptic longitude and latitude ($\Delta \lambda \times \mathrm{cos}(\beta)$, $\Delta \beta$). The applied offsets range between 
 $ -4.0\arcsec\; \mathrm{and} +3.9\arcsec$ and
$ -1.5\arcsec\; \mathrm{and} +2.5\arcsec$, respectively.

The resulting statistical errors are given in each coordinate as upper and lower 
bounds (columns \texttt{RA\_LOWERR}, \texttt{RA\_UPERR}, \texttt{DEC\_LOWERR}, \texttt{DEC\_UPERR}).  
An error estimate averaged over both dimensions and directions is given as
\begin{equation}\label{eq:radecerr}
    \begin{split}   
      \sigma_{\mathrm{RA}}  & = (\mathrm{RA\_LOWERR} + \mathrm{RA\_UPERR})/2 \\
       \sigma_{\mathrm{DEC}} & = (\mathrm{DEC\_LOWERR} + \mathrm{DEC\_UPERR})/2 \\
       \mathrm{RADEC\_ERR} & = \sqrt{(\sigma_{\mathrm{RA}})^2 + (\sigma_{\mathrm{DEC}})^2}\,,
    \end{split}
\end{equation}
in line with other X-ray catalogues \citep[e.g.][]{Webb2020}. 

\begin{figure*}[hptb]
\centering
\includegraphics[width=\columnwidth]{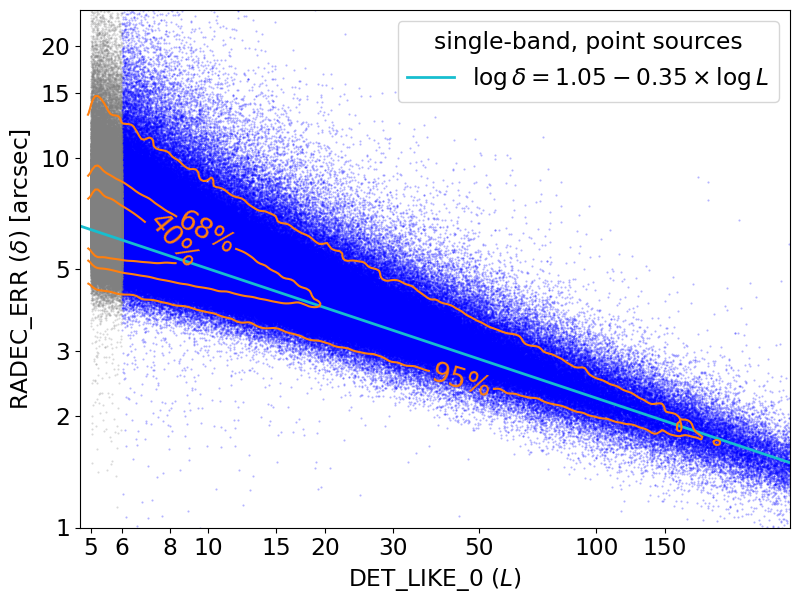}
\includegraphics[width=\columnwidth]{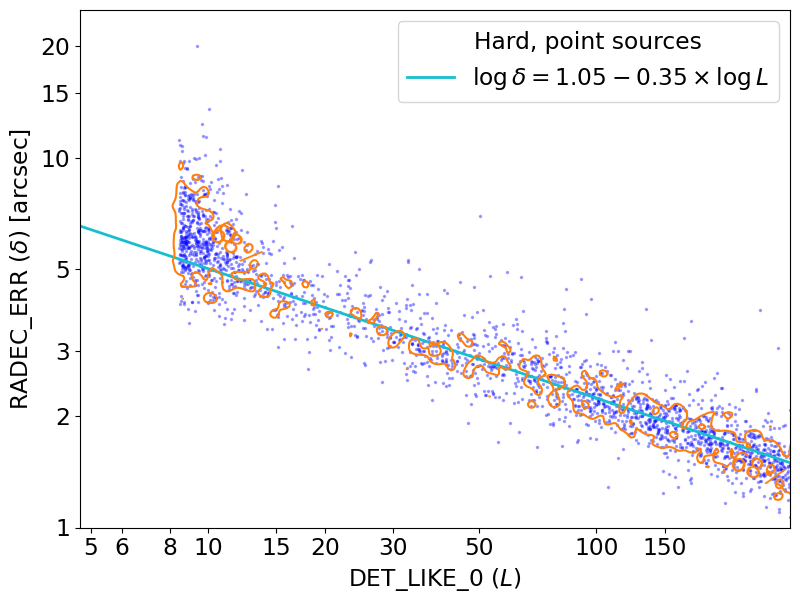}
\includegraphics[width=\columnwidth]{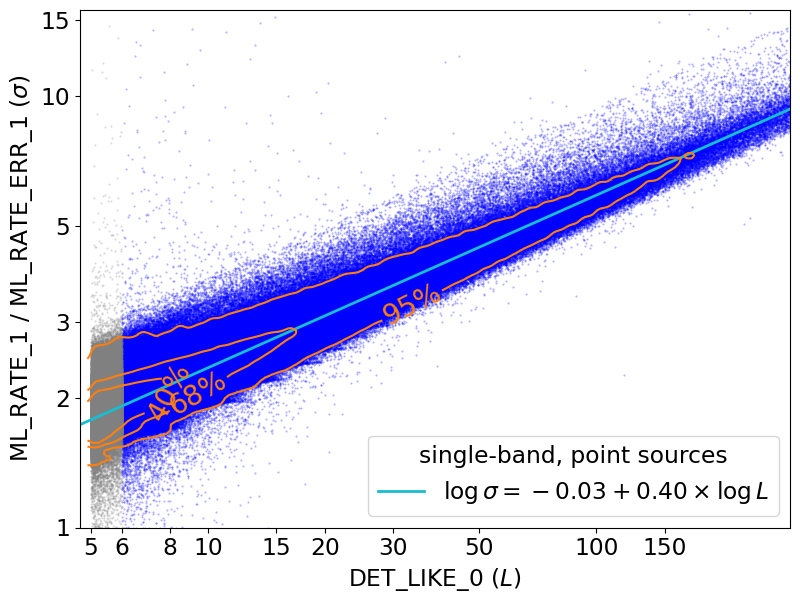}
\includegraphics[width=\columnwidth]{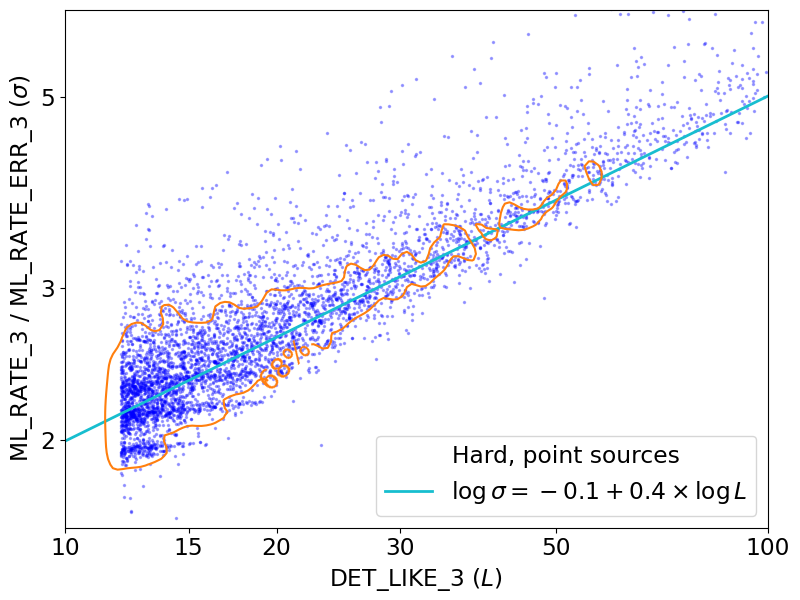}
\caption{Distributions of \texttt{RADEC\_ERR} (upper panels) and flux measurement significance (in terms of \texttt{ML\_RATE\_1}/\texttt{ML\_RATE\_ERR\_1}, lower panels) as a function of detection likelihood for point sources in the eRASS1 single-band detected catalogue (left panels) and in the {\it Hard} catalogue (right panels). In the left panels, the 40\%, 68\%, and 95\% contours are plotted in orange, while the cyan line indicates an empirical correlation that describes the mode of the distribution, as reported in the label. Sources from the Main catalogue are shown in blue, those from the Supplementary catalogue in grey. 
The right panels ({\it Hard} catalogue) have only the 68\% contour plotted. Note that the lower right panel displays the likelihood and flux significance in the 2.3--5~keV band.
\label{fig:DET_LIKE_RADEC_ERR}}
\end{figure*}

The upper panels of Figure~\ref{fig:DET_LIKE_RADEC_ERR} displays the distribution of \texttt{RADEC\_ERR} as a function of \DETLIKE{0} for the point sources in the {\it Main} and {\it Hard} catalogues, respectively.
For sources where the calculation of \texttt{RADEC\_ERR} failed (see Sect.~\ref{subsection:proc_issues}), we calculate \texttt{RADEC\_ERR} using the empirical correlation extracted from Fig.~\ref{fig:DET_LIKE_RADEC_ERR}. It should be noted here that \texttt{RADEC\_ERR} does not represent
the 68\% error radius for two parameters. Under the assumption of a circular error region, the averaged 1-dimensional 68\% error as required e.g. for the comparison with a Rayleigh distribution can be derived with
$\sigma = \mathrm{RADEC\_ERR}/\sqrt{2}$. We further elaborate on the astrometric accuracy of the eRASS1 catalogue in Sect.~\ref{subsec:astrometric_ext}, where we present a validation method based on a comparison with external catalogues, which reveals the extent of the systematic uncertainty beyond the statistical one described here.

{The lower panels of Fig.~\ref{fig:DET_LIKE_RADEC_ERR} display the distribution of flux significance as a function of detection likelihood for the point sources in the main and hard catalogues. For the large majority of the sources, the flux measurements have large uncertainties; in order to have at least 3-$\sigma$ flux measurements in the Main catalog, one could adopt an approximate threshold of \DETLIKE{0}$>$20.

\subsection{Flagging of problematic sources}\label{sec:spurious_flagging}

\begin{table*}[htbp]
    \caption{Spurious and problematic source flag description.}
    \label{tab:spurious_source_flag_descriptions}
    \centering
    %\begin{tabular}{p{3cm}|p{13cm}}
    \begin{tabular}{ll}
        \hline\hline
          Flag name & Description \\
        \hline
         \texttt{FLAG\_SP\_SNR} & Source may lie within an overdense region near a supernova remnant.\\
         \texttt{FLAG\_SP\_BPS} & Source may lie within an overdense region near a bright point source. \\
         \texttt{FLAG\_SP\_SCL} & Source may lie within an overdense region near a stellar cluster.\\
         \texttt{FLAG\_SP\_LGA} & Source may lie within an overdense region near a local large galaxy.\\
         \texttt{FLAG\_SP\_GC\_CONS} & Source may lie within an overdense region near a galaxy cluster (Sect.~\ref{sec:spurious_gal_cluster}).  \\
         \texttt{FLAG\_NO\_RADEC\_ERR} & Source contained no \texttt{RA\_DEC\_ERR} in the pre-processed version of the eSASS catalogue. \\
         \texttt{FLAG\_NO\_CTS\_ERR} & Source contained no \texttt{ML\_CTS\_ERR\_1} in the pre-processed version of the eSASS catalogue. \\
         \texttt{FLAG\_NO\_EXT\_ERR} & Source contained no \texttt{EXT\_ERR} in the pre-processed version of the eSASS catalogue. \\
        \hline
    \end{tabular}
    \label{tab:flags}
\end{table*}

The imperfect nature of the source detection process inevitably leads to contamination of the eRASS1 catalogue by spurious sources and/or inaccuracies in the derived source properties. The most clearly identifiable examples of spurious detections can be found within the vicinity of extremely bright X-ray point sources, such as Sco~X--1, or bright, large extended sources, like supernova remnants, nearby galaxies, or galaxy clusters (see Fig.~\ref{fig:erass1_spurious}), whereas less clear-cut cases can be found at the lowest detection likelihoods, and their contribution to the eRASS1 catalogue quantified via simulations. Optical loading of the CCDs could also introduce fake X-ray sources in the catalogue.

To warn users of a potential spurious origin for a detection, despite a possibly high-detection likelihood, we have flagged sources that are located within overdensities in the eRASS1 source catalogue associated with systems that could create problems for the automatic background estimation in the detection pipeline (supernova remnants, extremely bright X-ray point sources, Galactic star clusters, local galaxies, and galaxy clusters). We also flag catalogue entries matched to very bright optical stars, as we describe below.

\subsubsection{Identification of overdense regions}\label{sec:spurious_identification_overdense} 
In order to identify those regions on the sky where many potentially spurious sources are clustered, we performed an empirical search for regions with a suspiciously large number of detected sources compared to their surroundings:
after performing a uniform cut on the detected flux at  $F_{\rm 0.2-2.3~keV} > 5\times10^{-14}$\,erg\,s$^{-1}$\,cm$^{-2}$, to reduce dependence on the spatially varying exposure, we computed a density map of point-like and extended sources in the single-band catalogue, using a pixelisation of $0.25\,\mathrm{deg^2}$. A ``background'' source density map was then created by applying a median filter with a radius of 10\degr. By comparing the two maps, we identified all regions with a local source density more than twice the background. 
The exact shape of the overdensities was extracted by ``zooming in'' on each identified region, creating a smoothed histogram of the local source distribution, and selecting all regions with a density larger than three times the local background which contribute more than $20$ excess sources. 

While this procedure yields many overdensities caused by truly spurious source excesses, some resulting regions are expected to correspond to accumulations of real astrophysical sources. We thus manually identified the correspondence of each of the $\sim$80 localised overdensities to astrophysical sources, using the SIMBAD database \citep{Wenger2000}. Our overdense regions were then classified, and the enclosed sources flagged, according to their correspondence to 
i) diffuse emission associated with known supernova remnants (\texttt{FLAG\_SP\_SNR}), ii) excess emission in the vicinity of extremely bright point sources (\texttt{FLAG\_SP\_BPS}), 
iii) Galactic star clusters (\texttt{FLAG\_SP\_SCL}), iv) nearby galaxies (\texttt{FLAG\_SP\_LGA}).

This classification might be useful, for instance, if one were interested in studying the population of Milky Way point sources, as one would likely want to mask spurious sources caused by mis-classified diffuse emission from supernova remnants, but might not want to mask Galactic stellar clusters.  

\begin{figure*}
    \centering
    \includegraphics[scale=0.44]{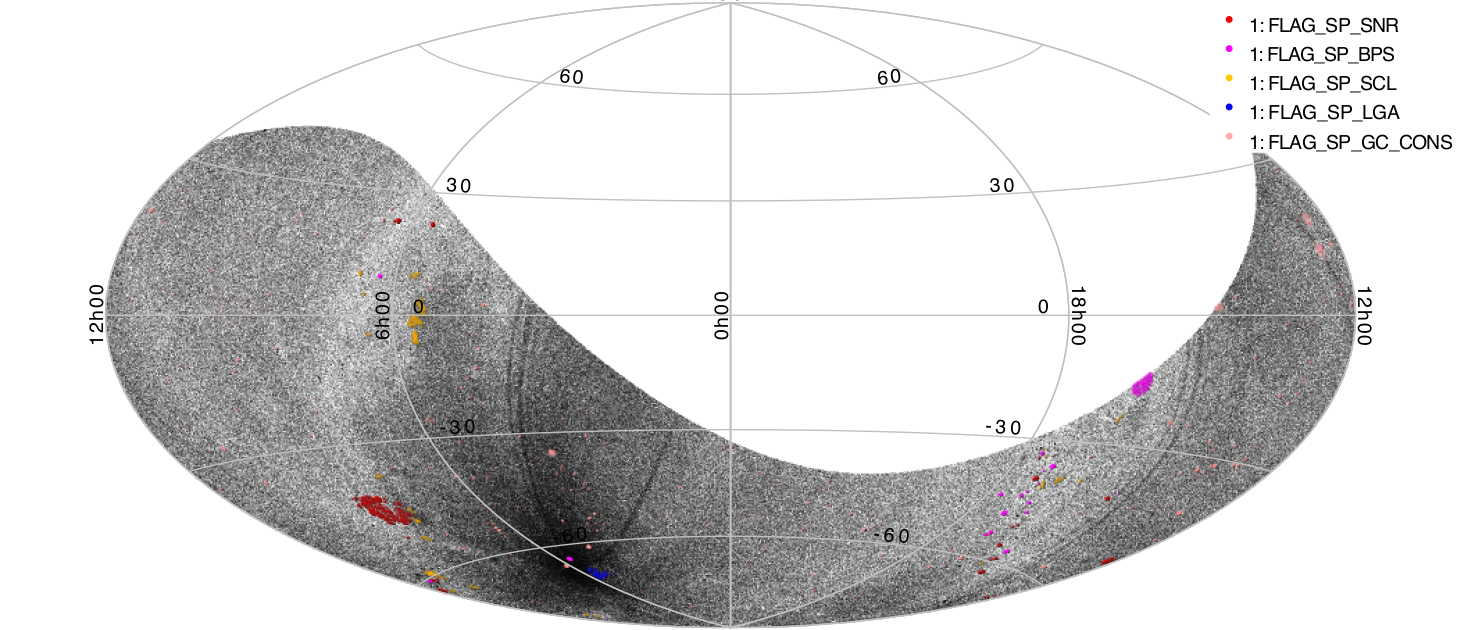}
    \includegraphics[scale=0.4]{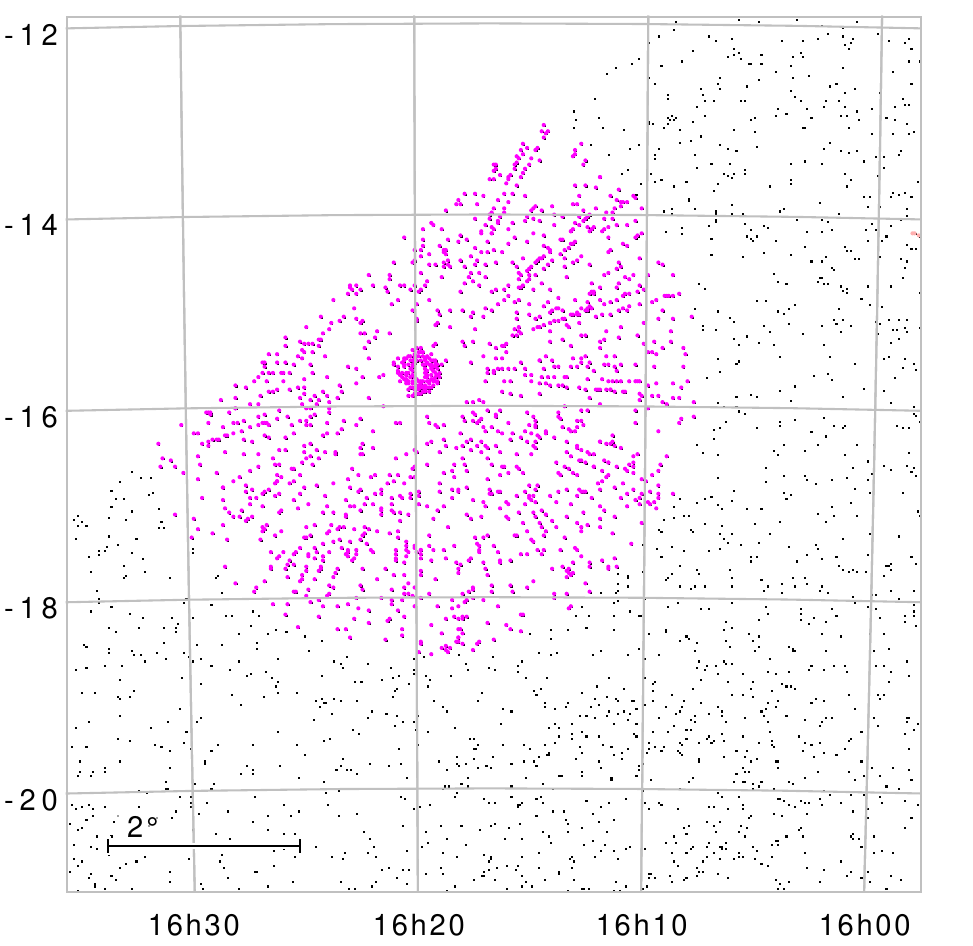}
    \includegraphics[scale=0.4]{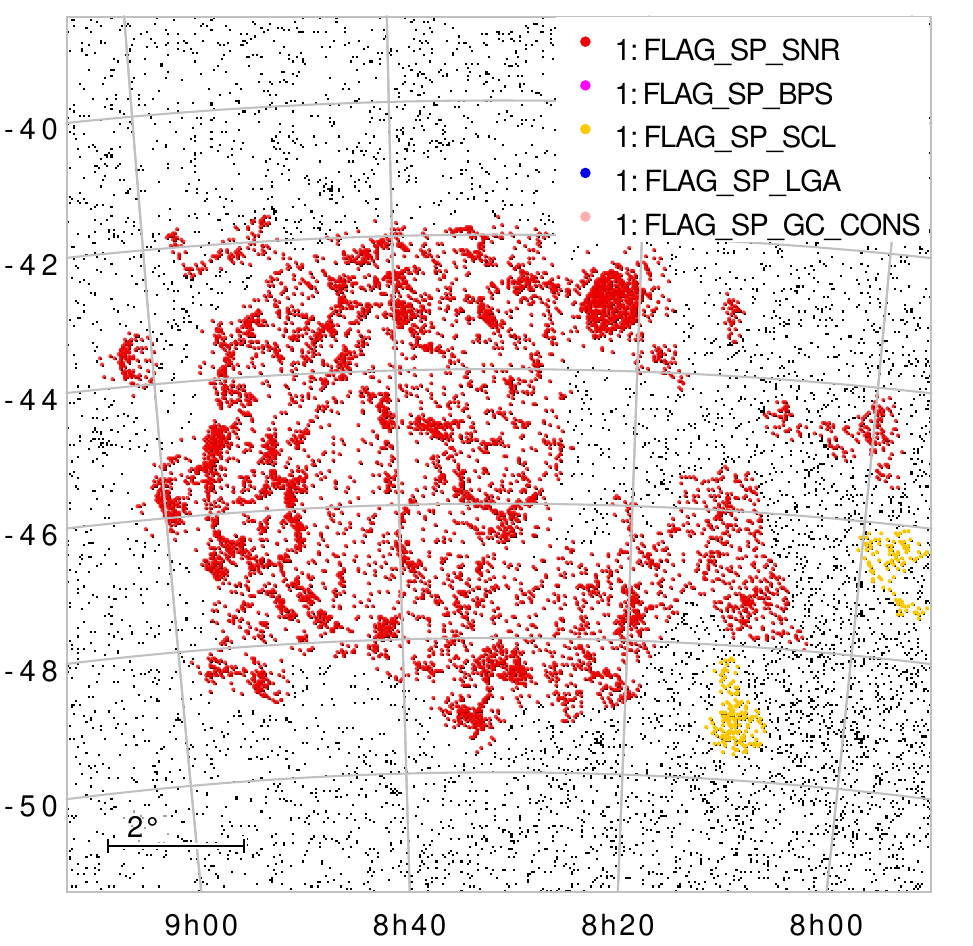}
    \caption{\textit{Top panel}: Aitoff projection of the eRASS1 1B catalogue in Equatorial coordinates (J2000), with each grey point representing a detected source within the catalogue, and the coloured points denoting sources that have been flagged as potentially spurious according to the scheme presented in Table~\ref{tab:spurious_source_flag_descriptions}. Darker stripes are due to larger sources density due to a higher exposure in those parts of the sky. \textit{Bottom left}: Zoom in plot of sources detected within the vicinity of Sco~X--1, with pink sources flagged as potentially spurious using the \texttt{FLAG\_SP\_BPS} column in the catalogue. \textit{Bottom right}: Similar for sources within the vicinity of the Vela SNR.}
    \label{fig:erass1_spurious}
\end{figure*}

\subsubsection{Galaxy cluster catalogues}\label{sec:spurious_gal_cluster}

Another flag is applied for possible spurious sources which are located close to known galaxy clusters. To do that, we make use of published X-ray cluster catalogues, including MCXC \citep{Piffaretti2011}, XXL365 \citep{Adami2018}, XCS \citep{Mehrtens2012}, eFEDS \citep{LiuA2022}, and X-CLASS \citep{Clerc2012}. Sources lying within $R=0.5\times R_{500}$ from the cluster center are flagged as {\tt FLAG\_SP\_GC\_CONS}. 
When $R_{500}$ is not provided in the published catalogue, we adopt a radius of $R=500$\,kpc. We note that, to avoid over-counting, optical cluster catalogues are not used in this step, because large offsets between clusters' X-ray centers and optical centers are frequently observed \citep[see, e.g.,][]{Seppi2022, Seppi2023}. Cluster catalogues selected on the basis of the Sunyaev-Zeldovich (SZ) effect are not included either, due to the relatively large positional uncertainties in SZ surveys. Therefore, the catalogue of known galaxy clusters we used in the above approach is a rather conservative and incomplete compilation, and the identified spurious sources should be considered as a supplement for the identification of overdensities. Further cleaning is performed only for galaxy cluster candidates in the extended source catalogue. An extended source is flagged as a possible spurious detection when it is too close to its neighbour. Visual inspections are also performed on the extended source catalogue to remove any remaining obvious spurious sources and correct for mis-classified cases. We refer the readers to Bulbul et al. (\citeyear{Bulbul2023}) for more details of the cleaning procedure we performed in the extended source catalogue.

\subsubsection{Summary of spurious sources flagging procedure}
The flagging procedure described above identifies cases where high local background levels render the automatic pipeline detection process unreliable. This is illustrated in Fig.~\ref{fig:bkg_hist_flags}, which shows the background rate distributions for sources with and without flags: the flagged sources are mostly found in regions with enhanced background rate.

\begin{figure}[ht]
    \centering
\includegraphics[width=\columnwidth]{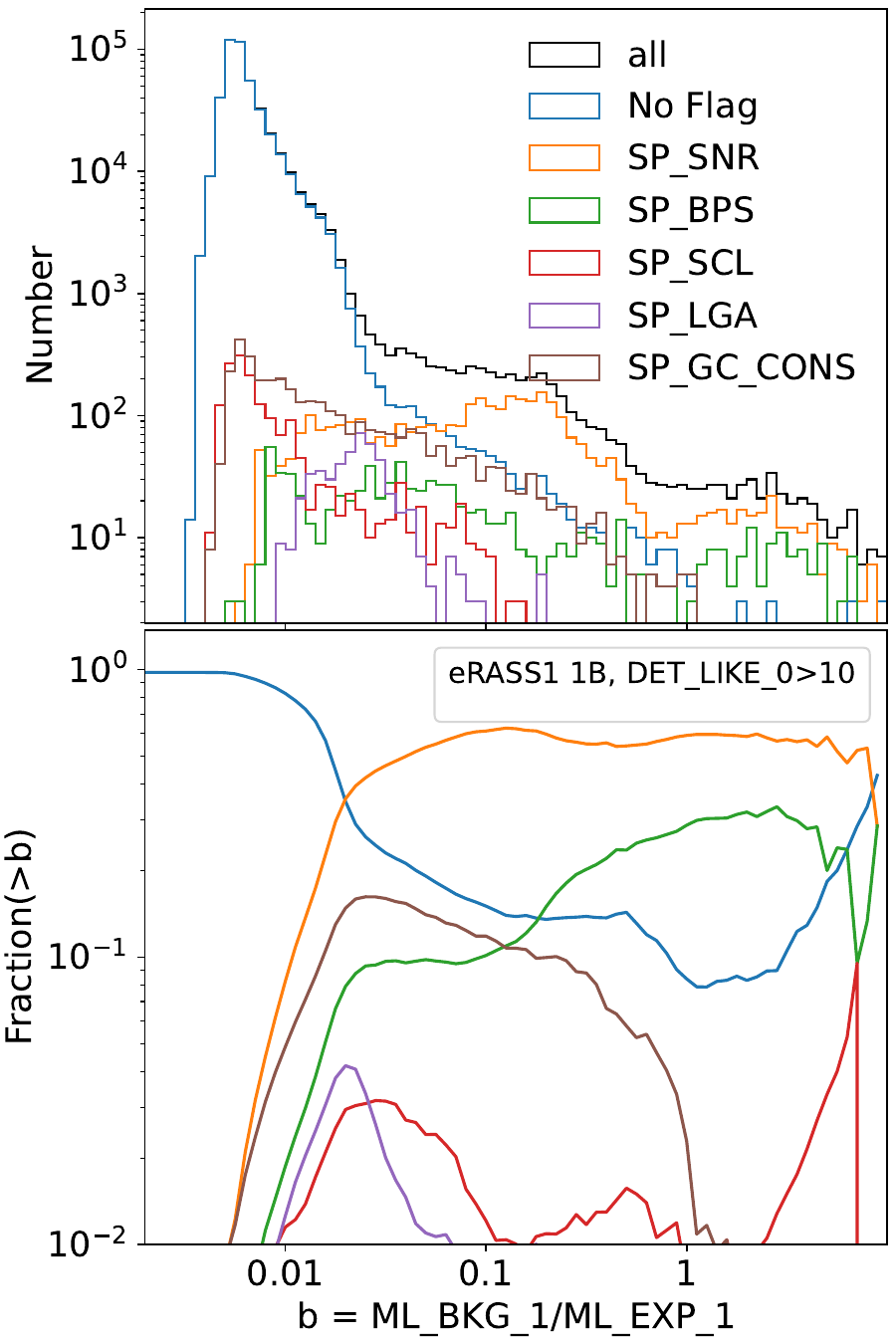}
    \caption{{\it Top panel:} distribution of local background rate level (\texttt{ML\_BKG\_1}/\texttt{ML\_EXP\_1}, in units of counts\ arcmin$^{-2}$\ s$^{-1}$) for sources with and without flags and \DETLIKE{0} $>10$. {\it Bottom panel:} the fraction of each corresponding sub-sample among all the sources above any background level.}
    \label{fig:bkg_hist_flags}
\end{figure}

A summary of the different spurious sources flags added is presented in Table~\ref{tab:spurious_source_flag_descriptions}, while Table~\ref{tab:source_flag_summary} reports the number of sources for each of the flag categories. After removing all those sources that are flagged by at least one of the potential spurious categories, the {\it Main} catalogue contains 890\,036 point sources, with a median sky density of approximately 37 deg$^{-2}$.  

\begin{table}
\caption{Number of sources flagged as potentially spurious.}
\label{tab:source_flag_summary}
\centering
\begin{tabular}{lrrrr}
\hline\hline
  \multicolumn{1}{l}{Flag} &
  \multicolumn{1}{c}{{\it M} PS} & 
  \multicolumn{1}{c}{{\it M} Ext.} &
  \multicolumn{1}{c}{{\it H} PS} & 
  \multicolumn{1}{c}{{\it H} Ext.} \\
  \hline
  SP\_SNR & 4381 & 2710 & 35 & 23 \\
  SP\_BPS & 1503 & 568 & 39 & 61 \\
  SP\_SCL & 2443 & 260 & 40 & 4 \\
  SP\_LGA & 465 & 413 & 20 & 2\\
  SP\_GC\_CONS & 4707 & 2143 & 34 & 155 \\
\hline
  Any SP Flag & 13485 & 6084 & 165 & 243\\
\hline\hline
  NO\_RADEC\_ERR & 2606 & 1173 & 15 & 20 \\
  NO\_EXT\_ERR & 8469 & 744 & 0 & 19 \\
  NO\_CTS\_ERR & 2532 & 746 & 9 & 0 \\
\hline
\end{tabular}
\tablefoot{`PS' stands for Point Sources, `Ext.' for extent-selected, and we separate the {\it Main} (`{\it M}') and the {\it Hard} (`{\it H}') catalogues. The `Any SP Flag' row indicates the number of sources that are flagged by any of the five identifiers of potential spurious sources in overdense regions, \texttt{SP\_SNR}, \texttt{SP\_BPS}, \texttt{SP\_SCL}, \texttt{SP\_LGA}, and \texttt{SP\_GC\_CONS}. The last three lines give the number of sources for which statistical error estimate from the pipeline failed (see Sect.~\ref{subsection:proc_issues} for more details).}
\end{table}

A note of caution is in place here: although steps are taken here to greatly reduce the number of high-detection-likelihood contaminants, it is still recommended that users double-check the relevant science images for their sources of interest before publication (e.g., if their point sources lies within a likely galaxy cluster but is not flagged as spurious here).

\subsubsection{Optical loading}
\label{subsec:optical_loading}

The eROSITA detectors are prone to optical loading, i.e. the accumulation of low energy (optical/UV) photons within a CCD pixel over the frame integration time of 50~ms, whereas the relevance of the effect is governed by the optical brightness and detector position of the respective source. If the summed energy exceeds the X-ray detection threshold, any optically bright source will start to generate false X-ray events. Due to the nature of the effect, these events appear predominantly at lower X-ray energies. The strength of the optical loading signal strongly increases with increasing source brightness; however, for very bright sources `saturation-like' effects may occur, when events are removed by the pattern filtering procedure. The optical loading signal further depends on the off-axis angle of the source, as the optical PSF is sharper close to FOV center and photons are focussed on a smaller detector area. i.e. fewer pixels. During eRASS data taking, any optically bright source passes several times on different scan paths through the FOVs and thereby a time dependent optical loading signal may be generated. If a sufficient number of X-ray events is created, source detection triggers and the object makes it into the eRASS catalog. Due to the required intrinsic brightness, stars are by far the main contributors to optical loading sources in our data.

Among the practical consequences for astrophysical studies are the presence of fake X-ray sources, pseudo X-ray variability and the distortion of real X-ray sources, as the source may of course be optically bright and also an intrinsic X-ray emitter. If the optically bright source is X-ray dark, or at least faint enough to fall well below the eRASS sensitivity limits, all the registered events would come  from optical loading. These fake X-ray sources always have very soft spectra and show pseudo variability. If the optically bright sources are also X-ray bright, these characteristics are likewise present, but the detected signal is a mixture of true X-ray photons plus optical loading events and contamination effects. Any potential disentangling between these effects depends on the individual source properties and on the detailed science objectives, but in general the X-ray properties of sources contaminated by optical loading are uncertain.

To best characterise the optical loading effect, X-ray dark sources are required and two suitable stellar populations are used here: main-sequence and mildly evolved stars with spectral types late-B to mid-A and red giants, i.e. K/M stars with luminosity class III-I. Cross-matching these with the eRASS1 catalogs shows that optical loading is expected to affect DR1 sources if they exceed certain brightness thresholds, and likely affected X-ray sources are flagged in the catalog. The adopted brightness limits are B, V, G $\le 4.5$~mag and J~$\le 3$~mag; if one or more criteria are fulfilled, then the source is tagged with {\tt FLAG\_OPT}. In total, 750 (17, 14) sources are flagged as potentially contaminated by optical loading in the {\it Main} ({\it Supplementary}, {\it Hard}) catalogues, respectively.
Input for cross-matching are the Tycho-2 \citep{2000A&A...355L..27H}, 2MASS \citep{2003tmc..book.....C}, Gaia DR3 \citep{2023A&A...674A...1G} catalogues plus Simbad database. Where proper motion information is available, optical catalog entries are updated to epoch 2020 positions,  and a uniform matching radius of 15\arcsec\ is used.
A full treatment depends on the individual source properties and is beyond the scope of this work, but users should be aware that specific attention is required when dealing with eROSITA data of brighter stars.

\subsection{Association of soft and hard band selected sources}
\label{sec:1B3B}

As discussed above, we performed source detection with two different settings: a single X-ray band (1B, 0.2--2.3~keV) and a three-band detection (3B, 0.2--0.6, 0.6--2.3, and 2.3--5~keV), 
subsequently down-selected based on the 2.3--5 keV band significance. 
These produce, respectively, soft and hard X-ray selected source samples. The data used in the 1B and 3B detection are nonetheless largely overlapping. The only photons that are included in the 3B detection but not in the 1B are those in the 2.3--5~keV band, where the instrumental effective area is relatively low and the background relatively high. For this reason, most of the {\it Hard} catalogue sources are expected to have a matching entry in the 1B catalogues. This was the case also in the eFEDS survey, where 90\% of hard band sources were found to have a counterpart in the main, soft X-ray selected catalogue (Nandra et al., in prep.). It is nonetheless of particular interest to identify those sources that are only detected in the hard band, as they signpost objects with extremely hard spectra, most obviously due to heavy obscuration. The association between sources in the two catalogues is not straightforward, however. 
The complexity arises due to various factors, such as the positional uncertainties, the morphological classification (extent measurement) and blending with nearby objects.
In this section, through a specific matching procedure between the 1B and 3B sources, we provide a ``weak'' association that can be used to select sources only detected in the hard band and a ``strong'' association that can be used to select entries in the X-ray catalogues with a high degree of confidence that the X-rays originate from the same astrophysical source.

The weak association is based only on positional information. Since for extended sources the value of the extent parameter \EXT (i.e. the best-fit core size of the beta model fitted to an extended source) is very broadly distributed, ranging from $\sim$10\arcsec\ to the maximum allowed value of 60\arcsec, to search for a counterpart to those sources we adopt a large matching radius of four times the \EXT value. For point sources (\EXT $=0$), we simply adopt a matching radius of 10\arcsec, which is 99 percentile of the point source RADEC\_ERR in the main catalog. For each 1B source, we search for 3B sources within its matching radius, and for each 3B source, we search for 1B sources within its matching radius. Then we merge the results, which is equivalent to adopting the larger radius among the two. 
As a result of such loose association criteria, one source could be associated with multiple sources, some of which are false matches and some of which have different morphological classifications. To classify these matches, we define P2P, E2E, P2E, and E2P matches to denote pairs of point source (P) and extended sources (E); the former and the latter letters (P or E) indicate the classifications in the 1B and 3B catalogues, respectively.
Hard band sources which do not have any counterpart in the soft band catalogues within the match radius are designated as hard band only sources.

The ``strong'' associations require more strict criteria.
First, we require that each pair of sources have the same morphological classification, i.e., we adopt only E2E or P2P pairs. 
With E2E associations, one source can be matched to multiple ones. We adopt only the nearest match in such cases, so that the involved sources are unique.
In the cases of P2P matches, one source is never matched to multiple counterparts, 
but in addition to a maximum separation of 10\arcsec, we further require 
i) $\mathrm{separation}/\mathrm{error}<3$, where error is the larger of the two \texttt{RADEC\_ERR}, and ii) the 1B and 3B measured 0.2--2.3~keV (combining band 1 and 2 in the 3B case) source counts cannot differ by $>50\%$. In this way, we enforce not only that the matched sources are sufficiently close in sky position, but that also have a consistent broadband brightness, such that sources with the same position but different extents (caused by de-blending uncertainties in crowded regions) can be excluded.
We store the counterpart unique source ID (\texttt{UID}) in the \texttt{UID\_Hard} column of the {\it Main} and {\it Supplementary} catalogues and the \texttt{UID\_1B} column of the Hard catalogue as indicators of strongly associated sources. 

For the weak associations, we also store the counterpart \texttt{UID} in the \texttt{UID\_Hard} or \texttt{UID\_1B} columns but multiplying it by $-1$. In this way, catalogue columns \texttt{UID\_Hard}$>0$ or \texttt{UID\_1B}$>0$ indicate that a source has a strong association
in the other catalogue, while catalogue columns \texttt{UID\_Hard}$<0$ or \texttt{UID\_1B}$<0$ indicate that a source may have a (weak-association) counterpart in the other catalogue. 
Finally, \texttt{UID\_1B}$=0$ marks the {\it Hard} sources that have no counterpart within their matching radius in the 1B catalogues. 
Following this procedure, we find $780$  ``Hard-only'' sources; of these, 764 are point-like (\EXT = 0) and 16 extended (\EXT $>0$). Figure~\ref{fig:hard_only_rate_rate} displays the count rate distribution of the hard-band selected sources in the 0.6--2.3 and 2.3--5 keV bands. The Hard-only sources identified above (red points) have significantly higher hardness than the others, by construction.

\begin{table*}[htp]
    \centering
    \caption{Summary of identification of Hard (3B) sources in the 1B (Main and Supplementary) catalogues.} 
    \begin{tabular}{lllllll}
    \hline\hline
    Hard             & Further  & Total & \multicolumn{2}{c}{``Weak'' associations} & ``Strong'' & Hard \\
    \cline{4-5}
classification                  & filtering& number      & same class       & diff. class      & associations & only\\
\hline
\multirow{2}{*}{\texttt{\EXT = 0}} & All      &       5087 & 4 P2P     & 44 E2P      &       4275 P2P & 764\\
\cline{2-7}
                                & FLAG=0   &       4922 & 3 P2P        & 20 E2P        &       4154 P2P & 745 \\
                       \hline
\multirow{2}{*}{\EXT > 0} & All      &        379 & 6 E2E      & 11 P2E      &        346 E2E & 16\\
\cline{2-7}
                                & FLAG=0   &        136 & -          & 11 P2E        &        110 E2E & 15          \\

      \hline
                
    \end{tabular}
    \tablefoot{FLAG=0 selects sources with all the five spurious flags in the Hard catalogue being 0. Here, P2P, E2P, P2E and E2E refer to possible combinations of matched pairs between the 1B and 3B catalogs, based on the morphology of the sources: point-like (P) or extended (E); the former and the latter letters (P or E) indicate the classifications in the 1B and 3B catalogues, respectively. See text for more details.}
    \label{tab:1B3Bmatch}
\end{table*}

The matrix of possible identifications of Hard sources in the 1B catalogues, following our association criteria, is given in Table~\ref{tab:1B3Bmatch}. For point sources, about 84\% have a strong association, 15\% are Hard-only and just 1\% have weak associations in the 1B catalogues. These  fractions do not change if one considers only sources without any spurious flag. Among extended sources in the {\it Hard} catalogue, on the other hand, a larger fraction is flagged as potentially spurious. Of the remaining ones (136 in total), about 80\% have a strong association with a 1B catalogue extended source, about 9\% have a weak association in the 1B catalogues, and 11\% (15 in total) are Hard-only. 
Such Hard-only extended sources are most probably mis-classified point sources; indeed, only 1/15 of these has extent likelihood \EXTLIKE $>6$ and extent \EXT $>30$\arcsec.  

\begin{figure}
    \centering
    \includegraphics[width=\columnwidth]{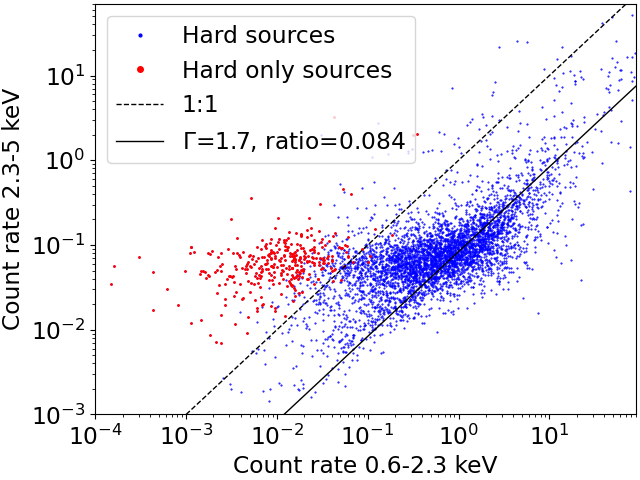}
    \caption{Distributions of 0.6--2.3~keV (x-axis) and 2.3--5~keV (y-axis) count rates of the hard-band selected sources. The "Hard-only" sources are marked in red colour. The solid line indicates the ratio (0.084) that corresponds to an unabsorbed power-law spectrum with a slope of $1.7$. The dashed line indicates a 1:1 ratio.}
    \label{fig:hard_only_rate_rate}
\end{figure}

\begin{figure*}[hptb]
  \centering
  \includegraphics[width=0.45\textwidth]{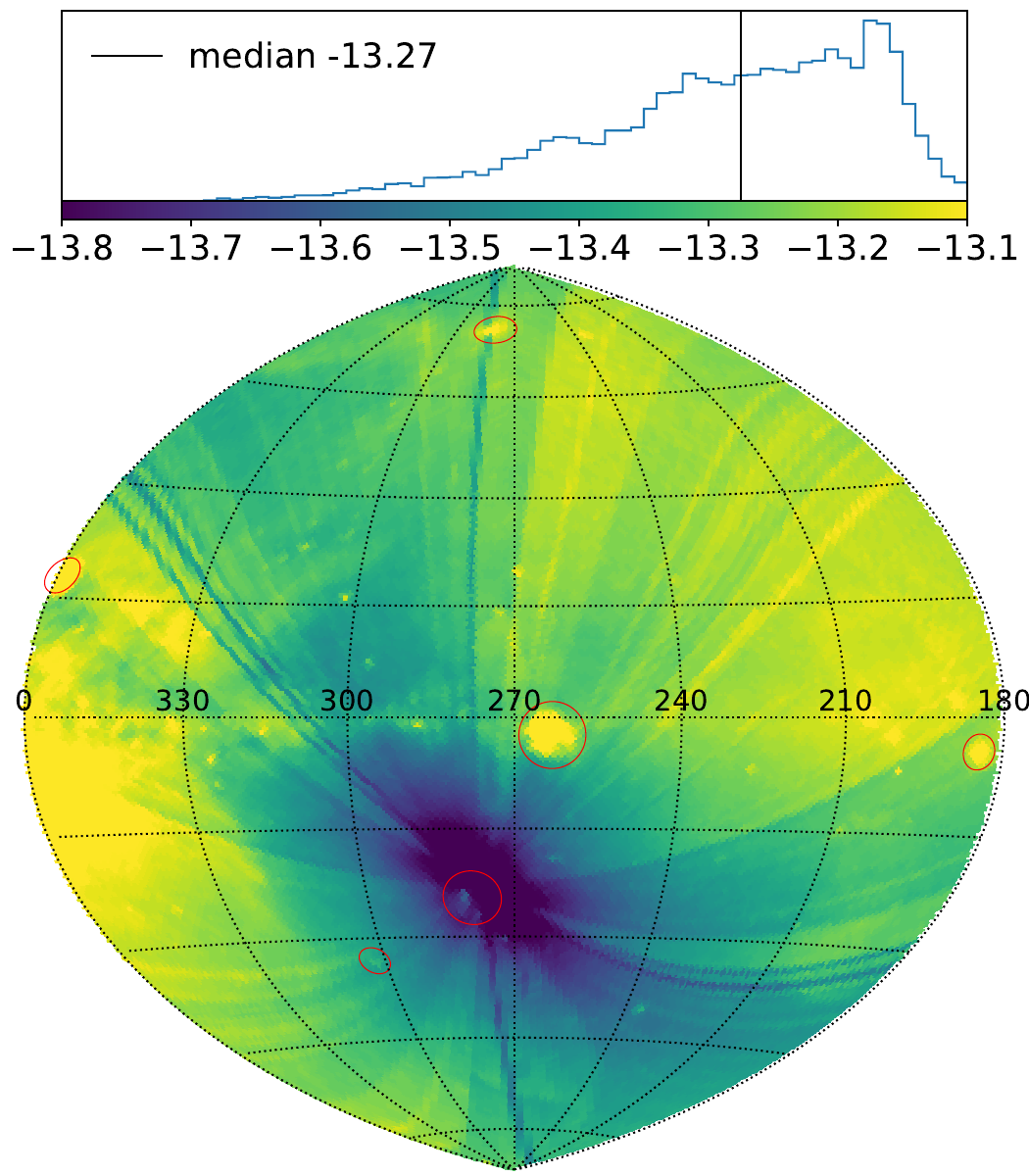}
\includegraphics[width=0.45\textwidth]{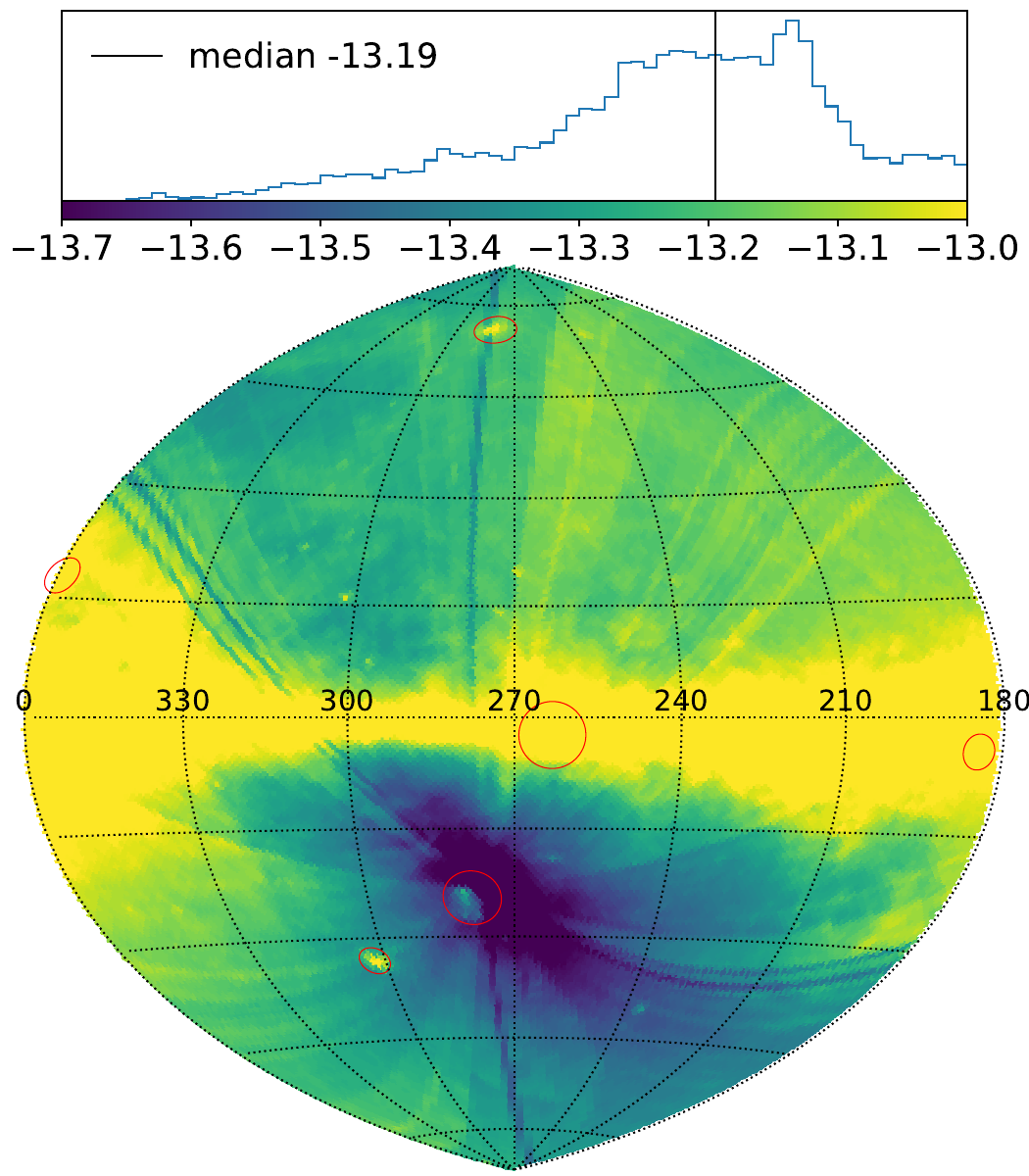}
  \caption{{\it Bottom panels:} Hammer-Aitoff projection maps, in Galactic coordinates, of the logarithm of the 0.5--2~keV flux limit  calculated as the flux at 50\% sky covering fraction (see text for details). The left and right panels correspond to values uncorrected and corrected for Galactic absorption, respectively. Six particular regions where the flux limits are increased by bright X-ray sources are marked with red circles; from the north (top) to the south (bottom) they are: the Virgo cluster, Sco~X--1, the Vela SNR, the Crab pulsar, the LMC, and the SMC. {\it Top panels:} histogram of the logarithm of the flux limit in erg\,s$^{-1}$\,cm$^{-2}$. The colour bar on the X-axis illustrates the intensity scale of the corresponding map in the bottom panels.
    \label{fig:fluxlimitmap}}
\end{figure*}

\subsection{The flux limit and completeness of eRASS1 in the 0.5--2 keV band}
\label{subsec:flux_limit}

To quantify the flux limit of the eRASS1 survey in the commonly adopted energy band 0.5--2~keV, we use the aperture-photometry-based method based on the {\tt apetool} task (see Sect.\,\ref{sec:methodsensmaps_apetool}). A local flux limit can be defined based on the local sensitivity curve as the flux where the probability
of collecting a sufficient number of source photons to constitute a detection at the given threshold 
reaches a particular value, for example 50\%. 
To calculate a sensitivity curve as a function of flux, we need an ECF to convert count rate to flux. Considering the nonuniform Galactic absorption, we assume a power-law spectral model with
$\Gamma = 2$ and the total Galactic absorption column density \citep{Willingale2013} at each source position, and calculate two versions of ECF for each source, one correcting the 0.5--2~keV flux
for absorption and the other not.

Forced photometry results at source positions in the 0.5--1 (band P2) and 1--2~keV (band P3) bands are available in the catalogue and can be combined into the 0.5--2~keV values. We sum the total counts (\texttt{APE\_CTS\_P2} + \texttt{APE\_CTS\_P3}) and background counts (\texttt{APE\_BKG\_P2} + \texttt{APE\_BKG\_P3}) to calculate \texttt{APE\_CTS\_S} and \texttt{APE\_BKG\_S}, where the suffix ``S'' (Soft) indicates the 0.5--2~keV band. To combine the vignetted exposure times in P2 and P3, we measure the weighted average as 
$$\mathrm{APE\_EXP\_S}=(1+w)/\left( \frac{1}{\mathrm{APE\_EXP\_P2}}+\frac{w}{\mathrm{APE\_EXP\_P3}} \right),$$ where $w$ is the relative weight factor, which depends on the ECF adopted to create the exposure maps in the P2, P3 and S bands. By comparing these three maps, we find a proper factor of $w=0.6268$. 
The distribution of sources in the space of background counts and exposure time shows a strong concentration along a linear correlation, which indicates the typical background (\texttt{APE\_BKG\_S}/\texttt{APE\_EXP\_S}$\sim 0.003$ counts/s). There is a small fraction ($\sim 1\%$) of obvious outliers below 0.002 counts/s, which indicates an underestimated background in low S/N regions. We adopt a minimum value of 0.002 counts/s to correct them. 
Having computed \texttt{APE\_CTS\_S}, \texttt{APE\_BKG\_S}, \texttt{APE\_EXP\_S}, and the ECF, we can calculate the Poissonian probability (\texttt{APE\_POIS\_S}) of each source and the sensitivity curve determined by the local background using the Python package ``scipy.special''\footnote{\url{docs.scipy.org/doc/scipy/reference/special.html}}, which provides the same function as in the \texttt{apetool} task. The values of \texttt{APE\_POIS\_S} are reported in the {\it Main} and {\it Supplementary} catalogues, so that a sample with a fixed FAP threshold can be defined by the user as needed.

We here adopt an aperture size for photometry corresponding to a radius that includes 75\% of the PSF photons (i.e.\ encircled energy fraction of 0.75), while the Poisson false detection probability threshold is set to $P_{\rm thresh}=4\times10^{-6}$ \citep[following][]{Georgakakis2008}. To project the flux limits to a map, we subdivide the sky into HEALPix\footnote{\url{https://healpix.sourceforge.io/}} pixels \citep{Zonca2019,Gorski2005}, adopting a HEALPix resolution order of 6. Through Voronoi tessellation, we also pixelise the sky into cells, each of which contains only one source.
In each HEALPix pixel, we average the sensitivity curves of all the sources and weigh each source by its Voronoi cell area.
Adopting the 50\% flux of the averaged sensitivity curve, we obtain a flux limit for each HEALPix pixel and thus a flux limit map of the hemisphere.

We create two distinct maps, as displayed in Fig.~\ref{fig:fluxlimitmap}. The left panel shows the flux limit of a (galactic) X-ray point source without consideration of the galactic absorption. The flux limit map follows the exposure map and the diffuse background map well. The right panel shows the case of an (extragalactic) X-ray source obscured by the total Galactic $N_\textrm{H}$.
In the absorption corrected case, the high $N_\textrm{H}$ in the Galactic plane boosts the flux limit. 
The hemisphere median flux limits are approximately $5\times 10^{-14}$ erg\,s$^{-1}$\,cm$^{-2}$ (before Galactic absorption correction) and $6\times 10^{-14}$ erg\,s$^{-1}$\,cm$^{-2}$ (after Galactic absorption correction), respectively. If we were to increase the probability threshold to 80\%, the corresponding flux limit would increase by about 50\%. 

We provide here some basic characteristics for two examples of sub-sample selection with well-defined statistical properties: a flux-limited one, and one above a fixed Poisson false-probability threshold.
A selection of point sources obeying $F_{\rm 0.5-2 keV} >5\times 10^{-14}$ erg\,s$^{-1}$\,cm$^{-2}$  (observed, i.e. not corrected for absorption) returns a sample with 207\,439 entries (after removing all those that are flagged by at least one of the potential spurious source categories), with a median sky density of approximately 9.8 deg$^{-2}$. 
On the other hand, a selection of point sources obeying \texttt{APE\_POIS\_S}$<4\times 10^{-6}$ returns a sample with 229\,266 entries (after removing all those that are flagged by at least one of the potential spurious source categories), with a median sky density of approximately 8.6 deg$^{-2}$.  

Finally, an estimate of the completeness of the eRASS1 catalogue is provided by the analysis of the detailed `digital twin' eRASS1 simulation presented in \citet{Seppi2022}. There it is shown how the point sources' completeness as a function of 0.5-2 keV flux changes with increasing exposure. Based on that work, Figure\ref{fig:completeness_seppi_pnt} shows the estimated completeness for point sources, expressed as the ratio of detected to simulated AGN from the analysis of the simulations as a function of the net exposure time (for the {\it Main} catalog with \texttt{DET\_LIKE\_0}$>6$).

\begin{figure}
    \centering
    \includegraphics[width=0.98\columnwidth]{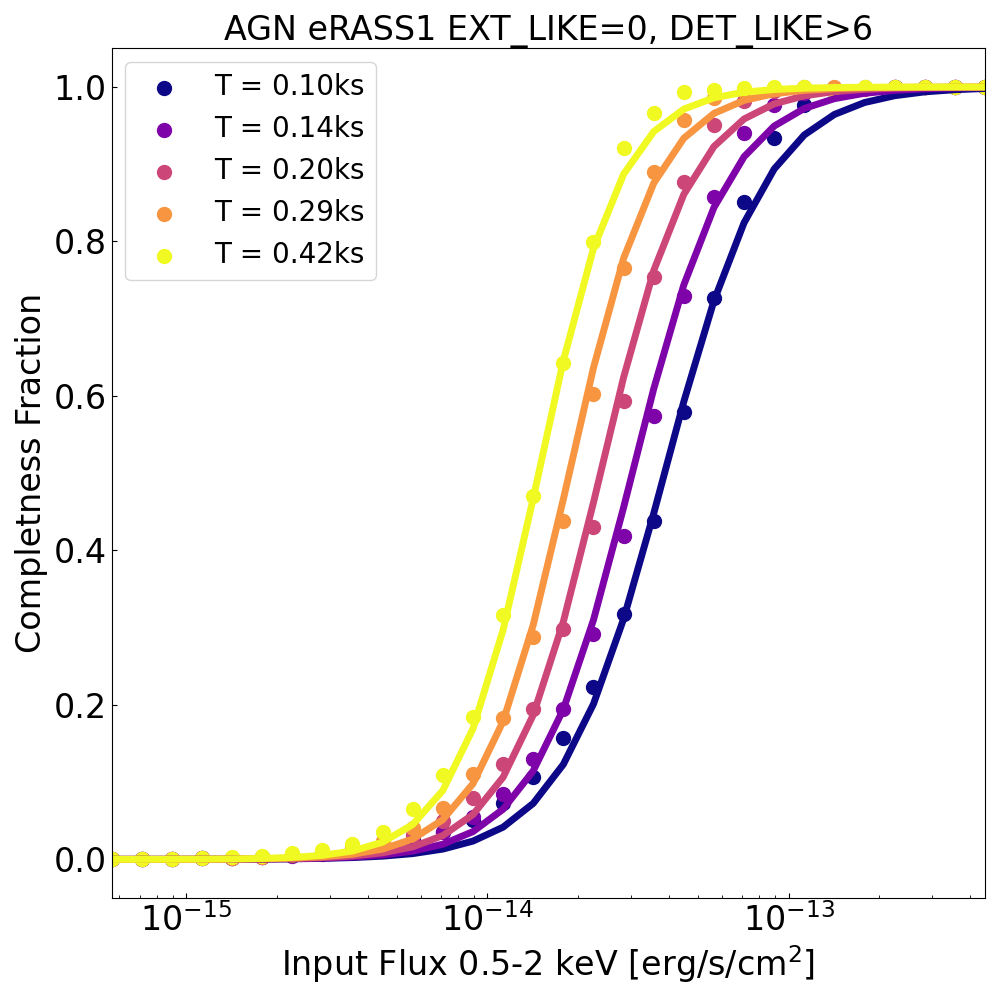}  
    \caption{Estimated completeness of eRASS1 from the simulations of \citet{Seppi2022} as a function of input 0.5-2 keV flux for different exposure times. The completeness is estimated considering only point sources (\texttt{EXT\_LIKE}$=0$) with \texttt{DET\_LIKE\_0}$>6$.}
    \label{fig:completeness_seppi_pnt}
\end{figure}

\subsection{Number counts of point sources}
\label{subsec:lognlogs}
Before eROSITA, deep extra-galactic X-ray surveys\footnote{Here by "deep" survey we adopt a quite generic definition that encompass all the deepest X-ray observations of blank fields of sizes less than a few tens of square degrees.} were generally confined to particular regions with areas of at most tens of square degrees. The largest contiguous X-ray survey before eRASS1 is the eFEDS \citep{Brunner2022}, which covers an area of 140 deg$^2$. In such (relatively) small extra-galactic regions, the point source number density can be considered uniform. Serendipitous surveys, built by combining narrow field exposures (with \cxo, \xmm, \swift) can provide even larger sky coverage, albeit very patchy. 
However, in eRASS1 it becomes apparent that the X-ray point source number density is nonuniform across the sky.
The first-order reasons for this are obviously the inhomogeneous Galactic absorption \citep[see e.g.][]{Ponti23b} and the non-uniform distribution of the Galactic X-ray population itself.
Even after excluding these Galactic features, the number density of distant AGN may not necessarily be uniform either, due to the inhomogeneous large-scale structure and the potential anisotropy \citep[e.g., a dipole structure; see ][]{Secrest2021} of the Universe.
More details about the eROSITA point-source number density maps are presented in Liu et al. (in preparation).
In this section, we present the point-source number count distributions averaged in a few wide Galactic latitude ranges.

\begin{figure*}
    \centering
    \includegraphics[scale=0.4]{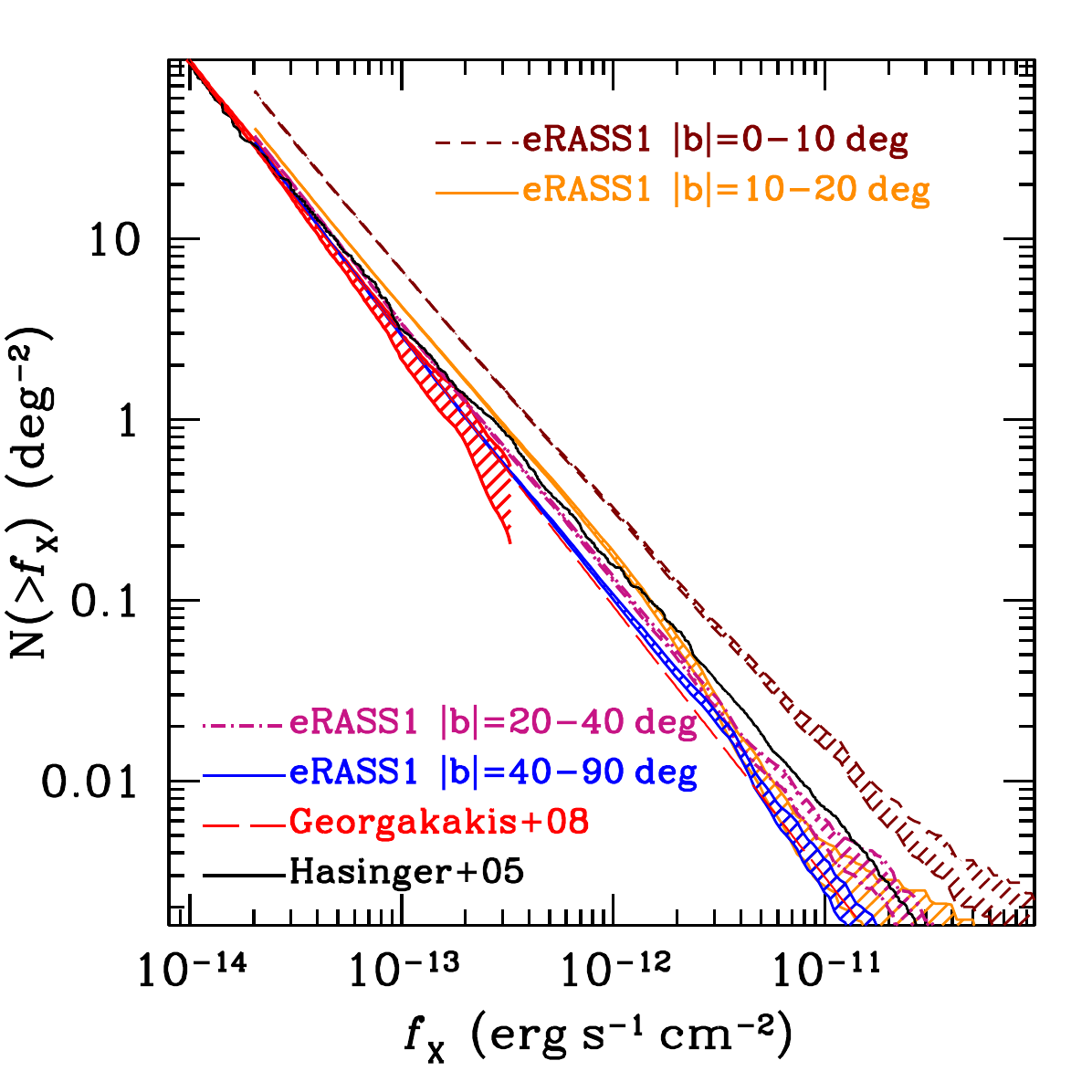} 
    \includegraphics[scale=0.4]{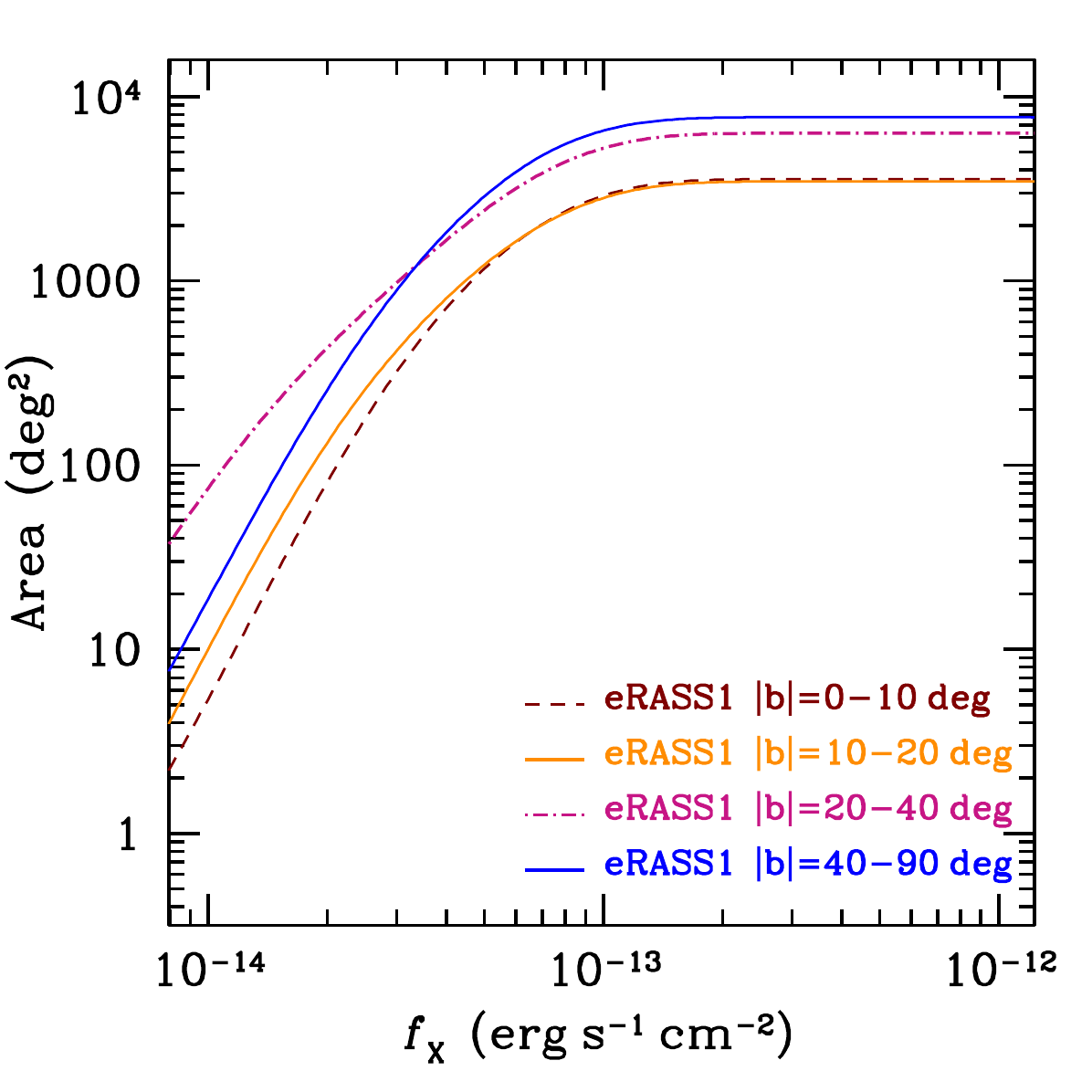}
    \caption{{\it Left}: Cumulative number counts as a function of flux for eRASS1 X-ray point sources selected in the 0.6--2.3\,keV band. The {\tt apetool} sensitivity maps and the eRASS1 aperture photometry for individual sources are used to construct the number counts in different Galactic latitude intervals, $|b|$=0--10\degr (brown dashed curves), 10--20\degr (yellow solid curves), 20--40\degr (purple dashed curves) and 40--90\degr (blue solid curves) following methods described in \cite{Georgakakis2008}. The shaded region associated with each of these curves corresponds to the 68\% uncertainty in the determination of number counts estimated using the bootstrap resampling method described in the text. The curves are plotted up to the flux of $2\times10^{-14}$ erg\,s$^{-1}$\,cm$^{-2}$ where the {\tt apetool} sensitivity curve corresponds to about 1\% of its maximum value. For comparison also shown are the 0.5--2\,keV number counts determined by \cite[][black solid line]{Hasinger2005} using \ROSAT, \xmm and \cxo surveys. The red shaded region and red dashed curve are the 0.5--2\,keV number counts estimated from \cxo extra-galactic survey fields  \citep{Georgakakis2008}. The extent of the red shaded region at fixed flux corresponds to the 1$\sigma$ uncertainty. For fluxes brighter than  $f_X$(0.5--2\,keV) $\approx 10^{-12}$\,erg\,s$^{-1}$\,cm$^{-2}$ the red dashed curve is an extrapolation of the best-fit double power determined by  \cite{Georgakakis2008}. {\it Right}: The corresponding sensitivity curve, i.e. area coverage vs. flux limit in the 0.6--2.3\,keV band, for different galactic latitude intervals. }
    \label{fig:apetool_lgNlgS_0p6_2p3}
\end{figure*}

We divide the hemisphere into four Galactic latitude levels, 0--10\degr, 10--20\degr, 20--40\degr, and 40--90\degr. As introduced in the previous Sect.\,\ref{subsec:flux_limit}, we calculate the point source number counts using the method described by \citet{Georgakakis2008} based on the products of {\tt apetool} for the 3B catalogue in the 0.6--2.3~keV band. Through the $N_\textrm{H}$-dependent ECF (see Table~\ref{tab:sens_ecf}), we convert the count rate of each source into absorption-corrected flux. The uncertainties of the cumulative X-ray number counts are estimated using a bootstrap re-sampling approach. For a given eROSITA sub-sample selected within a given Galactic latitude interval, the X-ray sources are randomly drawn with replacement to generate 100 new samples with the same size as the original one. The X-ray number counts are then generated for each of the 100 samples following the same approach as with the real data and the uncertainty at fixed flux is then estimated as the 1-$\sigma$ rms scatter of the 100 number count realisations.

The resulting cumulative number count distributions as a function of the absorption-corrected flux and the corresponding sensitivity curves are displayed in Fig.~\ref{fig:apetool_lgNlgS_0p6_2p3}. 
The number counts obtained in sky regions with $|b| >$ 20\degr\ (high Galactic latitudes) are consistent with each other and with the results of \citet{Georgakakis2008}, while a $\sim$30\% excess is seen in the counts at high fluxes reported in \citet{Hasinger2005}. These latter are based on a total of about 200 type-1 AGN selected from the ROSAT Bright Survey \citep[RBS;][]{Schwope2000}. The systematic offset with the eRASS1 results may be  related to e.g. uncertainties in the sensitivity calculations of the ROSAT sample, or the inclusion of more extended sources in RBS because of the larger ROSAT PSF.

There is a significant excess for the source population at low Galactic latitude.
The absorbed spectral model adopted in the flux calculation is only valid for un-obscured AGN, and the cosmic variance of AGN is of small amplitude and unrelated to Galactic latitude.
So the excess must be due to Galactic sources, for which the invalid absorption correction biases high the source fluxes and thus the number counts.
Therefore, the low-latitude samples trace the distribution of Galactic X-ray sources (see Freund et al., submitted).

\subsection{The resolved fraction of the Cosmic X-ray Background in eRASS1}
The cosmic X-ray background (CXB) radiation, discovered by Giacconi and collaborators at the dawn of X-ray astronomy in 1962, can be considered as the ultimate inventory of the energy released by high-energy processes throughout the history of the Universe. As the CXB radiation is dominated by accretion onto black holes, detailed modelling of the CXB over the years has accompanied our deeper understanding of the physical properties of AGN, and of their cosmological evolution. Indeed, deep extra-galactic X-ray surveys have resolved about 80-90\% of the CXB using synthesis models of the obscured and un-obscured AGN population \citep[e.g][]{Setti1989,Comastri1995,Treister2005,Hickox2006,Gilli2007,Ueda2014,Aird2015,Ananna2019}. 

Here we provide an estimate of the fraction of the CXB radiation resolved into individual sources by eRASS1 in the soft band (1--2 keV), where the eROSITA sensitivity is around its maximum, 
and the CXB contribution to the measured background emission dominates over both the foreground Galactic components and over the instrumental background \citep[see e.g.][]{Ponti2023}. We first adopt a mean value of the CXB intensity in the 1--2 keV band of 12.7 keV\,cm$^{-2}$\,s$^{-1}$\,sr$^{-1}$ by fitting the data compilation of \citet{Gilli2007}. We then compute the fraction of this intensity in the same energy range contributed by highly reliable eRASS1 sources with \DETLIKE{0} > 10 (excluding all flagged ones\footnote{We also exclude very bright, rare, sources with $F_{\rm 0.5-2~keV}> 10^{-12}$ erg\,s$^{-1}$\,cm$^{-2}$.}; see Table~\ref{tab:flags}) above two flux limit thresholds: $F_{\rm 0.5-2~keV}>5 \times 10^{-14}$ erg\,s$^{-1}$\,cm$^{-2}$ or $F_{\rm 0.5-2~keV}>1 \times 10^{-14}$ erg\,s$^{-1}$\,cm$^{-2}$. As discussed in Sect.~\ref{subsec:flux_limit} above, the former represents an approximately uniform flux limit achieved over the entire sky, while the latter includes fainter sources detected in the areas of increased exposure near the ecliptic poles. 

The two bottom panels of Fig.~\ref{fig:resolved_CXB} show a HEALPix map (order 4) of the 1--2 keV CXB intensity resolved fraction in the eROSITA-DE hemisphere. On the right hand side, the fraction computed using the eRASS1 flux limit appears more uniform, with the exception of the low-Galactic latitude regions, where X-ray stars in nearby ($<500$ pc) star-forming regions \citep[Sco-Cen association and Gould belt;][]{Schmitt2022} increase the cumulative average emission. On the left hand side, instead, a significantly higher fraction of the CXB is resolved into fainter sources close to the south ecliptic pole (SEP). Finally, in the top panels of Fig.~\ref{fig:resolved_CXB} we show the histogram of the number of HEALPix order 4 pixels at a given resolved fraction in the extra-galactic sky ($|b|>30\degr$) for the two cases. The solid lines are the median values: 0.19 and 0.24, respectively. Including only point sources (i.e. removing all those with \EXTLIKE $>0$) the median value of the resolved fraction is reduced by about 2\%, to 0.17 and 0.22, respectively. On the other hand, if we include all (non-flagged) sources with \DETLIKE{0} > 6 the median value for the CXB resolved fraction remain almost unchanged (0.20) for the uniform flux limit case ($F_{\rm 0.5-2~keV}>5 \times 10^{-14}$ erg\,s$^{-1}$\,cm$^{-2}$), but increases to 0.33 for $F_{\rm 0.5-2~keV}>1 \times 10^{-14}$ erg\,s$^{-1}$\,cm$^{-2}$, due to the large number of fainter sources detected with lower significance in the deeply exposed area near the SEP.

\begin{figure*}
    \centering
    \includegraphics[width=0.45\textwidth]{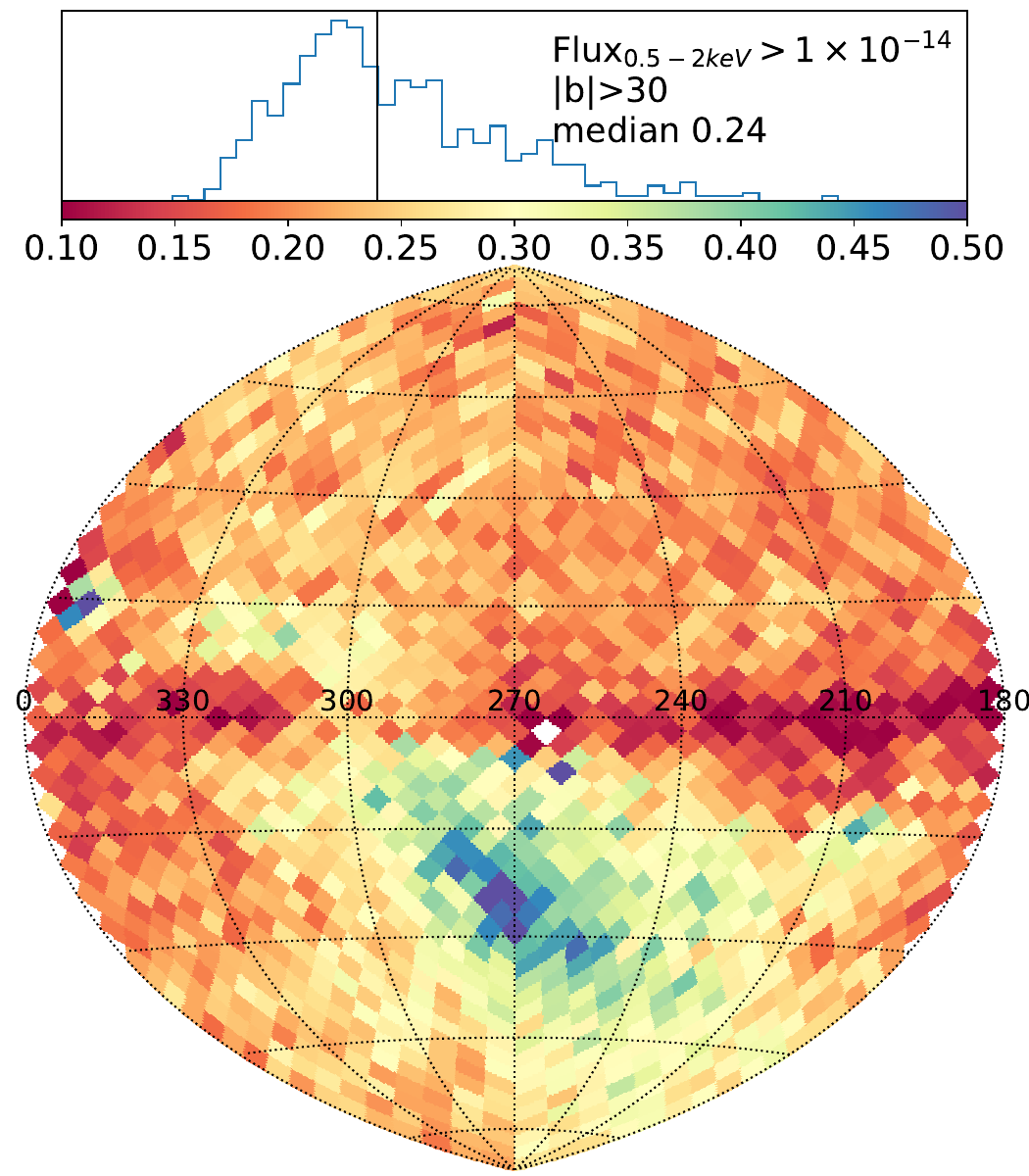}
        \includegraphics[width=0.45\textwidth]{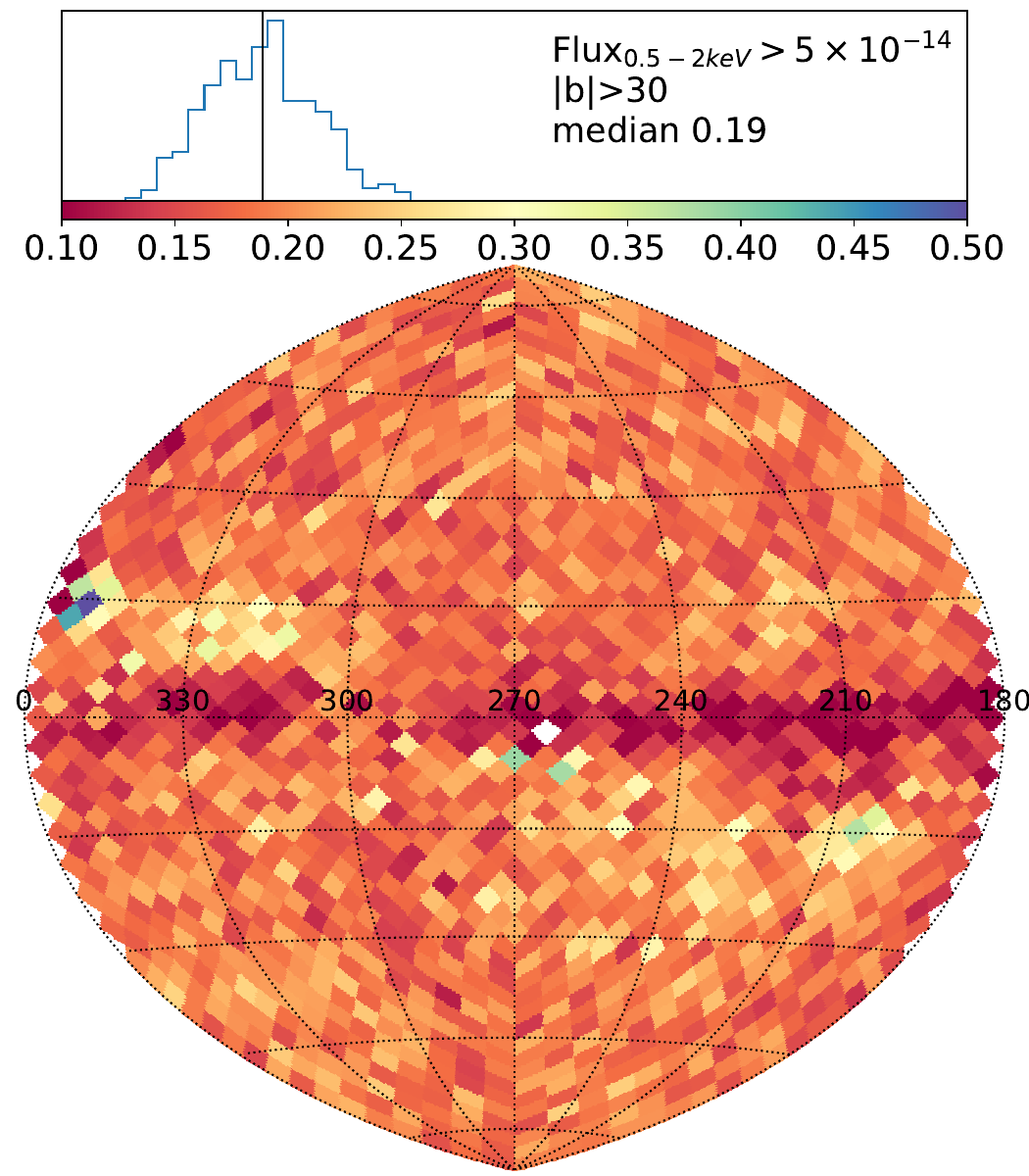}
    \caption{Resolved CXB fraction in eRASS1. {\it Bottom panels:} HEALPix order 4 map of the 1--2 keV CXB intensity resolved fraction (eROSITA-DE hemisphere, Hammer-Aitoff projection, in Galactic coordinates). On the {\it bottom left} we include all sources with \DETLIKE{0} $>10$, no flags and X-ray flux $F_{\rm 0.5-2~keV}>1 \times 10^{-14}$ erg\,s$^{-1}$\,cm$^{-2}$. On the {\it bottom right} we include all sources with \DETLIKE{0} $>10$, no flags and and X-ray flux $F_{\rm 0.5-2~keV}>5 \times 10^{-14}$ erg\,s$^{-1}$\,cm$^{-2}$ (i.e. the hemisphere median flux limit at 50\% completeness). {\it Top panels:} distribution of the resolved fraction in the extragalacitc sky ($|b|>30\degr$) in 13.4287 square-degree pixels for the two cases. The solid lines are the median values: 0.24 and 0.19, respectively.}
    \label{fig:resolved_CXB}
\end{figure*}

\subsection{Particular regions}
In the soft X-ray band, a few nearby sources of X-ray emission are very prominent, including the eROSITA bubbles \citep{Predehl2020}, the two Magellanic clouds, the Vela SNR, Virgo cluster, Crab Pulsar, and Sco X--1. 
They enhance the effective background for the detection of distant sources and thus appear clearly in the flux limit maps displayed in Fig.~\ref{fig:fluxlimitmap}.
As discussed in Sect.~\ref{sec:spurious_identification_overdense}, our source detection algorithm is not optimized for such regions with bright or extended source contamination. This results in an increased number of spurious sources caused by background fluctuations. There are large uncertainties in the measurement of background and source properties in such regions.

In addition to these, the SEP region also poses particular issues for source detection, because the exposure is deeper than the typical depth across the sky by more than two orders of magnitude. Our detection algorithm is not optimized for such deep exposures either. As discussed by \citet{LiuT2022}, one difficulty of source detection is de-blending nearby sources (or confusion). This problem is more severe where the fields are crowded, like in the regions near the ecliptic poles. Although the eRASS1 catalogues presented here are complete (i.e. includes also sources in the SEP region), an optimized SEP catalogue will be presented in Liu et al. (in preparation), where a more sophisticated data processing and source detection optimized for crowded regions with deep (and spatially variable) exposure will be introduced.

\section{Consistency checks with external catalogues}
\label{sec:cons_checks}

\subsection{Comparison with XMM source fluxes}
\label{sec:flux_compare}
Before eRASS1, the previous best all-sky X-ray catalogue was provided by the \ROSAT all-sky survey \citep[2RXS:][]{Boller2016}. 
In terms of pure source numbers, the largest X-ray catalogue previous to eRASS1 is the \xmm serendipitous catalogue \citep[4XMM:][]{Rosen2016,Webb2020}. Covering only a small fraction of the sky, 4XMM has a similar resolution to eROSITA and generally deeper exposure than eRASS1. Therefore, we use the 4XMM-DR12 catalogue 
for a consistency check of our eRASS1 sources. 

We selected $1250.8$k point sources from the eRASS1 1B catalogue and $432.9$k point sources (\texttt{SC\_SUM\_FLAG}=0 and \texttt{SC\_EXTENT}=0) from 4XMM-DR12 over the entire sky, and performed a simple positional cross-correlation between them, adopting a maximum separation of 10\arcsec. We found $16.5$k detection pairs. The majority of the X-ray sources are in the low S/N regime, where their fluxes suffer large uncertainties and Eddington bias. Taking the ratio of 0.2--2~keV flux to flux uncertainty as a measurement of flux measurement reliability, we selected sources with at least 1$\sigma$ reliability in both catalogues and compared their fluxes in the upper panel of Fig.~\ref{fig:4xmm}. 
For faint sources, the Eddington bias is still visible in terms of the overestimated eROSITA fluxes at the low-flux end.
For the brightest sources with 5$\sigma$ reliability, the \xmm and eROSITA measured fluxes are consistent, with the mode of the distribution of the flux ratio consistent with unity within 6\%. Assuming the \xmm fluxes in this band are correct, this can in turn be interpreted as an estimate of the flux calibration uncertainty of eROSITA in the 0.2-2 keV band.

We also compared the hard-band fluxes of the Hard sources with that measured in the 4XMM catalogue, as displayed in the lower panel of Fig.~\ref{fig:4xmm}. Here we are using slightly different bands (2.3-5 keV for eROSITA and 2-4.5 keV for \xmm), but we note here that assuming a power-law spectrum with a slope of $\Gamma=1.7$ the fluxes in these two bands are expected to be almost identical, with a difference $<1\%$. Because of the small sample size (only $131$ matches) and the small eROSITA source photon counts in the hard band, this comparison is dominated by selection biases, even though residual effective area calibration issues cannot be ruled out.

A small fraction of the sources show significant variability, almost all of which turn brighter in eRASS1. This is because, typically, \xmm observations are much deeper than eRASS1, and thus the sources that become fainter are filtered out by selection bias. The analysis of these variable sources will be presented elsewhere.

\begin{figure}
    \centering
    \includegraphics[width=\columnwidth]{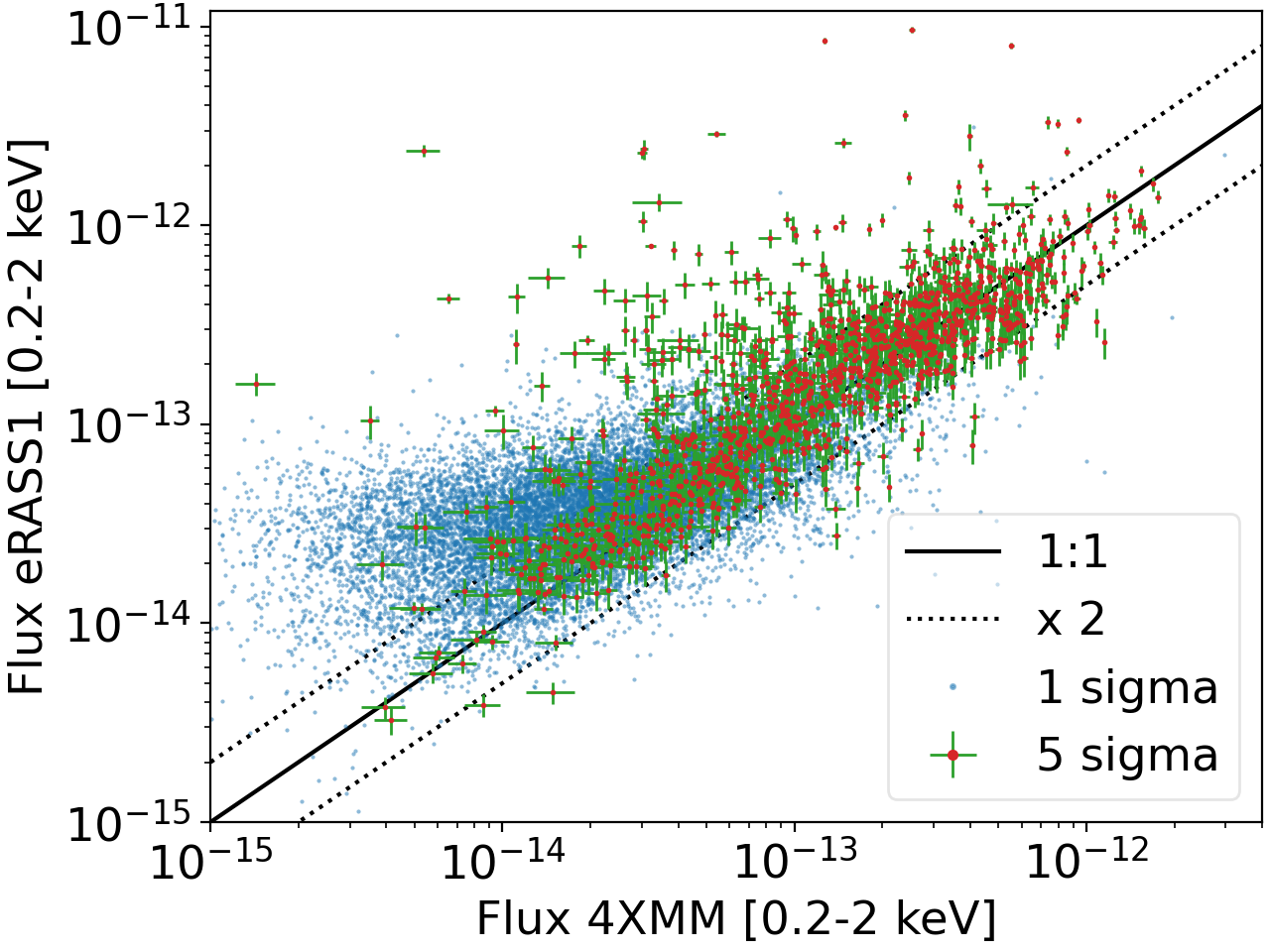}  
        \includegraphics[width=\columnwidth]{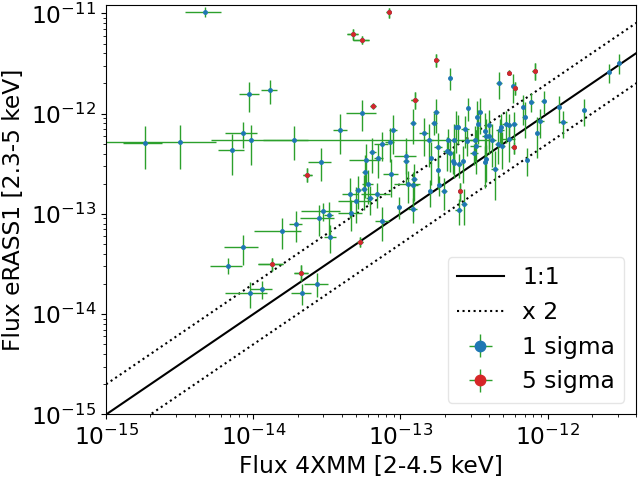}  
    \caption{Flux comparison between eRASS1 and 4XMM catalogues. The upper panel displays the 0.2--2~keV fluxes of point sources. The brightest sources with at least 5$\sigma$ flux measurements are plotted in red points with green 1-sigma uncertainties. The other sources are plotted in blue points without error bars. The black lines indicate 1:1 and deviation from 1:1 by a factor of 2.
    The lower panel shows a hard-band comparison for the Hard sources.}
    \label{fig:4xmm}
\end{figure}

\subsection{Astrometric validation}
\label{subsec:astrometric_ext}

This section presents a methodology that provides independent constraints on the X-ray positional uncertainty using external multi-wavelength source catalogues with known and accurate positions. 

We assume that each source {\it i} has a circular, Gaussian positional uncertainty with a standard deviation of $\sigma_i$.
 It is then possible to cross-correlate two catalogues and count the number of matches as a function of angular distance. An example of such an angular separation distribution is shown in Fig.~\ref{fig:age-radial} and consists of a linear part at large angular distances and a pronounced peak at small separations. The former represents random matches between X-ray sources and the external catalogue, the latter is related to true associations. The number of random matches at a given angular separation depends only on the sky density of the external catalogue and the total number of X-ray sources:
\begin{equation}\label{eq:radial-dist.rand}
    N_{\rm rand}(\theta) = N_X \cdot 2 \, \pi \cdot \theta \cdot \rho \cdot d\theta, 
\end{equation}
where  $N_{\rm rand}(\theta)$ is the number of associations for angular separations between $\theta$ and $\theta+d\theta$ and $\rho$ is the sky density of the external catalogue. The peak in Fig.~\ref{fig:age-radial} can be expressed as the superposition of $N_X$ Rayleigh distributions, each of which depends on the positional uncertainty of the corresponding X-ray source, $\sigma_i$, so
\begin{equation}\label{eq:radial-dist.assoc}
    N_{\rm assoc}(\theta) = F\,\sum_{i=1}^{N_X} P_{i}(\theta\, | \,\sigma_i), 
\end{equation}
where $P_{i}(\theta\, | \,\sigma_i)$  is the Rayleigh distribution that describes the probability of true associations separated by angular distance $\theta$ given the positional uncertainty of the X-ray sources $\sigma_i$. This parametrisation assumes that the positional errors of the external catalogue are much small than $\sigma_i$ and therefore can be ignored. The factor $F$, which takes values between zero and one, represents the fraction of X-ray sources expected to have counterparts in the external catalogue. In Fig.~\ref{fig:age-radial} the total number of sources at a given angular separation $\theta$ can be expressed as the sum of the terms given by Equations  \ref{eq:radial-dist.rand} and \ref{eq:radial-dist.assoc}. Put differently, it is possible to model the observationally determined angular separation distribution of Fig.~\ref{fig:age-radial} using the equations above and hence to infer parameters such as $F$, $\rho$ and the positional uncertainty. The number of matches at a given angular separation bin $\theta$ is a Poisson variate with expectation value $\lambda(\theta) = N_{\rm rand}(\theta) + N_{\rm assoc}(\theta)$. The likelihood of the observations in Fig.~\ref{fig:age-radial} can then be expressed as the product of the Poisson probabilities at each angular separation bin:
\begin{equation}\label{eq:radial-dist.like}
    \mathcal{L} = \prod_{j=1}^{N_\theta} \, \mathrm{Pois} \left( N_j \, | \, \lambda(\theta_j) \right ), 
\end{equation}
where the index $j$ is for all angular separation bins $N_\theta$ and $\mathrm{Pois}( N_j \, | \, \lambda(\theta_j))$ is the Poisson probability of $N_j$ matches given the expectation value $\lambda(\theta_j)$. We use the {\sc UltraNest} nested sampling algorithm \citep{Buchner2016,Buchner2019,Buchner2021} to perform Bayesian inference on the likelihood of Equation (\ref{eq:radial-dist.like}) and determine posteriors for the various model parameters. For this exercise we parameterise the positional uncertainties of individual sources as 
\begin{equation}\label{eq:radial-dist.poserror}
    \sigma_i = \mathrm{POS\_ERR} = \sqrt{A\cdot \sigma^2 + \sigma_{0}^2},
\end{equation}

\noindent where $\sigma =\mathrm{RADEC\_ERR}/\sqrt{2}$ and $\mathrm{RADEC\_ERR}$ is the catalogued positional uncertainty produced by \texttt{ermldet} (see Eq.~\ref{eq:radecerr}). In equation~\ref{eq:radial-dist.poserror} above, $\sigma_{0}$ represents systematic uncertainties and $A$ is a multiplicative factor that scales the \texttt{ermldet} uncertainties. Under these assumptions, the four model parameters that are inferred by modelling the distribution of Fig.~\ref{fig:age-radial} are $\rho$, $F$, $A$, $\sigma_{0}$. 
We caution that this parametrisation assumes that the external catalogue is assumed to have the same sky density, $\rho$, across the sky. Although the sample selection (see below) minimises such variations, it is inevitable that $\rho$ has an intrinsic scatter that is not accounted for in the current version of the methodology.

\begin{figure}
    \centering
    \includegraphics[scale=0.4]{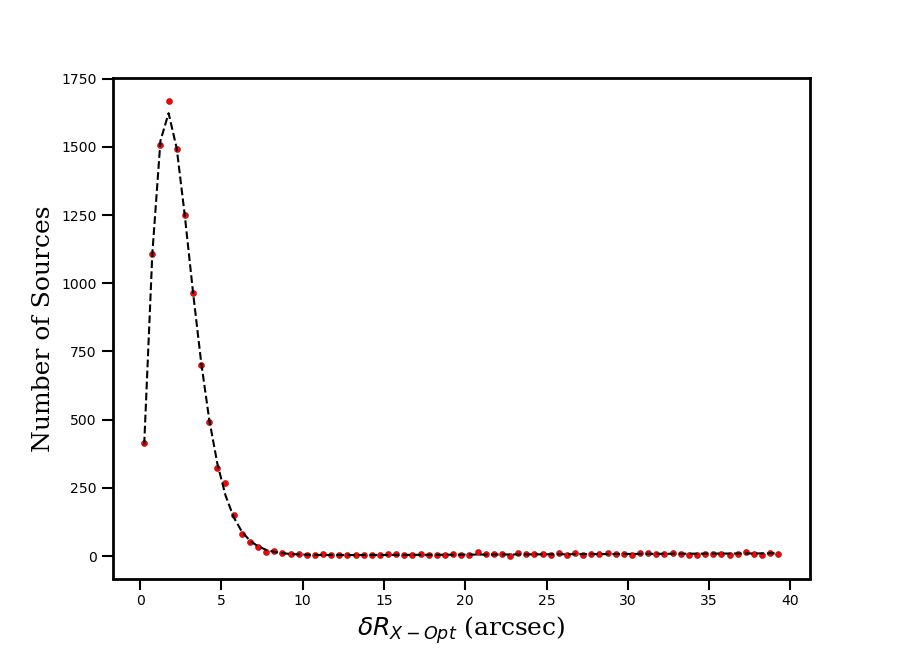}
    \caption{Distribution of the angular separation between eRASS1 X-ray source positions and the Gaia/unWISE QSOs. The red dots correspond to the observed number of associations at a given angular separation bin, $\delta R_{X-Opt}$. We use only eRASS1 X-ray sources with Galactic latitudes in the range $|b|=30-70$\degr (i.e. extragalactic sky), \texttt{DET\_LKE\_0} $>7$, with more than 30 counts, are not spatially extended (parameter \EXT $=0$) and are not identified as potentially spurious by the algorithms described in Sect.~\ref{sec:spurious_flagging}. The dashed line is the model described in the text for parameters fixed to the median of the corresponding posteriors.}
    \label{fig:age-radial}
\end{figure}

We use the catalogue of AGN from Gaia and unWISE Data \cite[Gaia/unWISE;][]{Shu2019} to cross-match  against the eRASS1 X-ray source catalogue. We only consider Gaia/unWISE sources with probability being a QSO $\mathrm{PROB\_RF}>0.8$ and $G$-band magnitude $<20.5$\,mag. The latter criterion is adopted to minimise variations in the sky density of QSO candidates because of the variable depth of the GAIA survey as a result of the scanning law of the mission. For this magnitude cut it is empirically found that the sky density of QSO candidates in the extra-galactic sky (Galactic latitudes $|b|>20\degr$) is nearly homogeneous. We further limit the eRASS1 catalogue to sources with \texttt{ermldet} detection likelihood \texttt{DET\_LIKE\_0} $>7$ (to increase the purity of the sample) that are not spatially extended (parameter \EXT $=0$) and are not identified as potentially spurious by the algorithms described in Sect.~\ref{sec:spurious_flagging}. 
We analyse separately eRASS1 sources with low ($<30$) and high ($>30$) number of net counts to accommodate a dependence of the positional uncertainty corrections on source brightness. Figure~\ref{fig:age-radial} shows the angular separation distribution from the cross-correlation of the high count eRASS1 sample with the Gaia/unWISE catalogue. The model with parameters set to the median of the corresponding posteriors is shown with the dashed line in that figure. The model represents the observations reasonably well. For the high-counts eRASS1 sub-sample, the parameters that are relevant to the positional error are estimated as $A=1.3\pm0.1$ and $\sigma_0=0.9\arcsec \pm0.1\arcsec$. 

For the low-count sub-sample, the 3$\sigma$ upper limit of the $\sigma_0$ term is about 0.2\arcsec\ and consistent with zero within uncertainties. This indicates that the positional error budget for low count sources is dominated by the multiplicative term of Eq. (\ref{eq:radial-dist.poserror}), as the typical positional uncertainties of sources at the low count regime are already quite large. For simplicity, we then adopt a single scaling function to determine the positional uncertainty from Eq.~\ref{eq:radial-dist.poserror}, which we report as \texttt{POS\_ERR} in the catalogue
\begin{equation}\label{eq:radial-dist.poserror_final}
    \mathrm{POS\_ERR} =  \sqrt{ 1.3 \cdot \sigma^2 + 0.9^2}.
\end{equation}
\noindent Indeed, for sources with < 30 counts the difference in positional error between the above more general expression in Eq.(\ref{eq:radial-dist.poserror_final}) and a modified version of this equation where the additive term is assumed to be zero is only $\sim$ 0.1--0.2\arcsec\ on average.

As a further test, we have also tried to calibrate the positional uncertainty of the eRASS1 sources by comparing the catalogue entries with both \cxo and \xmm serendipitous catalogues. The resulting values of the scaling terms $A$ and $\sigma_0$ in Eq.~\ref{eq:radial-dist.poserror} are consistent within the uncertainties with those derived above.

\subsection{On the effect of source blending}
With the relatively large PSF of eROSITA (see Section~\ref{sec:psf_cal}), source blending could become a issue in crowded regions of the sky. Empirical tests of the effect of blending (or lack thereof) on the catalog would require cross-match against X-ray catalogs where blending is expected to be negligible. Such tests, however, are potentially degenerate with source variability and non-uniformity (of background, exposure, etc.) of the matched catalogues. An alternative method is through simulations, where the full observation and detection process is simulated starting from a realistic sky source population. These have been performed for the eFEDS field \citep{LiuT2022} and for eRASS1 \citep{Seppi2022}. 
The biggest impact of source blending is on the detection of extended objects, that is, blending of point
sources leads to spurious extended sources in the catalog. Blending also causes incompleteness, since multiple individual sources are considered as only one. At the depth of eFEDS, the incompleteness caused by blending is
between 1\% and 3\%, dependent on source brightness and extent. For eRASS1, at the \texttt{DET\_LIKE\_0} threshold of the main catalog, the fraction of blended sources is less than 1\% \citep[see][Table 3]{Seppi2022}, which is definitely much smaller than the expected level of contamination from spurious detections. Only in the south ecliptic pole region, where the exposure is deeper than average
by a factor of more than hundred, this effect becomes more significant. This will be discussed in a separate paper (Liu et al. in prep.).

\section{eROSITA-DE Data Release 1 (DR1)}
\label{sec:dr1}

\begin{figure*}
\centering
\includegraphics[scale=0.39]{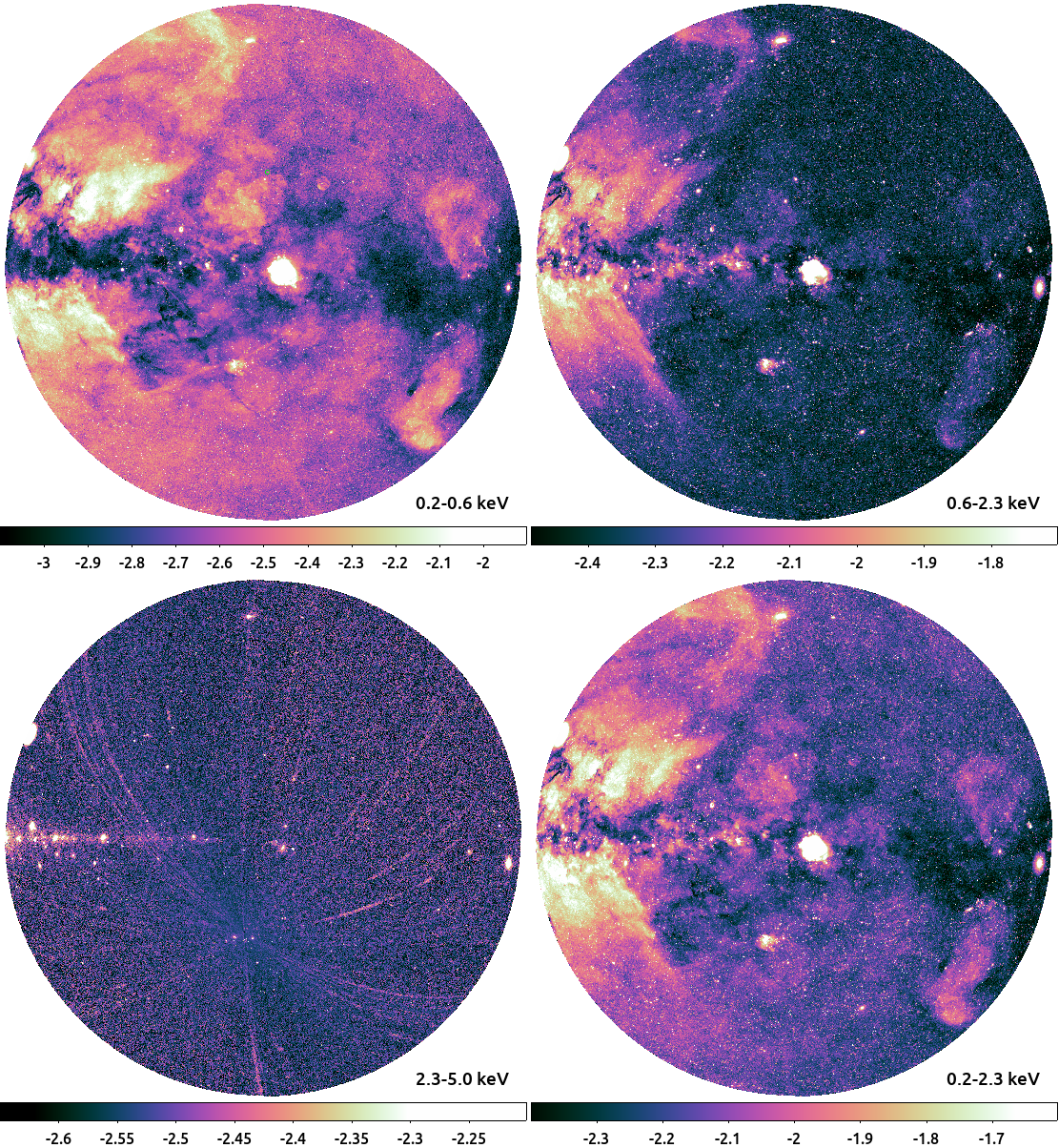}
\caption{Four examples of half-sky maps released as part of DR1. The maps are colour-coded by the logarithm of the count rate intensity (in cts s$^{-1}$ arcmin$^{-2}$), displayed in Zenith Equal Area (ZEA) projection in Galactic coordinates, with pixel size of 0.09 deg$^{2}$ and centered on ({\it l}, {\it b})=(270, 0). We show the 0.2--0.6 keV ({\it top left}), 0.6--2.3 keV ({\it top right}), 2.3--5 keV ({\it bottom left}) and 0.2--2.3 ({\it bottom right}) bands. Please note that each map has a different (logarithmic) colour scale.}
 \label{fig:maps}
\end{figure*}

The primary data products of the eROSITA-DE DR1\footnote{\url{https://erosita.mpe.mpg.de/dr1/}} consist of eRASS1 calibrated event files, which contain the information generated by the operating cameras during eRASS1 observations. 
The data taken during the CalPV phase, which were released as part of the Early Data Release, are not part of DR1.
The eSASS software package to help processing and analysing eROSITA data is also made public, with the name eSASS4DR1. 

As discussed above, the eSASS pipeline divides the all-sky observations into 4700 sky tiles for practical purposes of computational tractability (see Fig.~\ref{fig:erass1_tiles}). eROSITA-DE have proprietary rights on 2248 of these tiles, while 199 of them have shared rights between the German and Russian consortia. eROSITA-DE DR1 finally comprises 2447 sky tiles, of which 199 are only partially filled with eROSITA-DE data. This public eRASS1 data was processed with version 010 of the eSASS pipeline.

The released calibrated event lists of each sky tile contain photons:
\begin{itemize}
    \item with energies between 0.2--10 keV;
    \item flagged as \texttt{flag=0xE000F000}, i.e., good events from the nominal field of view, excluding bad pixels;
    \item with patterns pattern=15, i.e., including single, double, triple, and quadruple events.
\end{itemize}

Besides calibrated event lists, the DR1 team also releases the following products per sky tile:
\begin{itemize}
    \item Counts and count rate maps,
    \item Exposure maps,
    \item Background maps,
    \item Tables containing cumulative survey area as a function of limiting flux (for sources with \texttt{likemin > 5}),
    \item Sensitivity maps based on aperture photometry. Includes aperture-averaged exposure maps, aperture-integrated background maps, and area curves (survey area sensitive to a given count rate),
    \item Sensitivity maps (for sources with \texttt{likemin > 5}),
    \item Source data products.
\end{itemize} 

\begin{figure*}
    \centering
    \begin{tabular}{cc}
    \includegraphics[scale=0.38]{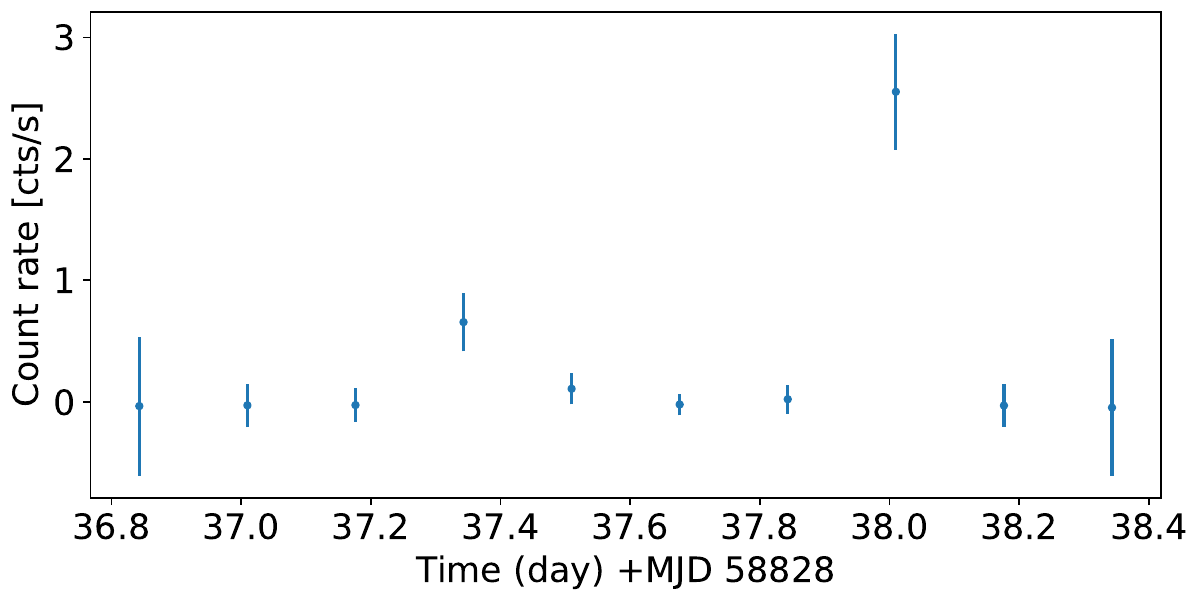} &
    \includegraphics[scale=0.38]{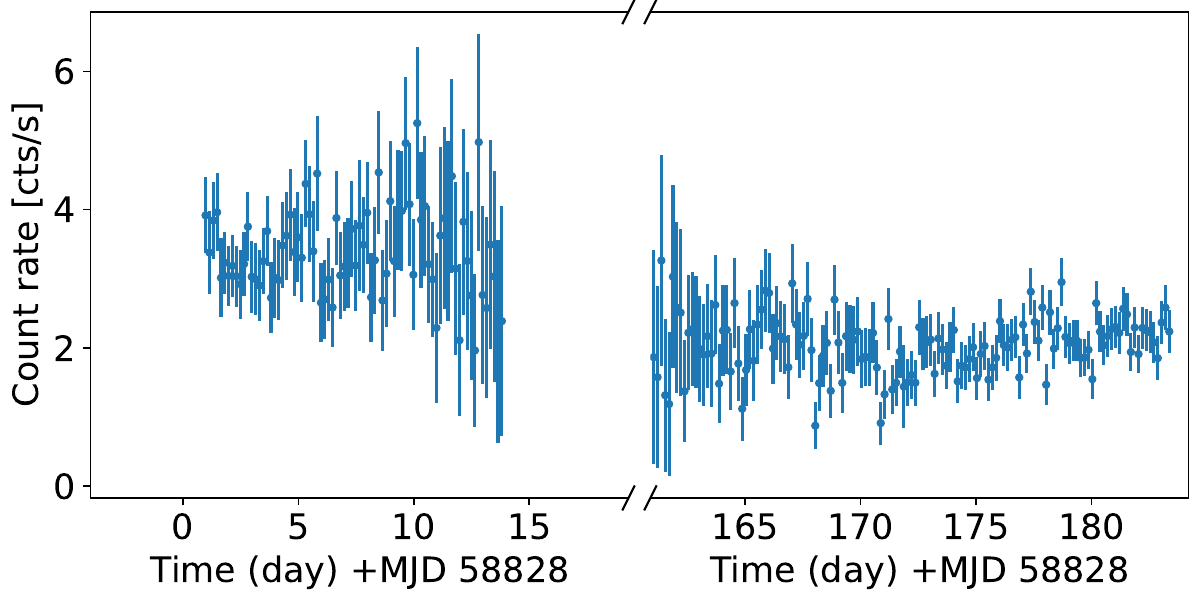}\\
    \includegraphics[scale=0.35]{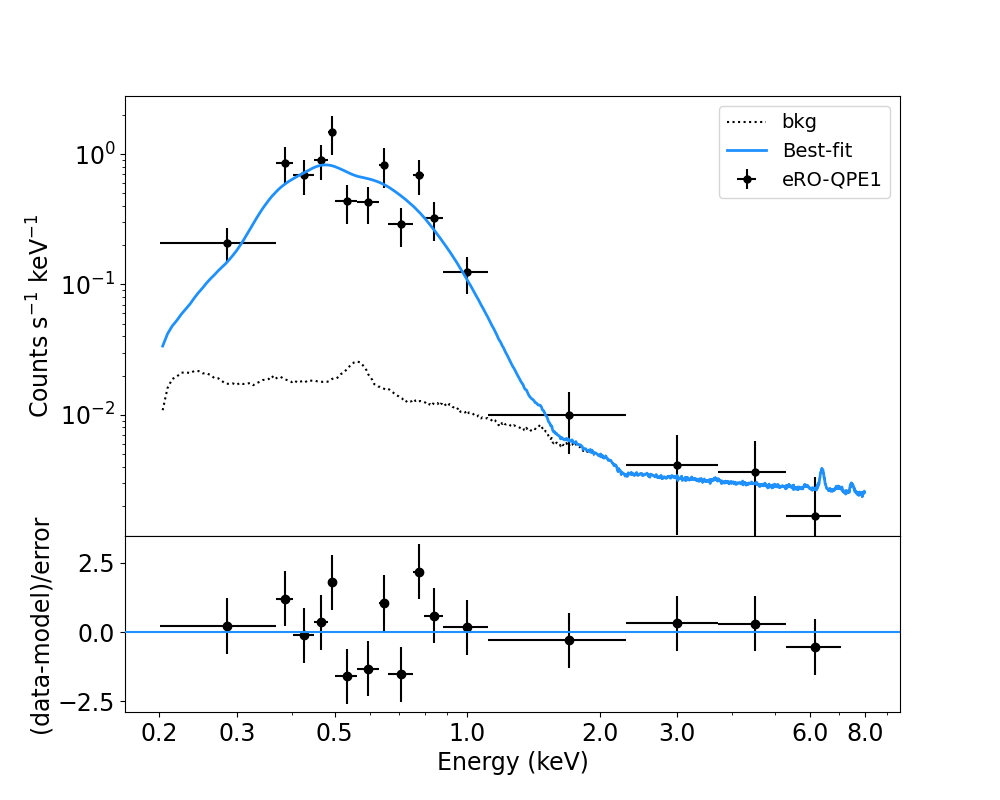} & 
    \includegraphics[scale=0.35]{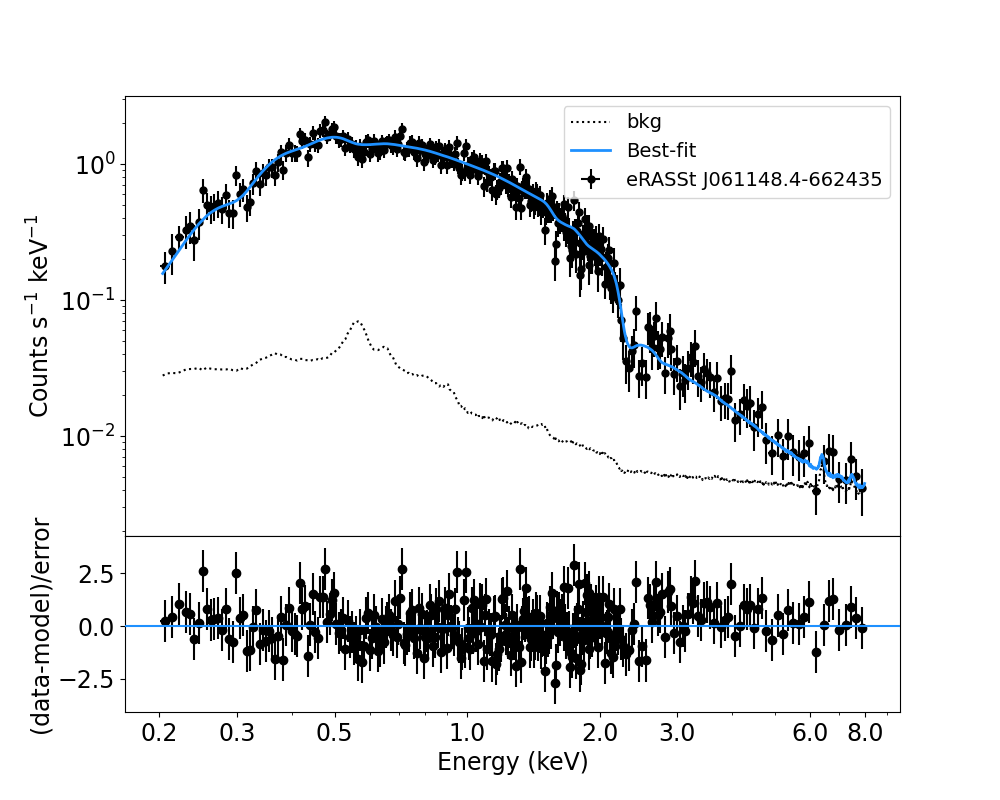} 
    \end{tabular}
    \caption{Two examples of light curves (top panels) and spectra (bottom panels) for two sources contained in the eRASS1 Main catalogue, generated from the released source products. On the left hand side, we show the light curve and spectrum of the first eROSITA Quasi Periodic Eruptor  \citep[eRO-QPE1;][]{Arcodia2021} (1eRASS J023147.1--102011; \texttt{DETUID}: eb01\_038099\_020\_ML00005\_002\_c010). The light curve ({\it top left}) is binned in individual eroday visits, separated by 4-hours intervals. The spectrum ({\it bottom left}) is taken from all data, and fitted with an absorbed black body (\texttt{zbbody}). On the right hand side, we show one of the brightest AGN in the SEP region (1eRASS J061148.2--662434; \texttt{DETUID}: em01\_093156\_020\_ML00001\_012\_c010; Bogensberger et al., in prep.). Given the high ecliptic latitude, the source is visited over two separate long periods of about 14 and 24 days, respectively. The light curve ({\it top right}) shown has the original binning in 10 seconds intervals. The spectrum ({\it bottom right}) is fitted with an absorbed power law. Note that both best-fit lines (light blue) are simple physical models aiming to guide the eye and do not intend to encompass all the potential detailed features present in the spectra. The background model is shown as black dotted lines in both figures. For more details on eROSITA spectral modelling and appropriate handling of background spectra, refer to Sect.~3 of \citet{LiuT2022b}. 
    }
    \label{fig:src_examples1}
\end{figure*}

The data are made available to users using a web-based tool called eRODat.
This interface allows one to interactively view the eROSITA sky, using Aladin Lite \citep{Boch2014} to show the eROSITA HiPS (Hierarchical Progressive Surveys) maps and source positions.
It also allows the user to identify the sky tiles associated with a given source position or a list of positions, and download the data products for those tiles.
In addition, the user can search for sources around a position from the eROSITA X-ray catalogues and view those sources in various surveys, view the details of the catalogue entries, or download the individual source products.
Data can be either downloaded by navigating the archive structure to obtain the required data files, following direct links as appropriate, or by adding the products to a virtual basket.
The contents of the basket can be download immediately as a single tar file or eRODat can generate a shell script to later download those products.

\begin{figure*}
    \centering
    \begin{tabular}{cc}
    \includegraphics[scale=0.38]{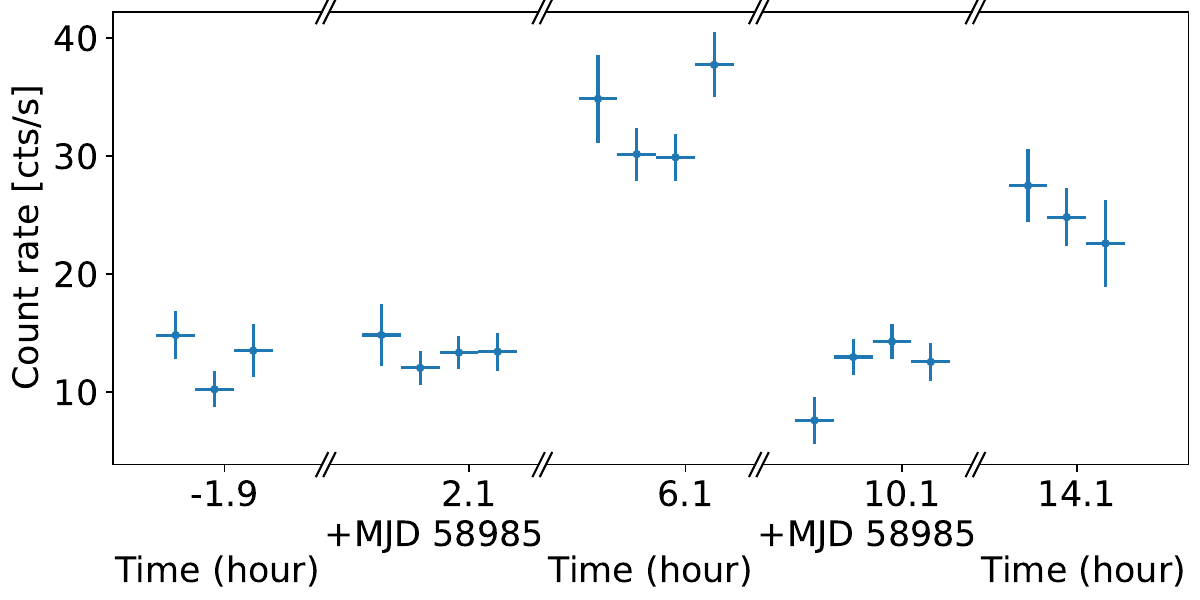} &
    \includegraphics[scale=0.38]{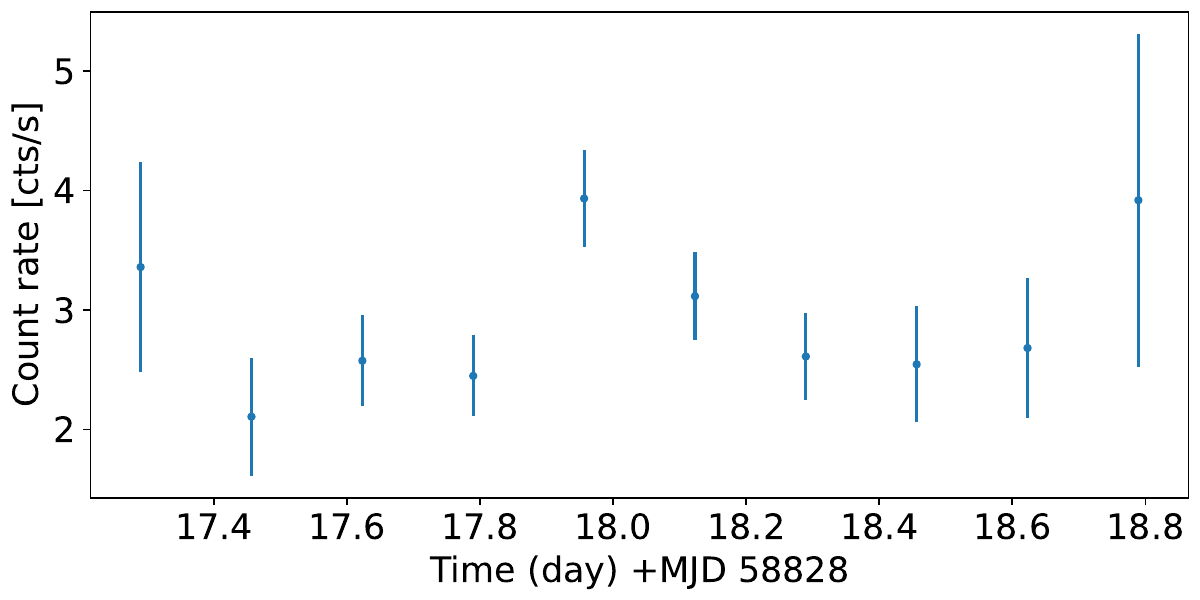}\\
    \includegraphics[scale=0.35]{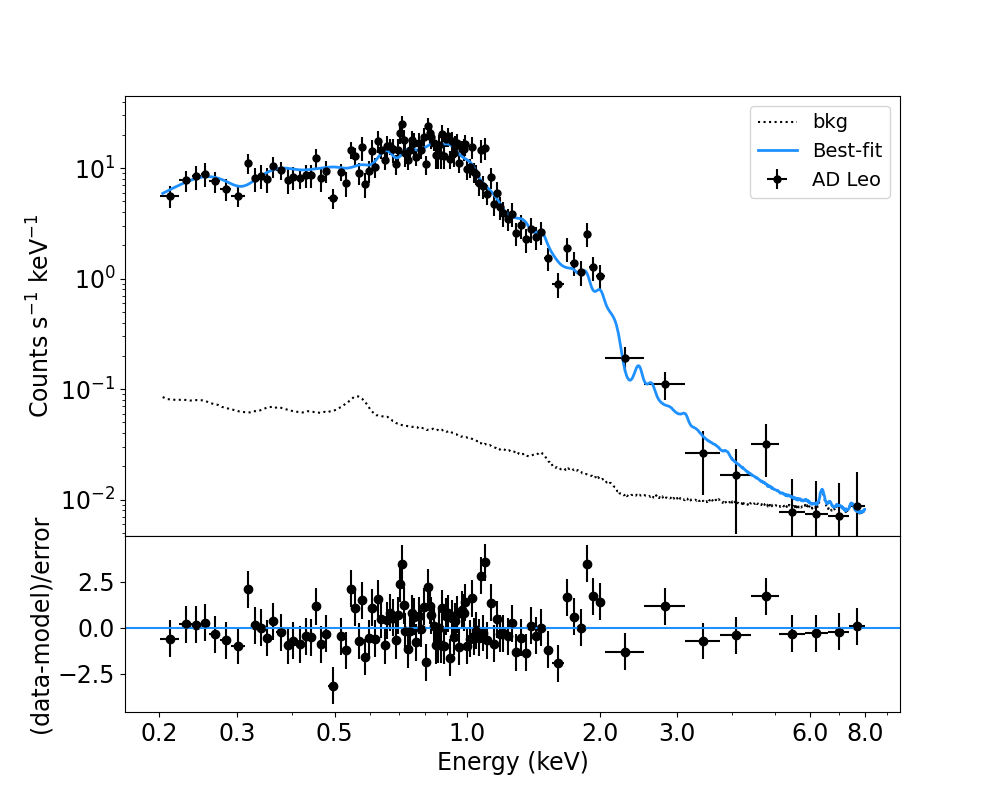} & 
    \includegraphics[scale=0.35]{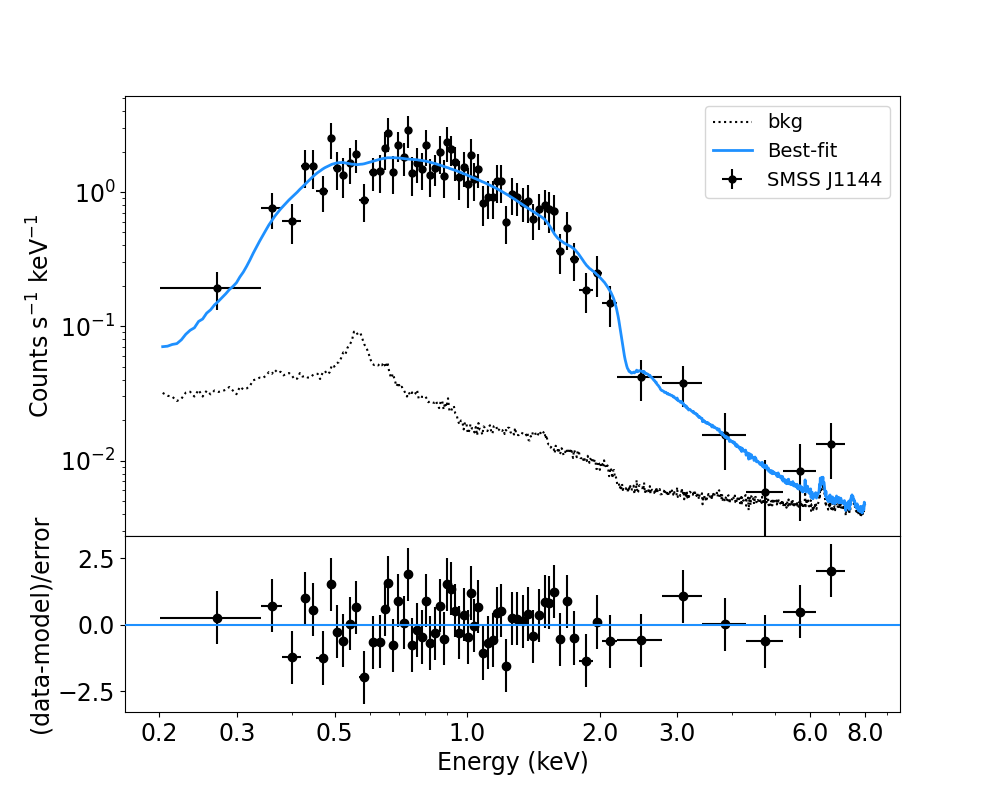} 
    \\
    \end{tabular}
    \caption{Two more examples of light curves (top panels) and spectra (bottom panels) for two sources contained in the eRASS1 Main catalogue, generated from the released source products. On the left hand side, we show the light curve and spectrum of the bright star AD Leo (1eRASS J101935.6+195211; \texttt{DETUID}: em01\_155069\_020\_ML00001\_011\_c010). The light curve ({\it top left}) is shown with the original 10s binning. The spectrum ({\it bottom left}) is fitted with a multi-temperature thermal collisionally-ionised plasma emission model (\texttt{apec+apec+apec}, with tied abundances). On the right hand side, we show the light curve and spectrum of SMSS J114447.770-430859.3, the most luminous QSO of the last 9 Gyr \citep{Onken2022,Kammoun2023} (1eRASS J114447.6--430858; \texttt{DETUID}: em01\_176132\_020\_ML00001\_002\_c010). The light curve ({\it top right}) is binned in individual eroday visits, while the spectrum ({\it bottom right}) is fitted with an absorbed power law. Note that both best-fit lines (light blue) are simple physical models aiming to guide the eye and do not intend to encompass all the potential detailed features present in the spectra. The background model is shown as black dotted lines in both figures. For more details on eROSITA spectral modelling and appropriate handling of background spectra, refer to Sect.~3 of \citet{LiuT2022b}.}
    \label{fig:src_examples2}
\end{figure*}

DR1 also includes the publication of the eROSITA upper flux limit server (Tub{\'i}n-Arenas et al. \citeyear{Tubin2023}). The upper limits are calculated using X-ray photometry on the eROSITA standard calibration data products (counts image, background image, and exposure time), following the Bayesian approach described by \citet{Kraft1991}. Pre-computed upper limits are available for every pixel position in the eROSITA-DE sky at a confidence interval of 99.87\% (this corresponds to a one-sided 3$\sigma$ level). These values are stored using the hierarchical indices (HEALPix) framework to enable a fast search. The products are delivered for the 1B energy band (0.2--2.3~keV), as well as in all sub-bands of the 3B DET run (soft: 0.2--0.6~keV, medium: 0.6--2.3~keV, hard: 2.3--5.0~keV, and the combined total band: 0.2--5.0~keV). The upper flux limit data will be available in two ways: either by downloading the pre-computed data products or through a web tool. 

\subsection{Half-sky maps}
\label{sec:maps}
Half-sky counts, count rate and exposure maps are offered (in HiPS format) in seven different energy bands:
\begin{itemize}
    \item 0.2--2.3 keV (Main)
    \item 0.2--0.6 keV
    \item 0.6--2.3 keV
    \item 2.3--5.0 keV (Hard)
    \item 0.2--0.5 keV
    \item 0.5--1.0 keV
    \item 1.0--2.0 keV,
\end{itemize}
while the background maps and the sensitivity maps based on aperture photometry are released only for the first four energy bands above. As an example, Fig.~\ref{fig:maps} shows the half-sky images in Zenith Equal Area projection colour coded by count rate intensity in four bands: 0.2--0.6 keV, 0.6--2.3 keV, 2.3--5 keV and 0.2--2.3 keV, respectively. These images have been created using all TMs (also those affected by light leak), so some artifacts are still present in the lowest-energy map. On the other hand, at the highest energy (2.3--5 keV) the count rate maps are dominated by the (unvignetted) particle background; small temporal fluctuations of such background (including rare Coronal Mass Ejections hitting \SRG) are imprinted on those maps as stripes at fixed ecliptic longitudes\footnote{A coronal mass ejections (CME) may just miss a detection at Earth, but may still affect eROSITA depending on \SRG's orbit location, e.g. in mid-February 2020.}.

Zheng et al. \citeyear{Zheng2023} have created custom-made half-sky eRASS1 maps in the standard \ROSAT broad bands (R1-R7), with the goal of comparing the eRASS1 and RASS large scale emission in great detail. By only restricting to non-light-leak TMs (1, 2, 3, 4 and 6), and by carefully subtracting the non-vignetted particle background using FWC data, they produce clean maps almost free from artifacts. For more details on topics such as flux estimation of large-scale diffuse emission for the non-light-leaked TMs, we refer the reader to that work.

\subsection{Source products}
\label{sect:srctool}

Together with the catalogues of detected sources, eROSITA DR1 also contains a number of source-specific products, which are generated by the eSASS task \texttt{srctool}. In particular, we release source-specific products for 200\,217 Main catalogue entries with \DETLIKE{0} > 20. These products include source and background spectra, their respective ARFs and RMFs, light curves, and source and background event lists. All these source products are offered per telescope module and combined. 
In this section we describe how these products are generated.

\texttt{srctool} creates spectra, background spectra, response matrices, ancillary response files, and light curves for an input catalogue of sources. The tool chooses a circular source extraction region to optimise the signal to noise ratio of the source spectrum, given the local background surface brightness and the shape of the PSF, clipping the radius to a minimum of 15\arcsec\ and maximum of 99\% of the PSF encircled energy fraction, assuming a circular PSF, and taking into account any excluded neighbouring contaminating sources. A detailed explanation of the algorithm, including the background definition procedures, can be found online\footnote{\url{https://erosita.mpe.mpg.de/dr1/eSASS4DR1/eSASS4DR1_tasks/srctool_doc.html}}.
The final products are produced for each telescope module separately and combined for the telescope module groups TM0 (all seven TMs), TM8 (only those TMs with on-chip filter) and TM9 (only those TMs without on-chip filter).
The products for each source are placed into separate archive files in order to save disk space usage and reduce the number of files in the archive.
At odds with the rest of the pipeline, \texttt{srctool} was configured to use input GTIs calculated from the \texttt{flaregti} task, rather than the standard GTIs.
Light curves were created in 10s time bins, discarding those bins when the source is not visible.
The light curves were created in energy bands 0.2--0.6, 0.6--2.3 and 2.3--5.0 keV.

As eROSITA scans across the sky during the survey, a source samples different parts of each telescope and detector over time.
The role of \texttt{srctool} is to account for the varying vignetting of the telescope, bad pixels and the effect of photons lost outside the extraction region due to the shape of the PSF.
These effects are calculated by stepping through time a set sample points spatially sampling the source.
The average corrections are accounted for in the ancillary response matrix for spectral fitting, while the time-variable effects are included in the outputted light curve rates.
To avoid double counting these corrections, they are not included in the exposure time of the output files, which instead contain the time eROSITA is looking at the source.
Those sources with an extent of zero from the source detection pipeline are assumed to be point sources, while extended sources are assumed to follow a beta model with the core radius given by the extent and an outer power-law slope fixed by the exponent $\beta_c=2/3$ \citep[see][Appendix A.5]{Brunner2022}.
Table~\ref{tab:esass_versions} in the Appendix summarises the changes to the \texttt{srctool} package for DR1.

In Figs.~\ref{fig:src_examples1} and \ref{fig:src_examples2} we present four examples of light curves and spectra of four point sources, generated from the released products.

\section{Summary and outlook}
\label{sec:summary}
In this paper we give an overview of the operations, observations, data reduction and analysis of the first eROSITA all-sky survey for the hemisphere data (western Galactic) whose proprietary rights lie with the eROSITA-DE Consortium. We present the catalogues of X-ray sources extracted with the standard eSASS pipeline and describe quality control tests we have performed on the catalogued sources (astrometry, photometry, fidelity, etc.). Finally, we summarise the content of the data release associated with the eRASS1 survey so to facilitate interested users from the scientific community at large to interact with the rich data sets generated by \SRG/eROSITA.

The scientific and information content of the eRASS1 survey is large, varied and difficult to summarize in a few words. While our focus here is mainly on the eRASS1 X-ray catalogues and on the properties of the X-ray sources that are included there, other works focus on other relevant aspects of the all-sky survey data content. For example, already soon after the completion of eRASS1, \citet{Predehl2020} reported the discovery in the eROSITA all-sky image of large-scale X-ray bubbles in the Milky Way halo (the so-called `eROSITA bubbles'). Following up on that work, Zheng et al. (\citeyear{Zheng2023}) have studied maps of the diffuse emission as a function of energy and compared them with the \ROSAT ones. Locatelli et al. (\citeyear{Locatelli2023}) have then used those maps to derive structural parameters of the Milky Way Circum-Galactic Medium (CGM) emission. Using dense molecular clouds of known distance as shadowing tools \citet{Yeung2023} have derived constraints on the properties of the Local Hot Bubble from its soft X-ray emission measured by eROSITA.

 Based on these analysis, here we try to give a simplified answer to the question of what the survey contains by providing a photon (i.e. calibrated events) budget of eRASS1 split into its major physical components. In order to do so, we separate source from background counts on the basis of the catalogues source net counts in different bands. As for the background and foreground emission, we adopt here the simplifying assumption that the total spectrum in the eFEDS extra-galactic field is representative of the all-sky, and we adopt the best fit model of the diffuse emission spectrum presented in \citet{Ponti2023} to allocate photons to physical components. The result of this exercise is shown in Table~\ref{tab:erass1_by_number}. A more detailed inventory, as a function of position in the sky, can be found in Zheng et al. (\citeyear{Zheng2023}).

Time-domain variability analysis of the X-ray sources in eRASS1 will be presented in Boller et al. (in prep.), while Grotova et al. (in prep.) focus on the population of extra-galactic nuclear transients potentially associated to the tidal disruption of stars by SMBHs, which eROSITA is extremely sensitive to \citep[see e.g.][]{Malyali2021,Sazonov2021,Malyali2023,Homan2023,LiuZ2023}.

The details of the cross-identification methodology and of the classification schemes for the various classes of X-ray emitters contained in the eRASS1 catalogues will be described elsewhere. Specifically, the following papers will release the counterpart identification of: clusters of galaxies (Bulbul et al., \citeyear{Bulbul2023}; Kluge et al., submitted); AGN (Salvato et al., in prep., Waddel et al., submitted); Blazars (H{\"a}mmerich et al., in prep.); Cataclysmic Variables (Schwope et al., in prep.); coronal emitting stars  (Freund et al., \citeyear{Freund2023}); X-Ray Binaries (Avakyan et al., in prep.); X-ray source population in the Magellanic Clouds system (Maitra et al., in prep., Kaltenbrunner et al., in prep.).

\begin{table}[]
 \caption{eRASS1 by (rough) numbers.}
    \centering
    \begin{tabular}{lcc}
        \hline\hline
        Component  & \# of photons [10$^6$]  & fraction\\
                   &  [0.2--2.0 keV]         &         \\
        \hline
        CXB & 54 & 0.32\\
        MW hot CGM & 33 & 0.19 \\
        FWC & 32 & 0.19\\
        LHB & 17 & 0.10\\
        SWCX & 14 & 0.08\\
        PS & 16 & 0.09\\
        Ext. & 4 (1.6) & 0.02 (0.01)\\
        \hline
        Total & 170 & 1.0\\
        \hline
     \end{tabular}
     \tablefoot{We list simplified estimates of the approximate number of calibrated photon counts (in millions) registered by eROSITA in eRASS1 in the 0.2--2.0 keV band in one hemisphere, split into separate physical components: Cosmic X-ray Background (CXB), Milky Way hot Circum-Galactic Medium (MW hot CGM), Instrumental background (FWC), Local Hot Bubble (LHB), Solar Wind Change Exchange (SWCX), Point Sources (PS), Extended Sources (Ext.). For the latter component, we quote in parenthesis the number of net counts from confirmed clusters of galaxies (Bulbul et al., \citeyear{Bulbul2023}). See text for more details.}
    \label{tab:erass1_by_number}
\end{table}

\begin{table*}
    \centering
    \caption{Comparison among main catalogs from previous X-ray missions operating, at least partly, in the `classical' X-ray energy range ($\sim$0.2--10 keV).}
    \label{tab:catalogues_comparison}
    \begin{tabular}{lrclcl}
        \hline\hline
        Catalogue [{\it Mission}] & $N_{\rm objects}$ & Time span & $f_{\rm Area}$ & Energy coverage & Reference \\
        \hline
         4U [\uhuru] & 339 & 1970--1972 & 0.97 & 2.0--6.0 keV & (1)\\
         3A [{\it Ariel-V}] & 250 & 1974--1979 & 1.0 & 2.0--10 keV & (2)\\
         A1 [{\it HEAO-1}] & 842 & 1977--1978 & 1.0 & 0.25--25 keV & (3) \\
         IPC [\einstein] & 4000 & 1978--1981 & 0.33 & 0.3--3.5 keV & (4)\\
         2RXS [\ROSAT] & 135\,000 & 1990 & 1.0 & 0.1--2.4 keV & (5)\\
         WGACAT [\ROSAT] & 84\,000 & 1991--1995 & 0.18 & 0.1--2.4 keV & (6) \\
         CSC2.1 [\cxo] & 400\,000 & 1999--2022 & 0.019 & 0.2--7.0 keV & (7)\\
         4XMM-DR12 [\xmm] & 630\,000 & 2000--2022 & 0.031 & 0.2--12 keV & (8)\\
         4XMM-DR12 Hard [\xmm]$^*$ & 456\,000 & 2000--2022 & 0.031 & 2--5 keV & (8)\\ 
         XMMSL2 [\xmm] & 72\,000 & 2001--2014 & 0.84 & 0.2--12 keV & (9)\\
         2SXPS [\swift] & 206\,000 & 2005--2018 & 0.092 & 0.3--10 keV & (10)\\
         eFEDS [{\it SRG}/eROSITA] & 27\,000 & 2019 & 0.033 & 0.2--2.3 keV & (11)\\
         eRASS1 Main [{\it SRG}/eROSITA] & 930\,000 & 2019--2020 & 0.5 & 0.2--2.3 keV & This work\\
         eRASS1 Hard [{\it SRG}/eROSITA] & 5466 & 2019--2020 & 0.5 & 2.3--5.0 keV & This work\\
         \hline
         \end{tabular}  
         \tablefoot{The column $N_{\rm objects}$ lists the approximate number of sources in each catalogue and $f_{\rm Area}$ is the fraction of the sky observed.
         \\ $^*$The 4XMM-DR12 Hard catalogue (not shown in the figure) is derived from the 4XMM-DR12 by taking all sources for which the 2-5 keV flux is larger than the quoted 2-5 keV flux error.\\
         References: (1): \citet{1978form}; (2): \citet{Warwick1981}; (3) \citet{1984wood}; (4) \url{heasarc.gsfc.nasa.gov/W3Browse/einstein/ipc.html}, \citet{Harris1990}; (5): \citet{Boller2016}; (6): \url{heasarc.gsfc.nasa.gov/W3Browse/rosat/wgacat.html}; (7) \url{cxc.cfa.harvard.edu/csc/about2.1.html}; (8) \citet{Webb2020}; (9) \url{www.cosmos.esa.int/web/xmm-newton/xmmsl2-ug}; (10) \citet{Evans2020}; (11) \citet{Brunner2022}.}
\end{table*}

\begin{figure*}
    \centering
    \includegraphics[scale=0.5]{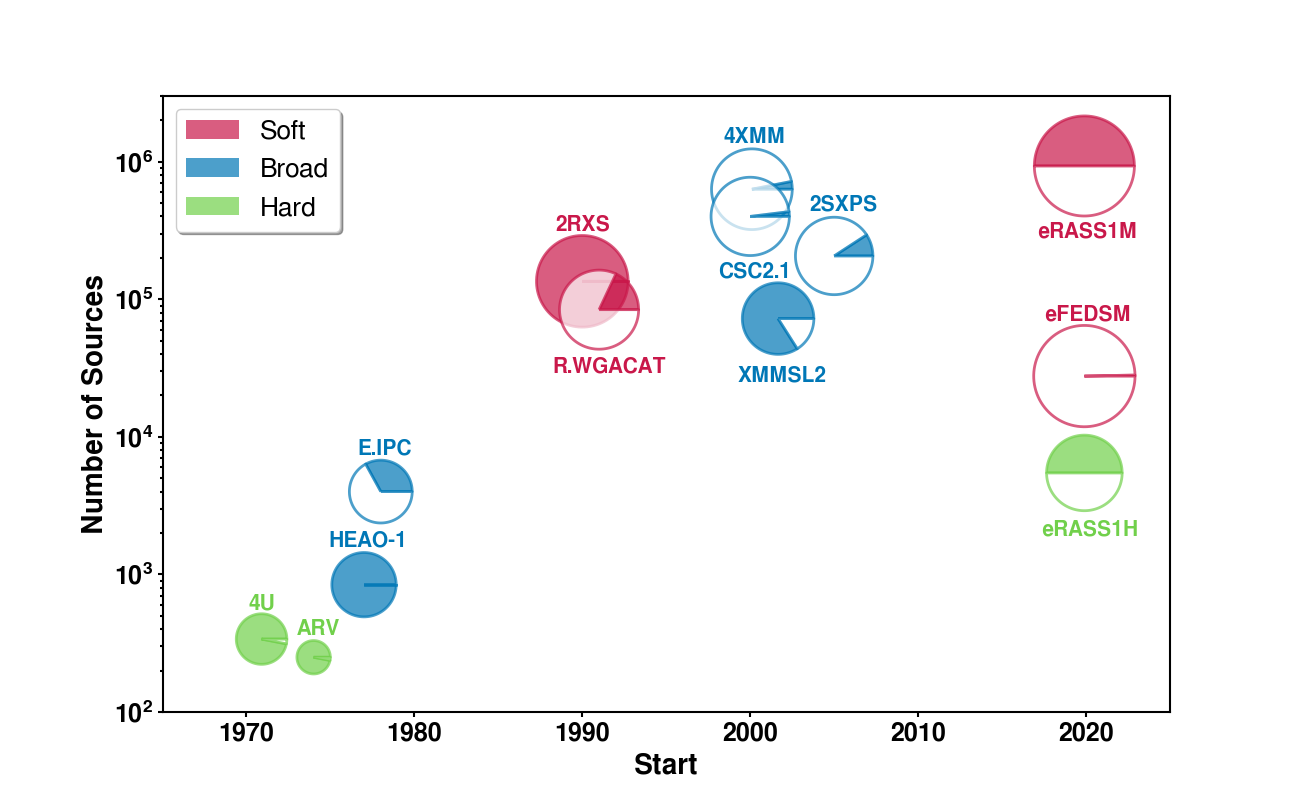}
    \caption{Graphical comparison of selected X-ray catalogs, based on the data presented in Table~\ref{tab:catalogues_comparison}. For each catalogue we plot the total number of objects vs. the time of the start of the corresponding data-taking period. Each catalogue is represented by a circle, whose radius is proportional to the logarithm of the ``discovery rate" (number of objects divided by the number of months of operations), while the shaded portion of the circle represents the fraction of the sky covered by the catalogue. The colour coding distinguishes telescopes operating mainly in the soft ($<2.5$ keV; red), hard ($>2$ keV; green) or broad (both soft and hard; blue) X-ray band. The datapoint corresponding to a putative 4XMM Hard catalog would sit very close to the full-band 4XMM one.}
    \label{fig:catalogs}
\end{figure*}

With almost one million entries, most of them never  detected before, the eRASS1 half-sky catalogues represent a major advance for our knowledge of the X-ray Universe. Table~\ref{tab:catalogues_comparison} and Fig.~\ref{fig:catalogs} present a comparison of eRASS1 with other catalogues from X-ray missions operating in the `classical' X-ray energy range 0.1--10 keV, and highlight the steady progress in X-ray survey capabilities culminating with the eROSITA catalogues we discuss here. Simply considering the union of all the unique objects catalogued by any previous X-ray mission (without removing possible overlaps), the eRASS1 main catalogue presented here increases the number of known X-ray sources in the published literature by more than  60\%.

The sensitive all-sky survey nature of the project implies that data are accumulated for a large variety of astronomical source classes, and for a plethora of possible science applications, well beyond the main mission-design-driving objectives; in other words, eROSITA data are endowed with tremendous {\it legacy} value. 
Indeed, existing all-sky and/or wide-area optical/IR surveys such as {\it Gaia}, SDSS, PanSTARRS, DES and Legacy Survey, HSC, VISTA/VHS, WISE already demonstrate the power of combining data sets for a deeper understanding of the high-energy Universe. Such a synergy extends also to longer wavelengths, at a time when the SKA precursors (LOFAR, MWA, ASKAP, MeerKAT) alongside the major observatories (APERTIF, JVLA) are surveying large swaths of the sky at unprecedented depth and speed. Beyond imaging, also large spectroscopic surveys have included among their main programs systematic follow-up of eRASS X-ray sources. Indeed, it is expected that SDSS-V \citep{Kollmeier2017} and 4MOST \citep{deJong2019,Merloni2019,Finoguenov2019}, by the end of their first survey period, will have accumulated hundreds of thousands of optical spectra of eROSITA X-ray sources.

Looking forward, future eROSITA data releases will comprise data from multiple all-sky surveys. Besides providing a deeper view of the X-ray sky, this will open up the systematic study of variability on months/year timescales. Extrapolating from the current data quality, and with the foreseen improvements of the calibration in sight, the next data releases will include high-fidelity catalogues with several million X-ray sources, closely following the expectations laid down in the early phases of the project \citep{Merloni2012}.

\begin{acknowledgements}
We thank the referee, Prof. F. Bauer, for the careful review and for the many constructive suggestions that significantly improved the final version of the paper.
This work is based on data from eROSITA, the soft X-ray instrument aboard \SRG, a joint Russian-German science mission supported by the Russian Space Agency (Roskosmos), in the interests of the Russian Academy of Sciences represented by its Space Research Institute (IKI), and the Deutsches Zentrum für Luft- und Raumfahrt (DLR). The \SRG spacecraft was built by Lavochkin Association (NPOL) and its subcontractors, and is operated by NPOL with support from the Max-Planck Institute for Extraterrestrial Physics (MPE).

The development and construction of the eROSITA X-ray instrument was led by MPE, with contributions from the Dr. Karl Remeis Observatory Bamberg, the University of Hamburg Observatory, the Leibniz Institute for Astrophysics Potsdam (AIP), and the Institute for Astronomy and Astrophysics of the University of T\"ubingen, with the support of DLR and the Max Planck Society. The Argelander Institute for Astronomy of the University of Bonn and the Ludwig Maximilians Universit\"at Munich also participated in the science preparation for eROSITA. 

The eROSITA data shown here were processed using the eSASS/NRTA software system developed by the German eROSITA consortium.

      Part of this work was supported by the German
      \emph{Deut\-sche For\-schungs\-ge\-mein\-schaft, DFG\/} project
      number Ts~17/2--1.

E.~Bulbul, A.~Liu, V.~Ghirardini, C.~Garrel and X.~Zhang acknowledge financial support from the European Research Council (ERC) Consolidator Grant under the European Union’s Horizon 2020 research and innovation programme (grant agreement CoG DarkQuest No 101002585). 

M. Brusa, A.~Georgakakis and B. Musiimenta acknowledge funding from the European Union’s Horizon 2020 research and innovation program under the Marie Skłodowska-Curie grant agreement No 860744. 

Z.~Igo, C.~Aydar, J.~Wolf acknowledge support by the Deutsche Forschungsgemeinschaft (DFG, German Research Foundation) under Germany’s Excellence Strategy - EXC-2094 - 390783311.

G.~Ponti acknowledges funding from the European Research Council (ERC) under the European Union’s Horizon 2020 research and innovation programme (grant agreement No 865637).

P.~Li is supported by the Alexander von Humboldt Foundation.

D.~Tubin-Arenas acknowledges support by DLR grant FKZ 50 OR 2203.

This research has made use of the SIMBAD database, operated at CDS, Strasbourg, France. 

\end{acknowledgements}

\bibliographystyle{aa} % style aa.bst
	\bibliography{erass1bib} % your references Yourfile.bib     

\begin{thebibliography}{165}
\expandafter\ifx\csname natexlab\endcsname\relax\def\natexlab#1{#1}\fi

\bibitem[{{Adami} {et~al.}(2018){Adami}, {Giles}, {Koulouridis}, {Pacaud},
  {Caretta}, {Pierre}, {Eckert}, {Ramos-Ceja}, {Gastaldello}, {Fotopoulou},
  {Guglielmo}, {Lidman}, {Sadibekova}, {Iovino}, {Maughan}, {Chiappetti},
  {Alis}, {Altieri}, {Baldry}, {Bottini}, {Birkinshaw}, {Bremer}, {Brown},
  {Cucciati}, {Driver}, {Elmer}, {Ettori}, {Evrard}, {Faccioli}, {Granett},
  {Grootes}, {Guzzo}, {Hopkins}, {Horellou}, {Lef{\`e}vre}, {Liske}, {Malek},
  {Marulli}, {Maurogordato}, {Owers}, {Paltani}, {Poggianti}, {Polletta},
  {Plionis}, {Pollo}, {Pompei}, {Ponman}, {Rapetti}, {Ricci}, {Robotham},
  {Tuffs}, {Tasca}, {Valtchanov}, {Vergani}, {Wagner}, {Willis}, \& {XXL
  Consortium}}]{Adami2018}
{Adami}, C., {Giles}, P., {Koulouridis}, E., {et~al.} 2018, \aap, 620, A5

\bibitem[{{Aird} {et~al.}(2015){Aird}, {Coil}, {Georgakakis}, {Nandra},
  {Barro}, \& {P{\'e}rez-Gonz{\'a}lez}}]{Aird2015}
{Aird}, J., {Coil}, A.~L., {Georgakakis}, A., {et~al.} 2015, \mnras, 451, 1892

\bibitem[{{Alexander} \& {Hickox}(2012)}]{Alexander2012}
{Alexander}, D.~M. \& {Hickox}, R.~C. 2012, \nar, 56, 93

\bibitem[{{Allen} {et~al.}(2011){Allen}, {Evrard}, \& {Mantz}}]{Allen2011}
{Allen}, S.~W., {Evrard}, A.~E., \& {Mantz}, A.~B. 2011, \araa, 49, 409

\bibitem[{{Ananna} {et~al.}(2019){Ananna}, {Treister}, {Urry}, {Ricci},
  {Kirkpatrick}, {LaMassa}, {Buchner}, {Civano}, {Tremmel}, \&
  {Marchesi}}]{Ananna2019}
{Ananna}, T.~T., {Treister}, E., {Urry}, C.~M., {et~al.} 2019, \apj, 871, 240

\bibitem[{{Arcangeli} {et~al.}(2017){Arcangeli}, {Borghi}, {Br{\"a}uninger},
  {Citterio}, {Ferrario}, {Friedrich}, {Grisoni}, {Marioni}, {Predehl},
  {Rossi}, {Ritucci}, {Valsecchi}, \& {Vernani}}]{Arcangeli2017}
{Arcangeli}, L., {Borghi}, G., {Br{\"a}uninger}, H., {et~al.} 2017, in Society
  of Photo-Optical Instrumentation Engineers (SPIE) Conference Series, Vol.
  10565, \procspie, 1056558

\bibitem[{{Arcodia} {et~al.}(2021){Arcodia}, {Merloni}, {Nandra}, {Buchner},
  {Salvato}, {Pasham}, {Remillard}, {Comparat}, {Lamer}, {Ponti}, {Malyali},
  {Wolf}, {Arzoumanian}, {Bogensberger}, {Buckley}, {Gendreau}, {Gromadzki},
  {Kara}, {Krumpe}, {Markwardt}, {Ramos-Ceja}, {Rau}, {Schramm}, \&
  {Schwope}}]{Arcodia2021}
{Arcodia}, R., {Merloni}, A., {Nandra}, K., {et~al.} 2021, \nat, 592, 704

\bibitem[{{Arnaud}(2005)}]{Arnaud2005}
{Arnaud}, M. 2005, in Background Microwave Radiation and Intracluster
  Cosmology, ed. F.~Melchiorri \& Y.~Rephaeli (Enrico Fermi International
  School of Physics Course CLIX), astro-ph/0508159

\bibitem[{{Bahcall}(1977)}]{Bahcall1977}
{Bahcall}, N.~A. 1977, \araa, 15, 505

\bibitem[{{Ballet}(1999)}]{Ballet1999}
{Ballet}, J. 1999, \aaps, 135, 371

\bibitem[{{Becker} \& {Truemper}(1997)}]{BEcker1997}
{Becker}, W. \& {Truemper}, J. 1997, \aap, 326, 682

\bibitem[{{Blasi}(2013)}]{Blasi2013}
{Blasi}, P. 2013, \aapr, 21, 70

\bibitem[{{Boch} \& {Fernique}(2014)}]{Boch2014}
{Boch}, T. \& {Fernique}, P. 2014, in Astronomical Society of the Pacific
  Conference Series, Vol. 485, Astronomical Data Analysis Software and Systems
  XXIII, ed. N.~{Manset} \& P.~{Forshay}, 277

\bibitem[{{Boller} {et~al.}(2016){Boller}, {Freyberg}, {Tr{\"u}mper}, {Haberl},
  {Voges}, \& {Nandra}}]{Boller2016}
{Boller}, T., {Freyberg}, M.~J., {Tr{\"u}mper}, J., {et~al.} 2016, \aap, 588,
  A103

\bibitem[{{Borgani}(2008)}]{Borgani2006}
{Borgani}, S. 2008, in Lecture Notes in Physics, Berlin Springer Verlag, Vol.
  740, A Pan-Chromatic View of Clusters of Galaxies and the Large-Scale
  Structure, ed. M.~{Plionis}, O.~{L{\'o}pez-Cruz}, \& D.~{Hughes}, 287

\bibitem[{{Borgani} \& {Kravtsov}(2011)}]{Borgani2009}
{Borgani}, S. \& {Kravtsov}, A. 2011, Advanced Science Letters, 4, 204

\bibitem[{{Brandt} \& {Alexander}(2015)}]{Brandt2015}
{Brandt}, W.~N. \& {Alexander}, D.~M. 2015, \aapr, 23, 1

\bibitem[{{Brandt} \& {Hasinger}(2005)}]{Brandt2005}
{Brandt}, W.~N. \& {Hasinger}, G. 2005, \araa, 43, 827

\bibitem[{{Brunner} {et~al.}(2022){Brunner}, {Liu}, {Lamer}, {Georgakakis},
  {Merloni}, {Brusa}, {Bulbul}, {Dennerl}, {Friedrich}, {Liu}, {Maitra},
  {Nandra}, {Ramos-Ceja}, {Sanders}, {Stewart}, {Boller}, {Buchner}, {Clerc},
  {Comparat}, {Dwelly}, {Eckert}, {Finoguenov}, {Freyberg}, {Ghirardini},
  {Gueguen}, {Haberl}, {Kreykenbohm}, {Krumpe}, {Osterhage}, {Pacaud},
  {Predehl}, {Reiprich}, {Robrade}, {Salvato}, {Santangelo}, {Schrabback},
  {Schwope}, \& {Wilms}}]{Brunner2022}
{Brunner}, H., {Liu}, T., {Lamer}, G., {et~al.} 2022, \aap, 661, A1

\bibitem[{{Buchner}(2016)}]{Buchner2016}
{Buchner}, J. 2016, Statistics and Computing, 26, 383

\bibitem[{{Buchner}(2019)}]{Buchner2019}
{Buchner}, J. 2019, \pasp, 131, 108005

\bibitem[{{Buchner}(2021)}]{Buchner2021}
{Buchner}, J. 2021, The Journal of Open Source Software, 6, 3001

\bibitem[{{Bulbul} {et~al.}(2022){Bulbul}, {Liu}, {Pasini}, {Comparat},
  {Hoang}, {Klein}, {Ghirardini}, {Salvato}, {Merloni}, {Seppi}, {Wolf},
  {Anderson}, {Bahar}, {Brusa}, {Br{\"u}ggen}, {Buchner}, {Dwelly},
  {Ibarra-Medel}, {Ider Chitham}, {Liu}, {Nandra}, {Ramos-Ceja}, {Sanders}, \&
  {Shen}}]{Bulbul2022}
{Bulbul}, E., {Liu}, A., {Pasini}, T., {et~al.} 2022, \aap, 661, A10

\bibitem[{{Bulbul} \& {Liu}(2024)}]{Bulbul2023}
{Bulbul}, E. \& {Liu}, A. e.~a. 2024, \aap, in press

\bibitem[{{Camilloni} {et~al.}(2023){Camilloni}, {Becker}, {Predehl},
  {Dennerl}, {Freyberg}, {Mayer}, \& {Sasaki}}]{Camilloni2023}
{Camilloni}, F., {Becker}, W., {Predehl}, P., {et~al.} 2023, \aap, 673, A45

\bibitem[{{Campana} {et~al.}(2021){Campana}, {Combes}, {Elbaz}, {Forveille},
  {Kotak}, {Pentericci}, \& {Shore}}]{Campana2021}
{Campana}, S., {Combes}, F., {Elbaz}, D., {et~al.} 2021, \aap, 647, E1

\bibitem[{{Cavaliere} \& {Fusco-Femiano}(1978)}]{Cavaliere1978}
{Cavaliere}, A. \& {Fusco-Femiano}, R. 1978, \aap, 70, 677

\bibitem[{{Cen} \& {Ostriker}(1999)}]{Cen1999}
{Cen}, R. \& {Ostriker}, J.~P. 1999, \apj, 514, 1

\bibitem[{{Churazov} {et~al.}(2023){Churazov}, {Khabibullin}, {Bykov},
  {Lyskova}, \& {Sunyaev}}]{Churazov2023}
{Churazov}, E., {Khabibullin}, I., {Bykov}, A.~M., {Lyskova}, N., \& {Sunyaev},
  R. 2023, \aap, 670, A156

\bibitem[{{Clerc} {et~al.}(2012){Clerc}, {Sadibekova}, {Pierre}, {Pacaud}, {Le
  F{\`e}vre}, {Adami}, {Altieri}, \& {Valtchanov}}]{Clerc2012}
{Clerc}, N., {Sadibekova}, T., {Pierre}, M., {et~al.} 2012, \mnras, 423, 3561

\bibitem[{{Comastri} {et~al.}(1995){Comastri}, {Setti}, {Zamorani}, \&
  {Hasinger}}]{Comastri1995}
{Comastri}, A., {Setti}, G., {Zamorani}, G., \& {Hasinger}, G. 1995, \aap, 296,
  1

\bibitem[{{Comparat} {et~al.}(2020){Comparat}, {Eckert}, {Finoguenov},
  {Schmidt}, {Sanders}, {Nagai}, {Lau}, {K�fer}, {Pacaud}, {Clerc},
  {Reiprich}, {Bulbul}, {Chitham}, {Chiang}, {Ghirardini}, {Gonzalez-Perez},
  {Gozaliasl}, {Fitzpatrick}, {Klypin}, {Merloni}, {Nandra}, {Liu}, {Prada},
  {Ramos-Ceja}, {Salvato}, {Seppi}, {Tempel}, \& {Yepes}}]{Comparat2020}
{Comparat}, J., {Eckert}, D., {Finoguenov}, A., {et~al.} 2020, The Open Journal
  of Astrophysics, 3, 13

\bibitem[{{Comparat} {et~al.}(2019){Comparat}, {Merloni}, {Salvato}, {Nandra},
  {Boller}, {Georgakakis}, {Finoguenov}, {Dwelly}, {Buchner}, {Del Moro},
  {Clerc}, {Wang}, {Zhao}, {Prada}, {Yepes}, {Brusa}, {Krumpe}, \&
  {Liu}}]{Comparat2019}
{Comparat}, J., {Merloni}, A., {Salvato}, M., {et~al.} 2019, \mnras, 487, 2005

\bibitem[{{Coutinho} {et~al.}(2022){Coutinho}, {Ramos-Ceja}, {Dennerl},
  {Haberl}, {Meidinger}, {Merloni}, {Predehl}, {Stewart}, {Freyberg},
  {Bornemann}, {Brunner}, {Burwitz}, {Czesla}, {Eder}, {Friedrich}, {Gaida},
  {Gueguen}, {Hartner}, {Kink}, {Kreykenbohm}, {Lamer}, {Maitra}, {Mernik},
  {Mueller}, {Nandra}, {Pfeffermann}, \& {Robrade}}]{Coutinho2022}
{Coutinho}, D., {Ramos-Ceja}, M.~E., {Dennerl}, K., {et~al.} 2022, in Society
  of Photo-Optical Instrumentation Engineers (SPIE) Conference Series, Vol.
  12181, Society of Photo-Optical Instrumentation Engineers (SPIE) Conference
  Series, ed. J.-W.~A. {den Herder}, S.~{Nikzad}, \& K.~{Nakazawa}, 121811A

\bibitem[{{Cutri} {et~al.}(2003){Cutri}, {Skrutskie}, {van Dyk}, {Beichman},
  {Carpenter}, {Chester}, {Cambresy}, {Evans}, {Fowler}, {Gizis}, {Howard},
  {Huchra}, {Jarrett}, {Kopan}, {Kirkpatrick}, {Light}, {Marsh}, {McCallon},
  {Schneider}, {Stiening}, {Sykes}, {Weinberg}, {Wheaton}, {Wheelock}, \&
  {Zacarias}}]{2003tmc..book.....C}
{Cutri}, R.~M., {Skrutskie}, M.~F., {van Dyk}, S., {et~al.} 2003, {2MASS All
  Sky Catalog of point sources.}

\bibitem[{{Cutri} {et~al.}(2021){Cutri}, {Wright}, {Conrow}, {Fowler},
  {Eisenhardt}, {Grillmair}, {Kirkpatrick}, {Masci}, {McCallon}, {Wheelock},
  {Fajardo-Acosta}, {Yan}, {Benford}, {Harbut}, {Jarrett}, {Lake}, {Leisawitz},
  {Ressler}, {Stanford}, {Tsai}, {Liu}, {Helou}, {Mainzer}, {Gettngs},
  {Gonzalez}, {Hoffman}, {Marsh}, {Padgett}, {Skrutskie}, {Beck}, {Papin}, \&
  {Wittman}}]{Cutri2014}
{Cutri}, R.~M., {Wright}, E.~L., {Conrow}, T., {et~al.} 2021, VizieR Online
  Data Catalog, II/328

\bibitem[{{Dauser} {et~al.}(2019){Dauser}, {Falkner}, {Lorenz}, {Kirsch},
  {Peille}, {Cucchetti}, {Schmid}, {Brand}, {Oertel}, {Smith}, \&
  {Wilms}}]{Dauser2019}
{Dauser}, T., {Falkner}, S., {Lorenz}, M., {et~al.} 2019, \aap, 630, A66

\bibitem[{{Dav{\'e}} {et~al.}(2001){Dav{\'e}}, {Cen}, {Ostriker}, {Bryan},
  {Hernquist}, {Katz}, {Weinberg}, {Norman}, \& {O'Shea}}]{Dave2001}
{Dav{\'e}}, R., {Cen}, R., {Ostriker}, J.~P., {et~al.} 2001, \apj, 552, 473

\bibitem[{{Davis}(2001)}]{Davis2001}
{Davis}, J.~E. 2001, \apj, 562, 575

\bibitem[{{de Jong} {et~al.}(2019){de Jong}, {Agertz}, {Berbel}, {Aird},
  {Alexander}, {Amarsi}, {Anders}, {Andrae}, {Ansarinejad}, {Ansorge},
  {Antilogus}, {Anwand-Heerwart}, {Arentsen}, {Arnadottir}, {Asplund}, {Auger},
  {Azais}, {Baade}, {Baker}, {Baker}, {Balbinot}, {Baldry}, {Banerji},
  {Barden}, {Barklem}, {Barth{\'e}l{\'e}my-Mazot}, {Battistini}, {Bauer},
  {Bell}, {Bellido-Tirado}, {Bellstedt}, {Belokurov}, {Bensby}, {Bergemann},
  {Bestenlehner}, {Bielby}, {Bilicki}, {Blake}, {Bland-Hawthorn}, {Boeche},
  {Boland}, {Boller}, {Bongard}, {Bongiorno}, {Bonifacio}, {Boudon}, {Brooks},
  {Brown}, {Brown}, {Br{\"u}ggen}, {Brynnel}, {Brzeski}, {Buchert},
  {Buschkamp}, {Caffau}, {Caillier}, {Carrick}, {Casagrande}, {Case}, {Casey},
  {Cesarini}, {Cescutti}, {Chapuis}, {Chiappini}, {Childress}, {Christlieb},
  {Church}, {Cioni}, {Cluver}, {Colless}, {Collett}, {Comparat}, {Cooper},
  {Couch}, {Courbin}, {Croom}, {Croton}, {Daguis{\'e}}, {Dalton}, {Davies},
  {Davis}, {de Laverny}, {Deason}, {Dionies}, {Disseau}, {Doel}, {D{\"o}scher},
  {Driver}, {Dwelly}, {Eckert}, {Edge}, {Edvardsson}, {Youssoufi}, {Elhaddad},
  {Enke}, {Erfanianfar}, {Farrell}, {Fechner}, {Feiz}, {Feltzing}, {Ferreras},
  {Feuerstein}, {Feuillet}, {Finoguenov}, {Ford}, {Fotopoulou}, {Fouesneau},
  {Frenk}, {Frey}, {Gaessler}, {Geier}, {Gentile Fusillo}, {Gerhard},
  {Giannantonio}, {Giannone}, {Gibson}, {Gillingham},
  {Gonz{\'a}lez-Fern{\'a}ndez}, {Gonzalez-Solares}, {Gottloeber}, {Gould},
  {Grebel}, {Gueguen}, {Guiglion}, {Haehnelt}, {Hahn}, {Hansen}, {Hartman},
  {Hauptner}, {Hawkins}, {Haynes}, {Haynes}, {Heiter}, {Helmi}, {Aguayo},
  {Hewett}, {Hinton}, {Hobbs}, {Hoenig}, {Hofman}, {Hook}, {Hopgood},
  {Hopkins}, {Hourihane}, {Howes}, {Howlett}, {Huet}, {Irwin}, {Iwert},
  {Jablonka}, {Jahn}, {Jahnke}, {Jarno}, {Jin}, {Jofre}, {Johl}, {Jones},
  {J{\"o}nsson}, {Jordan}, {Karovicova}, {Khalatyan}, {Kelz}, {Kennicutt},
  {King}, {Kitaura}, {Klar}, {Klauser}, {Kneib}, {Koch}, {Koposov},
  {Kordopatis}, {Korn}, {Kosmalski}, {Kotak}, {Kovalev}, {Kreckel}, {Kripak},
  {Krumpe}, {Kuijken}, {Kunder}, {Kushniruk}, {Lam}, {Lamer}, {Laurent},
  {Lawrence}, {Lehmitz}, {Lemasle}, {Lewis}, {Li}, {Lidman}, {Lind}, {Liske},
  {Lizon}, {Loveday}, {Ludwig}, {McDermid}, {Maguire}, {Mainieri}, {Mali},
  {Mandel}, {Mandel}, {Mannering}, {Martell}, {Martinez Delgado}, {Matijevic},
  {McGregor}, {McMahon}, {McMillan}, {Mena}, {Merloni}, {Meyer}, {Michel},
  {Micheva}, {Migniau}, {Minchev}, {Monari}, {Muller}, {Murphy},
  {Muthukrishna}, {Nandra}, {Navarro}, {Ness}, {Nichani}, {Nichol}, {Nicklas},
  {Niederhofer}, {Norberg}, {Obreschkow}, {Oliver}, {Owers}, {Pai},
  {Pankratow}, {Parkinson}, {Paschke}, {Paterson}, {Pecontal}, {Parry},
  {Phillips}, {Pillepich}, {Pinard}, {Pirard}, {Piskunov}, {Plank},
  {Pl{\"u}schke}, {Pons}, {Popesso}, {Power}, {Pragt}, {Pramskiy}, {Pryer},
  {Quattri}, {Queiroz}, {Quirrenbach}, {Rahurkar}, {Raichoor}, {Ramstedt},
  {Rau}, {Recio-Blanco}, {Reiss}, {Renaud}, {Revaz}, {Rhode}, {Richard},
  {Richter}, {Rix}, {Robotham}, {Roelfsema}, {Romaniello}, {Rosario},
  {Rothmaier}, {Roukema}, {Ruchti}, {Rupprecht}, {Rybizki}, {Ryde}, {Saar},
  {Sadler}, {Sahl{\'e}n}, {Salvato}, {Sassolas}, {Saunders}, {Saviauk},
  {Sbordone}, {Schmidt}, {Schnurr}, {Scholz}, {Schwope}, {Seifert}, {Shanks},
  {Sheinis}, {Sivov}, {Sk{\'u}lad{\'o}ttir}, {Smartt}, {Smedley}, {Smith},
  {Smith}, {Sorce}, {Spitler}, {Starkenburg}, {Steinmetz}, {Stilz}, {Storm},
  {Sullivan}, {Sutherland}, {Swann}, {Tamone}, {Taylor}, {Teillon}, {Tempel},
  {ter Horst}, {Thi}, {Tolstoy}, {Trager}, {Traven}, {Tremblay}, {Tresse},
  {Valentini}, {van de Weygaert}, {van den Ancker}, {Veljanoski}, {Venkatesan},
  {Wagner}, {Wagner}, {Walcher}, {Waller}, {Walton}, {Wang}, {Winkler},
  {Wisotzki}, {Worley}, {Worseck}, {Xiang}, {Xu}, {Yong}, {Zhao}, {Zheng},
  {Zscheyge}, \& {Zucker}}]{deJong2019}
{de Jong}, R.~S., {Agertz}, O., {Berbel}, A.~A., {et~al.} 2019, The Messenger,
  175, 3

\bibitem[{{Dennerl} {et~al.}(2020){Dennerl}, {Andritschke}, {Br{\"a}uninger},
  {Burkert}, {Burwitz}, {Emberger}, {Freyberg}, {Friedrich}, {Gaida},
  {Granato}, {Hartner}, {von Kienlin}, {Meidinger}, {Menz}, \&
  {Predehl}}]{Dennerl2020}
{Dennerl}, K., {Andritschke}, R., {Br{\"a}uninger}, H., {et~al.} 2020, in
  Society of Photo-Optical Instrumentation Engineers (SPIE) Conference Series,
  Vol. 11444, Society of Photo-Optical Instrumentation Engineers (SPIE)
  Conference Series, 114444Q

\bibitem[{{Elvis}(2020)}]{Elvis2020}
{Elvis}, M. 2020, Nature Astronomy, 4, 23

\bibitem[{{Evans} {et~al.}(2020){Evans}, {Page}, {Osborne}, {Beardmore},
  {Willingale}, {Burrows}, {Kennea}, {Perri}, {Capalbi}, {Tagliaferri}, \&
  {Cenko}}]{Evans2020}
{Evans}, P.~A., {Page}, K.~L., {Osborne}, J.~P., {et~al.} 2020, \apjs, 247, 54

\bibitem[{{Fabian}(2012)}]{Fabian2012}
{Fabian}, A.~C. 2012, \araa, 50, 455

\bibitem[{{Fender} {et~al.}(2004){Fender}, {Belloni}, \& {Gallo}}]{Fender2004}
{Fender}, R.~P., {Belloni}, T.~M., \& {Gallo}, E. 2004, \mnras, 355, 1105

\bibitem[{{Finoguenov} {et~al.}(2019){Finoguenov}, {Merloni}, {Comparat},
  {Nandra}, {Salvato}, {Tempel}, {Raichoor}, {Richard}, {Kneib}, {Pillepich},
  {Sahl{\'e}n}, {Popesso}, {Norberg}, {McMahon}, \& {4MOST
  Collaboration}}]{Finoguenov2019}
{Finoguenov}, A., {Merloni}, A., {Comparat}, J., {et~al.} 2019, The Messenger,
  175, 39

\bibitem[{{Forman} {et~al.}(1978){Forman}, {Jones}, {Cominsky}, {Julien},
  {Murray}, {Peters}, {Tananbaum}, \& {Giacconi}}]{1978form}
{Forman}, W., {Jones}, C., {Cominsky}, L., {et~al.} 1978, \apjs, 38, 357

\bibitem[{{Foster} {et~al.}(2022){Foster}, {Poppenhaeger}, {Ilic}, \&
  {Schwope}}]{Foster2022}
{Foster}, G., {Poppenhaeger}, K., {Ilic}, N., \& {Schwope}, A. 2022, \aap, 661,
  A23

\bibitem[{{Freund} \& {Czesla}(2024)}]{Freund2023}
{Freund}, S. \& {Czesla}, S. e.~a. 2024, \aap, in press

\bibitem[{{Friedrich} {et~al.}(2012){Friedrich}, {Br{\"a}uninger}, {Budau},
  {Burkert}, {Burwitz}, {Dennerl}, {Eder}, {Freyberg}, {Gaida}, {Hartner},
  {Menz}, {Pfeffermann}, {Predehl}, {Roh{\'e}}, \& {Schreib}}]{Friedrich2012}
{Friedrich}, P., {Br{\"a}uninger}, H., {Budau}, B., {et~al.} 2012, in Society
  of Photo-Optical Instrumentation Engineers (SPIE) Conference Series, Vol.
  8443, Space Telescopes and Instrumentation 2012: Ultraviolet to Gamma Ray,
  ed. T.~{Takahashi}, S.~S. {Murray}, \& J.-W.~A. {den Herder}, 84431S

\bibitem[{{Friedrich} {et~al.}(2008){Friedrich}, {Br{\"a}uninger}, {Budau},
  {Burkert}, {Eder}, {Freyberg}, {Hartner}, {M{\"u}hlegger}, {Predehl},
  {Erhard}, {Gutruf}, {Jugler}, {Kampf}, {Borghi}, {Citterio}, {Rossi},
  {Valsecchi}, {Vernani}, \& {Zimmermann}}]{Friedrich2008}
{Friedrich}, P., {Br{\"a}uninger}, H., {Budau}, B., {et~al.} 2008, in Society
  of Photo-Optical Instrumentation Engineers (SPIE) Conference Series, Vol.
  7011, \procspie, 70112T

\bibitem[{F{\"u}rmetz {et~al.}(2008)F{\"u}rmetz, Pfeffermann, Predehl,
  Roh{\'e}, \& Tiedemann}]{Fuermetz2008}
F{\"u}rmetz, M., Pfeffermann, E., Predehl, P., Roh{\'e}, C., \& Tiedemann, L.
  2008, in Space Telescopes and Instrumentation 2008: Ultraviolet to Gamma Ray,
  Vol. 7011, International Society for Optics and Photonics, 70113Y

\bibitem[{{Gaia Collaboration} {et~al.}(2023){Gaia Collaboration}, {Vallenari},
  {Brown}, {Prusti}, {de Bruijne}, {Arenou}, {Babusiaux}, {Biermann},
  {Creevey}, {Ducourant}, {Evans}, {Eyer}, {Guerra}, {Hutton}, {Jordi},
  {Klioner}, {Lammers}, {Lindegren}, {Luri}, {Mignard}, {Panem}, {Pourbaix},
  {Randich}, {Sartoretti}, {Soubiran}, {Tanga}, {Walton}, {Bailer-Jones},
  {Bastian}, {Drimmel}, {Jansen}, {Katz}, {Lattanzi}, {van Leeuwen}, {Bakker},
  {Cacciari}, {Casta{\~n}eda}, {De Angeli}, {Fabricius}, {Fouesneau},
  {Fr{\'e}mat}, {Galluccio}, {Guerrier}, {Heiter}, {Masana}, {Messineo},
  {Mowlavi}, {Nicolas}, {Nienartowicz}, {Pailler}, {Panuzzo}, {Riclet}, {Roux},
  {Seabroke}, {Sordo}, {Th{\'e}venin}, {Gracia-Abril}, {Portell}, {Teyssier},
  {Altmann}, {Andrae}, {Audard}, {Bellas-Velidis}, {Benson}, {Berthier},
  {Blomme}, {Burgess}, {Busonero}, {Busso}, {C{\'a}novas}, {Carry}, {Cellino},
  {Cheek}, {Clementini}, {Damerdji}, {Davidson}, {de Teodoro}, {Nu{\~n}ez
  Campos}, {Delchambre}, {Dell'Oro}, {Esquej}, {Fern{\'a}ndez-Hern{\'a}ndez},
  {Fraile}, {Garabato}, {Garc{\'\i}a-Lario}, {Gosset}, {Haigron}, {Halbwachs},
  {Hambly}, {Harrison}, {Hern{\'a}ndez}, {Hestroffer}, {Hodgkin}, {Holl},
  {Jan{\ss}en}, {Jevardat de Fombelle}, {Jordan}, {Krone-Martins}, {Lanzafame},
  {L{\"o}ffler}, {Marchal}, {Marrese}, {Moitinho}, {Muinonen}, {Osborne},
  {Pancino}, {Pauwels}, {Recio-Blanco}, {Reyl{\'e}}, {Riello}, {Rimoldini},
  {Roegiers}, {Rybizki}, {Sarro}, {Siopis}, {Smith}, {Sozzetti}, {Utrilla},
  {van Leeuwen}, {Abbas}, {{\'A}brah{\'a}m}, {Abreu Aramburu}, {Aerts},
  {Aguado}, {Ajaj}, {Aldea-Montero}, {Altavilla}, {{\'A}lvarez}, {Alves},
  {Anders}, {Anderson}, {Anglada Varela}, {Antoja}, {Baines}, {Baker},
  {Balaguer-N{\'u}{\~n}ez}, {Balbinot}, {Balog}, {Barache}, {Barbato},
  {Barros}, {Barstow}, {Bartolom{\'e}}, {Bassilana}, {Bauchet}, {Becciani},
  {Bellazzini}, {Berihuete}, {Bernet}, {Bertone}, {Bianchi}, {Binnenfeld},
  {Blanco-Cuaresma}, {Blazere}, {Boch}, {Bombrun}, {Bossini}, {Bouquillon},
  {Bragaglia}, {Bramante}, {Breedt}, {Bressan}, {Brouillet}, {Brugaletta},
  {Bucciarelli}, {Burlacu}, {Butkevich}, {Buzzi}, {Caffau}, {Cancelliere},
  {Cantat-Gaudin}, {Carballo}, {Carlucci}, {Carnerero}, {Carrasco},
  {Casamiquela}, {Castellani}, {Castro-Ginard}, {Chaoul}, {Charlot}, {Chemin},
  {Chiaramida}, {Chiavassa}, {Chornay}, {Comoretto}, {Contursi}, {Cooper},
  {Cornez}, {Cowell}, {Crifo}, {Cropper}, {Crosta}, {Crowley}, {Dafonte},
  {Dapergolas}, {David}, {David}, {de Laverny}, {De Luise}, {De March}, {De
  Ridder}, {de Souza}, {de Torres}, {del Peloso}, {del Pozo}, {Delbo},
  {Delgado}, {Delisle}, {Demouchy}, {Dharmawardena}, {Di Matteo}, {Diakite},
  {Diener}, {Distefano}, {Dolding}, {Edvardsson}, {Enke}, {Fabre}, {Fabrizio},
  {Faigler}, {Fedorets}, {Fernique}, {Fienga}, {Figueras}, {Fournier},
  {Fouron}, {Fragkoudi}, {Gai}, {Garcia-Gutierrez}, {Garcia-Reinaldos},
  {Garc{\'\i}a-Torres}, {Garofalo}, {Gavel}, {Gavras}, {Gerlach}, {Geyer},
  {Giacobbe}, {Gilmore}, {Girona}, {Giuffrida}, {Gomel}, {Gomez},
  {Gonz{\'a}lez-N{\'u}{\~n}ez}, {Gonz{\'a}lez-Santamar{\'\i}a},
  {Gonz{\'a}lez-Vidal}, {Granvik}, {Guillout}, {Guiraud},
  {Guti{\'e}rrez-S{\'a}nchez}, {Guy}, {Hatzidimitriou}, {Hauser}, {Haywood},
  {Helmer}, {Helmi}, {Sarmiento}, {Hidalgo}, {Hilger}, {H{\l}adczuk}, {Hobbs},
  {Holland}, {Huckle}, {Jardine}, {Jasniewicz}, {Jean-Antoine Piccolo},
  {Jim{\'e}nez-Arranz}, {Jorissen}, {Juaristi Campillo}, {Julbe}, {Karbevska},
  {Kervella}, {Khanna}, {Kontizas}, {Kordopatis}, {Korn}, {K{\'o}sp{\'a}l},
  {Kostrzewa-Rutkowska}, {Kruszy{\'n}ska}, {Kun}, {Laizeau}, {Lambert},
  {Lanza}, {Lasne}, {Le Campion}, {Lebreton}, {Lebzelter}, {Leccia}, {Leclerc},
  {Lecoeur-Taibi}, {Liao}, {Licata}, {Lindstr{\o}m}, {Lister}, {Livanou},
  {Lobel}, {Lorca}, {Loup}, {Madrero Pardo}, {Magdaleno Romeo}, {Managau},
  {Mann}, {Manteiga}, {Marchant}, {Marconi}, {Marcos}, {Marcos Santos},
  {Mar{\'\i}n Pina}, {Marinoni}, {Marocco}, {Marshall}, {Martin Polo},
  {Mart{\'\i}n-Fleitas}, {Marton}, {Mary}, {Masip}, {Massari},
  {Mastrobuono-Battisti}, {Mazeh}, {McMillan}, {Messina}, {Michalik}, {Millar},
  {Mints}, {Molina}, {Molinaro}, {Moln{\'a}r}, {Monari}, {Mongui{\'o}},
  {Montegriffo}, {Montero}, {Mor}, {Mora}, {Morbidelli}, {Morel}, {Morris},
  {Muraveva}, {Murphy}, {Musella}, {Nagy}, {Noval}, {Oca{\~n}a}, {Ogden},
  {Ordenovic}, {Osinde}, {Pagani}, {Pagano}, {Palaversa}, {Palicio},
  {Pallas-Quintela}, {Panahi}, {Payne-Wardenaar}, {Pe{\~n}alosa Esteller},
  {Penttil{\"a}}, {Pichon}, {Piersimoni}, {Pineau}, {Plachy}, {Plum}, {Poggio},
  {Pr{\v{s}}a}, {Pulone}, {Racero}, {Ragaini}, {Rainer}, {Raiteri}, {Rambaux},
  {Ramos}, {Ramos-Lerate}, {Re Fiorentin}, {Regibo}, {Richards}, {Rios Diaz},
  {Ripepi}, {Riva}, {Rix}, {Rixon}, {Robichon}, {Robin}, {Robin}, {Roelens},
  {Rogues}, {Rohrbasser}, {Romero-G{\'o}mez}, {Rowell}, {Royer}, {Ruz Mieres},
  {Rybicki}, {Sadowski}, {S{\'a}ez N{\'u}{\~n}ez}, {Sagrist{\`a} Sell{\'e}s},
  {Sahlmann}, {Salguero}, {Samaras}, {Sanchez Gimenez}, {Sanna},
  {Santove{\~n}a}, {Sarasso}, {Schultheis}, {Sciacca}, {Segol}, {Segovia},
  {S{\'e}gransan}, {Semeux}, {Shahaf}, {Siddiqui}, {Siebert}, {Siltala},
  {Silvelo}, {Slezak}, {Slezak}, {Smart}, {Snaith}, {Solano}, {Solitro},
  {Souami}, {Souchay}, {Spagna}, {Spina}, {Spoto}, {Steele},
  {Steidelm{\"u}ller}, {Stephenson}, {S{\"u}veges}, {Surdej}, {Szabados},
  {Szegedi-Elek}, {Taris}, {Taylor}, {Teixeira}, {Tolomei}, {Tonello}, {Torra},
  {Torra}, {Torralba Elipe}, {Trabucchi}, {Tsounis}, {Turon}, {Ulla}, {Unger},
  {Vaillant}, {van Dillen}, {van Reeven}, {Vanel}, {Vecchiato}, {Viala},
  {Vicente}, {Voutsinas}, {Weiler}, {Wevers}, {Wyrzykowski}, {Yoldas}, {Yvard},
  {Zhao}, {Zorec}, {Zucker}, \& {Zwitter}}]{2023A&A...674A...1G}
{Gaia Collaboration}, {Vallenari}, A., {Brown}, A.~G.~A., {et~al.} 2023, \aap,
  674, A1

\bibitem[{{Georgakakis} {et~al.}(2008){Georgakakis}, {Nandra}, {Laird}, {Aird},
  \& {Trichas}}]{Georgakakis2008}
{Georgakakis}, A., {Nandra}, K., {Laird}, E.~S., {Aird}, J., \& {Trichas}, M.
  2008, \mnras, 388, 1205

\bibitem[{{Ghirardini} {et~al.}(2021){Ghirardini}, {Bulbul}, {Hoang}, {Klein},
  {Okabe}, {Biffi}, {Br{\"u}ggen}, {Ramos-Ceja}, {Comparat}, {Oguri},
  {Shimwell}, {Basu}, {Bonafede}, {Botteon}, {Brunetti}, {Cassano}, {de
  Gasperin}, {Dennerl}, {Gatuzz}, {Gastaldello}, {Intema}, {Merloni}, {Nandra},
  {Pacaud}, {Predehl}, {Reiprich}, {Robrade}, {R{\"o}ttgering}, {Sanders}, {van
  Weeren}, \& {Williams}}]{Ghirardini2021}
{Ghirardini}, V., {Bulbul}, E., {Hoang}, D.~N., {et~al.} 2021, \aap, 647, A4

\bibitem[{{Giacconi} {et~al.}(1979){Giacconi}, {Branduardi}, {Briel},
  {Epstein}, {Fabricant}, {Feigelson}, {Forman}, {Gorenstein}, {Grindlay},
  {Gursky}, {Harnden}, {Henry}, {Jones}, {Kellogg}, {Koch}, {Murray},
  {Schreier}, {Seward}, {Tananbaum}, {Topka}, {Van Speybroeck}, {Holt},
  {Becker}, {Boldt}, {Serlemitsos}, {Clark}, {Canizares}, {Markert}, {Novick},
  {Helfand}, \& {Long}}]{1979giacc}
{Giacconi}, R., {Branduardi}, G., {Briel}, U., {et~al.} 1979, \apj, 230, 540

\bibitem[{{Giacconi} {et~al.}(1962){Giacconi}, {Gursky}, {Paolini}, \&
  {Rossi}}]{Giacconi1962}
{Giacconi}, R., {Gursky}, H., {Paolini}, F.~R., \& {Rossi}, B.~B. 1962, \prl,
  9, 439

\bibitem[{{Giacconi} {et~al.}(1971){Giacconi}, {Kellogg}, {Gorenstein},
  {Gursky}, \& {Tananbaum}}]{1971giacc}
{Giacconi}, R., {Kellogg}, E., {Gorenstein}, P., {Gursky}, H., \& {Tananbaum},
  H. 1971, \apjl, 165, L27

\bibitem[{{Gilli} {et~al.}(2007){Gilli}, {Comastri}, \& {Hasinger}}]{Gilli2007}
{Gilli}, R., {Comastri}, A., \& {Hasinger}, G. 2007, \aap, 463, 79

\bibitem[{Gokus {et~al.}(2020)Gokus, Rau, Wilms, Ducci, Koenig, Weber, Boller,
  \& Malyali}]{gokus_srgt_2020}
Gokus, A., Rau, A., Wilms, J., {et~al.} 2020, The Astronomer's Telegram, 13657,
  1

\bibitem[{{G{\'o}rski} {et~al.}(2005){G{\'o}rski}, {Hivon}, {Banday},
  {Wandelt}, {Hansen}, {Reinecke}, \& {Bartelmann}}]{Gorski2005}
{G{\'o}rski}, K.~M., {Hivon}, E., {Banday}, A.~J., {et~al.} 2005, \apj, 622,
  759

\bibitem[{{Haberl}(2007)}]{Haberl2007}
{Haberl}, F. 2007, \apss, 308, 181

\bibitem[{{Haberl} {et~al.}(2022){Haberl}, {Maitra}, {Carpano}, {Dai},
  {Doroshenko}, {Dennerl}, {Freyberg}, {Sasaki}, {Udalski}, {Postnov}, \&
  {Shakura}}]{Haberl2022}
{Haberl}, F., {Maitra}, C., {Carpano}, S., {et~al.} 2022, \aap, 661, A25

\bibitem[{Haberl {et~al.}(2020)Haberl, Wilms, Gokus, Kreykenbohm, Weber,
  Koenig, Maitra, Carpano, \& Vasilopoulos}]{haberl_srgerosita_2020}
Haberl, F., Wilms, J., Gokus, A., {et~al.} 2020, The Astronomer's Telegram,
  13828, 1

\bibitem[{{Haiman} {et~al.}(2005){Haiman}, {Allen}, {Bahcall}, {Bautz},
  {Boehringer}, {Borgani}, {Bryan}, {Cabrera}, {Canizares}, {Citterio},
  {Evrard}, {Finoguenov}, {Griffiths}, {Hasinger}, {Henry}, {Jahoda},
  {Jernigan}, {Kahn}, {Lamb}, {Majumdar}, {Mohr}, {Molendi}, {Mushotzky},
  {Pareschi}, {Peterson}, {Petre}, {Predehl}, {Rasmussen}, {Ricker}, {Ricker},
  {Rosati}, {Sanderson}, {Stanford}, {Voit}, {Wang}, {White}, \&
  {White}}]{2005haim}
{Haiman}, Z., {Allen}, S., {Bahcall}, N., {et~al.} 2005, arXiv e-prints, astro

\bibitem[{{Harris}(1990)}]{Harris1990}
{Harris}, D.~E. 1990, {The Einstein Observatory Catalog of IPC X-ray Sources}

\bibitem[{{Hasinger} {et~al.}(2005){Hasinger}, {Miyaji}, \&
  {Schmidt}}]{Hasinger2005}
{Hasinger}, G., {Miyaji}, T., \& {Schmidt}, M. 2005, \aap, 441, 417

\bibitem[{{Hickox} {et~al.}(2009){Hickox}, {Jones}, {Forman}, {Murray},
  {Kochanek}, {Eisenstein}, {Jannuzi}, {Dey}, {Brown}, {Stern}, {Eisenhardt},
  {Gorjian}, {Brodwin}, {Narayan}, {Cool}, {Kenter}, {Caldwell}, \&
  {Anderson}}]{Hickox2009}
{Hickox}, R.~C., {Jones}, C., {Forman}, W.~R., {et~al.} 2009, \apj, 696, 891

\bibitem[{{Hickox} \& {Markevitch}(2006)}]{Hickox2006}
{Hickox}, R.~C. \& {Markevitch}, M. 2006, \apj, 645, 95

\bibitem[{{H{\o}g} {et~al.}(2000){H{\o}g}, {Fabricius}, {Makarov}, {Urban},
  {Corbin}, {Wycoff}, {Bastian}, {Schwekendiek}, \&
  {Wicenec}}]{2000A&A...355L..27H}
{H{\o}g}, E., {Fabricius}, C., {Makarov}, V.~V., {et~al.} 2000, \aap, 355, L27

\bibitem[{{Homan} {et~al.}(2023){Homan}, {Krumpe}, {Markowitz}, {Saha},
  {Gokus}, {Partington}, {Lamer}, {Malyali}, {Liu}, {Rau}, {Grotova},
  {Cackett}, {Buckley}, {Ciroi}, {Di Mille}, {Gendreau}, {Gromadzki},
  {Krishnan}, {Schramm}, \& {Steiner}}]{Homan2023}
{Homan}, D., {Krumpe}, M., {Markowitz}, A., {et~al.} 2023, \aap, 672, A167

\bibitem[{{Hopkins} {et~al.}(2008){Hopkins}, {Hernquist}, {Cox}, \&
  {Kere{\v{s}}}}]{Hopkins2008}
{Hopkins}, P.~F., {Hernquist}, L., {Cox}, T.~J., \& {Kere{\v{s}}}, D. 2008,
  \apjs, 175, 356

\bibitem[{{Kammoun} {et~al.}(2023){Kammoun}, {Igo}, {Miller}, {Fabian},
  {Reynolds}, {Merloni}, {Barret}, {Nardini}, {Petrucci}, {Piconcelli},
  {Barnier}, {Buchner}, {Dwelly}, {Grotova}, {Krumpe}, {Liu}, {Nandra}, {Rau},
  {Salvato}, {Urrutia}, \& {Wolf}}]{Kammoun2023}
{Kammoun}, E.~S., {Igo}, Z., {Miller}, J.~M., {et~al.} 2023, \mnras

\bibitem[{{Klein} {et~al.}(2022){Klein}, {Oguri}, {Mohr}, {Grandis},
  {Ghirardini}, {Liu}, {Liu}, {Bulbul}, {Wolf}, {Comparat}, {Ramos-Ceja},
  {Buchner}, {Chiu}, {Clerc}, {Merloni}, {Miyatake}, {Miyazaki}, {Okabe},
  {Ota}, {Pacaud}, {Salvato}, \& {Driver}}]{Klein2022}
{Klein}, M., {Oguri}, M., {Mohr}, J.~J., {et~al.} 2022, \aap, 661, A4

\bibitem[{Koenig {et~al.}(2020)Koenig, Wilms, Kreykenbohm, Weber, Bogensberger,
  Rau, Merloni, Maitra, Carpano, \& Ji}]{koenig_srgerosita_2020}
Koenig, O., Wilms, J., Kreykenbohm, I., {et~al.} 2020, The Astronomer's
  Telegram, 13765, 1

\bibitem[{{Kollmeier} {et~al.}(2017){Kollmeier}, {Zasowski}, {Rix}, {Johns},
  {Anderson}, {Drory}, {Johnson}, {Pogge}, {Bird}, {Blanc}, {Brownstein},
  {Crane}, {De Lee}, {Klaene}, {Kreckel}, {MacDonald}, {Merloni}, {Ness},
  {O'Brien}, {Sanchez-Gallego}, {Sayres}, {Shen}, {Thakar}, {Tkachenko},
  {Aerts}, {Blanton}, {Eisenstein}, {Holtzman}, {Maoz}, {Nandra}, {Rockosi},
  {Weinberg}, {Bovy}, {Casey}, {Chaname}, {Clerc}, {Conroy}, {Eracleous},
  {G{\"a}nsicke}, {Hekker}, {Horne}, {Kauffmann}, {McQuinn}, {Pellegrini},
  {Schinnerer}, {Schlafly}, {Schwope}, {Seibert}, {Teske}, \& {van
  Saders}}]{Kollmeier2017}
{Kollmeier}, J.~A., {Zasowski}, G., {Rix}, H.-W., {et~al.} 2017, arXiv
  e-prints, arXiv:1711.03234

\bibitem[{{K{\"o}nig} {et~al.}(2022){K{\"o}nig}, {Wilms}, {Arcodia}, {Dauser},
  {Dennerl}, {Doroshenko}, {Haberl}, {H{\"a}mmerich}, {Kirsch}, {Kreykenbohm},
  {Lorenz}, {Malyali}, {Merloni}, {Rau}, {Rauch}, {Sala}, {Schwope},
  {Suleimanov}, {Weber}, \& {Werner}}]{Koenig2022}
{K{\"o}nig}, O., {Wilms}, J., {Arcodia}, R., {et~al.} 2022, \nat, 605, 248

\bibitem[{{Kormendy} \& {Ho}(2013)}]{Kormendy2013}
{Kormendy}, J. \& {Ho}, L.~C. 2013, \araa, 51, 511

\bibitem[{{Kraft} {et~al.}(1991){Kraft}, {Burrows}, \& {Nousek}}]{Kraft1991}
{Kraft}, R.~P., {Burrows}, D.~N., \& {Nousek}, J.~A. 1991, \apj, 374, 344

\bibitem[{{Lammer} {et~al.}(2003){Lammer}, {Selsis}, {Ribas}, {Guinan},
  {Bauer}, \& {Weiss}}]{Lammer2003}
{Lammer}, H., {Selsis}, F., {Ribas}, I., {et~al.} 2003, \apjl, 598, L121

\bibitem[{{Liu} {et~al.}(2022{\natexlab{a}}){Liu}, {Bulbul}, {Ghirardini},
  {Liu}, {Klein}, {Clerc}, {{\"O}zsoy}, {Ramos-Ceja}, {Pacaud}, {Comparat},
  {Okabe}, {Bahar}, {Biffi}, {Brunner}, {Br{\"u}ggen}, {Buchner}, {Ider
  Chitham}, {Chiu}, {Dolag}, {Gatuzz}, {Gonzalez}, {Hoang}, {Lamer}, {Merloni},
  {Nandra}, {Oguri}, {Ota}, {Predehl}, {Reiprich}, {Salvato}, {Schrabback},
  {Sanders}, {Seppi}, \& {Thibaud}}]{LiuA2022}
{Liu}, A., {Bulbul}, E., {Ghirardini}, V., {et~al.} 2022{\natexlab{a}}, \aap,
  661, A2

\bibitem[{{Liu} {et~al.}(2023{\natexlab{a}}){Liu}, {Bulbul}, {Ramos-Ceja},
  {Sanders}, {Ghirardini}, {Bahar}, {Yeung}, {Gatuzz}, {Freyberg}, {Garrel},
  {Zhang}, {Merloni}, \& {Nandra}}]{LiuA2023}
{Liu}, A., {Bulbul}, E., {Ramos-Ceja}, M.~E., {et~al.} 2023{\natexlab{a}},
  \aap, 670, A96

\bibitem[{{Liu} {et~al.}(2022{\natexlab{b}}){Liu}, {Buchner}, {Nandra},
  {Merloni}, {Dwelly}, {Sanders}, {Salvato}, {Arcodia}, {Brusa}, {Wolf},
  {Georgakakis}, {Boller}, {Krumpe}, {Lamer}, {Waddell}, {Urrutia}, {Schwope},
  {Robrade}, {Wilms}, {Dauser}, {Comparat}, {Toba}, {Ichikawa}, {Iwasawa},
  {Shen}, \& {Medel}}]{LiuT2022b}
{Liu}, T., {Buchner}, J., {Nandra}, K., {et~al.} 2022{\natexlab{b}}, \aap, 661,
  A5

\bibitem[{{Liu} {et~al.}(2022{\natexlab{c}}){Liu}, {Merloni}, {Comparat},
  {Nandra}, {Sanders}, {Lamer}, {Buchner}, {Dwelly}, {Freyberg}, {Malyali},
  {Georgakakis}, {Salvato}, {Brunner}, {Brusa}, {Klein}, {Ghirardini}, {Clerc},
  {Pacaud}, {Bulbul}, {Liu}, {Schwope}, {Robrade}, {Wilms}, {Dauser},
  {Ramos-Ceja}, {Reiprich}, {Boller}, \& {Wolf}}]{LiuT2022}
{Liu}, T., {Merloni}, A., {Comparat}, J., {et~al.} 2022{\natexlab{c}}, \aap,
  661, A27

\bibitem[{{Liu} {et~al.}(2023{\natexlab{b}}){Liu}, {Malyali}, {Krumpe},
  {Homan}, {Goodwin}, {Grotova}, {Kawka}, {Rau}, {Merloni}, {Anderson},
  {Miller-Jones}, {Markowitz}, {Ciroi}, {Di Mille}, {Schramm}, {Tang},
  {Buckley}, {Gromadzki}, {Jin}, \& {Buchner}}]{LiuZ2023}
{Liu}, Z., {Malyali}, A., {Krumpe}, M., {et~al.} 2023{\natexlab{b}}, \aap, 669,
  A75

\bibitem[{{Locatelli} \& {Ponti}(2023)}]{Locatelli2023}
{Locatelli}, N. \& {Ponti}, G. e.~a. 2023, \aap, in press
  [\eprint[arXiv]{2310.10715}]

\bibitem[{{Madsen} {et~al.}(2021){Madsen}, {Burwitz}, {Forster}, {Grant},
  {Guainazzi}, {Kashyap}, {Marshall}, {Miller}, {Natalucci}, {Plucinsky}, \&
  {Terada}}]{Madsen2021}
{Madsen}, K.~K., {Burwitz}, V., {Forster}, K., {et~al.} 2021, arXiv e-prints,
  arXiv:2111.01613

\bibitem[{{Maitra} {et~al.}(2022){Maitra}, {Haberl}, {Sasaki}, {Maggi},
  {Dennerl}, \& {Freyberg}}]{Maitra2022}
{Maitra}, C., {Haberl}, F., {Sasaki}, M., {et~al.} 2022, \aap, 661, A30

\bibitem[{{Malyali} {et~al.}(2023){Malyali}, {Liu}, {Rau}, {Grotova},
  {Merloni}, {Goodwin}, {Anderson}, {Miller-Jones}, {Kawka}, {Arcodia},
  {Buchner}, {Nandra}, {Homan}, \& {Krumpe}}]{Malyali2023}
{Malyali}, A., {Liu}, Z., {Rau}, A., {et~al.} 2023, \mnras, 520, 3549

\bibitem[{{Malyali} {et~al.}(2021){Malyali}, {Rau}, {Merloni}, {Nandra},
  {Buchner}, {Liu}, {Gezari}, {Sollerman}, {Shappee}, {Trakhtenbrot}, {Arcavi},
  {Ricci}, {van Velzen}, {Goobar}, {Frederick}, {Kawka}, {Tartaglia}, {Burke},
  {Hiramatsu}, {Schramm}, {van der Boom}, {Anderson}, {Miller-Jones}, {Bellm},
  {Drake}, {Duev}, {Fremling}, {Graham}, {Masci}, {Rusholme}, {Soumagnac}, \&
  {Walters}}]{Malyali2021}
{Malyali}, A., {Rau}, A., {Merloni}, A., {et~al.} 2021, \aap, 647, A9

\bibitem[{{Marshall} {et~al.}(2021){Marshall}, {Chen}, {Drake}, {Guainazzi},
  {Kashyap}, {Meng}, {Plucinsky}, {Ratzlaff}, {van Dyk}, \&
  {Wang}}]{Marshall2021}
{Marshall}, H.~L., {Chen}, Y., {Drake}, J.~J., {et~al.} 2021, \aj, 162, 254

\bibitem[{{Mayer} {et~al.}(2023){Mayer}, {Becker}, {Predehl}, \&
  {Sasaki}}]{Mayer2023}
{Mayer}, M. G.~F., {Becker}, W., {Predehl}, P., \& {Sasaki}, M. 2023, \aap,
  676, A68

\bibitem[{{Mehrtens} {et~al.}(2012){Mehrtens}, {Romer}, {Hilton},
  {Lloyd-Davies}, {Miller}, {Stanford}, {Hosmer}, {Hoyle}, {Collins}, {Liddle},
  {Viana}, {Nichol}, {Stott}, {Dubois}, {Kay}, {Sahl{\'e}n}, {Young}, {Short},
  {Christodoulou}, {Watson}, {Davidson}, {Harrison}, {Baruah}, {Smith},
  {Burke}, {Mayers}, {Deadman}, {Rooney}, {Edmondson}, {West}, {Campbell},
  {Edge}, {Mann}, {Sabirli}, {Wake}, {Benoist}, {da Costa}, {Maia}, \&
  {Ogando}}]{Mehrtens2012}
{Mehrtens}, N., {Romer}, A.~K., {Hilton}, M., {et~al.} 2012, \mnras, 423, 1024

\bibitem[{{Meidinger} {et~al.}(2014){Meidinger}, {Andritschke}, {Bornemann},
  {Coutinho}, {Emberger}, {H{\"a}lker}, {Kink}, {Mican}, {M{\"u}ller},
  {Pietschner}, {Predehl}, \& {Reiffers}}]{Meidinger2014}
{Meidinger}, N., {Andritschke}, R., {Bornemann}, W., {et~al.} 2014, in Society
  of Photo-Optical Instrumentation Engineers (SPIE) Conference Series, Vol.
  9144, \procspie, 91441W

\bibitem[{{Meidinger} {et~al.}(2021){Meidinger}, {Andritschke}, {Dennerl},
  {Emberger}, {Eraerds}, {H{\"a}lker}, {Hartner}, {Pietschner}, \&
  {Reiffers}}]{Meidinger2021}
{Meidinger}, N., {Andritschke}, R., {Dennerl}, K., {et~al.} 2021, Journal of
  Astronomical Telescopes, Instruments, and Systems, 7, 025004

\bibitem[{{Merloni} {et~al.}(2019){Merloni}, {Alexander}, {Banerji}, {Boller},
  {Comparat}, {Dwelly}, {Fotopoulou}, {McMahon}, {Nandra}, {Salvato}, {Croom},
  {Finoguenov}, {Krumpe}, {Lamer}, {Rosario}, {Schwope}, {Shanks}, {Steinmetz},
  {Wisotzki}, \& {Worseck}}]{Merloni2019}
{Merloni}, A., {Alexander}, D.~A., {Banerji}, M., {et~al.} 2019, The Messenger,
  175, 42

\bibitem[{{Merloni} {et~al.}(2012){Merloni}, {Predehl}, {Becker},
  {B{\"o}hringer}, {Boller}, {Brunner}, {Brusa}, {Dennerl}, {Freyberg},
  {Friedrich}, {Georgakakis}, {Haberl}, {Hasinger}, {Meidinger}, {Mohr},
  {Nandra}, {Rau}, {Reiprich}, {Robrade}, {Salvato}, {Santangelo}, {Sasaki},
  {Schwope}, {Wilms}, \& {German eROSITA Consortium}}]{Merloni2012}
{Merloni}, A., {Predehl}, P., {Becker}, W., {et~al.} 2012, arXiv e-prints,
  arXiv:1209.3114

\bibitem[{{Nevalainen} \& {Molendi}(2023)}]{Nevalainen2023}
{Nevalainen}, J. \& {Molendi}, S. 2023, \aap, 676, A142

\bibitem[{{Nicastro} {et~al.}(2018){Nicastro}, {Kaastra}, {Krongold},
  {Borgani}, {Branchini}, {Cen}, {Dadina}, {Danforth}, {Elvis}, {Fiore},
  {Gupta}, {Mathur}, {Mayya}, {Paerels}, {Piro}, {Rosa-Gonzalez}, {Schaye},
  {Shull}, {Torres-Zafra}, {Wijers}, \& {Zappacosta}}]{Nicastro2018}
{Nicastro}, F., {Kaastra}, J., {Krongold}, Y., {et~al.} 2018, \nat, 558, 406

\bibitem[{{Norman}(2005)}]{Norman2005}
{Norman}, M.~L. 2005, in Background Microwave Radiation and Intracluster
  Cosmology, ed. F.~Melchiorri \& Y.~Rephaeli (Enrico Fermi International
  School of Physics Course CLIX), astro-ph/0511451

\bibitem[{{Onken} {et~al.}(2022){Onken}, {Lai}, {Wolf}, {Lucy}, {Hon},
  {Tisserand}, {Sokoloski}, {Luna}, {Manick}, {Fan}, \& {Bian}}]{Onken2022}
{Onken}, C.~A., {Lai}, S., {Wolf}, C., {et~al.} 2022, \pasa, 39, e037

\bibitem[{{{\"O}zel} \& {Freire}(2016)}]{Ozel2016}
{{\"O}zel}, F. \& {Freire}, P. 2016, \araa, 54, 401

\bibitem[{{Pasini} {et~al.}(2022){Pasini}, {Br{\"u}ggen}, {Hoang},
  {Ghirardini}, {Bulbul}, {Klein}, {Liu}, {Shimwell}, {Hardcastle}, {Williams},
  {Botteon}, {Gastaldello}, {van Weeren}, {Merloni}, {de Gasperin}, {Bahar},
  {Pacaud}, \& {Ramos-Ceja}}]{Pasini2022}
{Pasini}, T., {Br{\"u}ggen}, M., {Hoang}, D.~N., {et~al.} 2022, \aap, 661, A13

\bibitem[{{Peebles}(1980)}]{Peebles1980}
{Peebles}, P.~J.~E. 1980, {The large-scale structure of the universe}

\bibitem[{{Piffaretti} {et~al.}(2011){Piffaretti}, {Arnaud}, {Pratt},
  {Pointecouteau}, \& {Melin}}]{Piffaretti2011}
{Piffaretti}, R., {Arnaud}, M., {Pratt}, G.~W., {Pointecouteau}, E., \&
  {Melin}, J.~B. 2011, \aap, 534, A109

\bibitem[{{Pillepich} {et~al.}(2012){Pillepich}, {Porciani}, \&
  {Reiprich}}]{Pillepich2012}
{Pillepich}, A., {Porciani}, C., \& {Reiprich}, T.~H. 2012, \mnras, 422, 44

\bibitem[{{Pizzolato} {et~al.}(2003){Pizzolato}, {Maggio}, {Micela},
  {Sciortino}, \& {Ventura}}]{Pizzolato2003}
{Pizzolato}, N., {Maggio}, A., {Micela}, G., {Sciortino}, S., \& {Ventura}, P.
  2003, \aap, 397, 147

\bibitem[{{Plucinsky} {et~al.}(2017){Plucinsky}, {Beardmore}, {Foster},
  {Haberl}, {Miller}, {Pollock}, \& {Sembay}}]{Plucinsky2017}
{Plucinsky}, P.~P., {Beardmore}, A.~P., {Foster}, A., {et~al.} 2017, \aap, 597,
  A35

\bibitem[{{Ponti} {et~al.}(2023{\natexlab{a}}){Ponti}, {Sanders}, {Locatelli},
  {Zheng}, {Zhang}, {Yeung}, {Freyberg}, {Dennerl}, {Comparat}, {Merloni}, {Di
  Teodoro}, {Sasaki}, \& {Reiprich}}]{Ponti23b}
{Ponti}, G., {Sanders}, J.~S., {Locatelli}, N., {et~al.} 2023{\natexlab{a}},
  \aap, 670, A99

\bibitem[{{Ponti} {et~al.}(2023{\natexlab{b}}){Ponti}, {Zheng}, {Locatelli},
  {Bianchi}, {Zhang}, {Anastasopoulou}, {Comparat}, {Dennerl}, {Freyberg},
  {Haberl}, {Merloni}, {Reiprich}, {Salvato}, {Sanders}, {Sasaki}, {Strong}, \&
  {Yeung}}]{Ponti2023}
{Ponti}, G., {Zheng}, X., {Locatelli}, N., {et~al.} 2023{\natexlab{b}}, \aap,
  674, A195

\bibitem[{{Poppenhaeger} {et~al.}(2021){Poppenhaeger}, {Ketzer}, \&
  {Mallonn}}]{Poppenhaeger2021}
{Poppenhaeger}, K., {Ketzer}, L., \& {Mallonn}, M. 2021, \mnras, 500, 4560

\bibitem[{{Predehl} {et~al.}(2021){Predehl}, {Andritschke}, {Arefiev},
  {Babyshkin}, {Batanov}, {Becker}, {B{\"o}hringer}, {Bogomolov}, {Boller},
  {Borm}, {Bornemann}, {Br{\"a}uninger}, {Br{\"u}ggen}, {Brunner}, {Brusa},
  {Bulbul}, {Buntov}, {Burwitz}, {Burkert}, {Clerc}, {Churazov}, {Coutinho},
  {Dauser}, {Dennerl}, {Doroshenko}, {Eder}, {Emberger}, {Eraerds},
  {Finoguenov}, {Freyberg}, {Friedrich}, {Friedrich}, {F{\"u}rmetz},
  {Georgakakis}, {Gilfanov}, {Granato}, {Grossberger}, {Gueguen}, {Gureev},
  {Haberl}, {H{\"a}lker}, {Hartner}, {Hasinger}, {Huber}, {Ji}, {Kienlin},
  {Kink}, {Korotkov}, {Kreykenbohm}, {Lamer}, {Lomakin}, {Lapshov}, {Liu},
  {Maitra}, {Meidinger}, {Menz}, {Merloni}, {Mernik}, {Mican}, {Mohr},
  {M{\"u}ller}, {Nandra}, {Nazarov}, {Pacaud}, {Pavlinsky}, {Perinati},
  {Pfeffermann}, {Pietschner}, {Ramos-Ceja}, {Rau}, {Reiffers}, {Reiprich},
  {Robrade}, {Salvato}, {Sanders}, {Santangelo}, {Sasaki}, {Scheuerle},
  {Schmid}, {Schmitt}, {Schwope}, {Shirshakov}, {Steinmetz}, {Stewart},
  {Str{\"u}der}, {Sunyaev}, {Tenzer}, {Tiedemann}, {Tr{\"u}mper}, {Voron},
  {Weber}, {Wilms}, \& {Yaroshenko}}]{Predehl2021}
{Predehl}, P., {Andritschke}, R., {Arefiev}, V., {et~al.} 2021, \aap, 647, A1

\bibitem[{{Predehl} {et~al.}(2020){Predehl}, {Sunyaev}, {Becker}, {Brunner},
  {Burenin}, {Bykov}, {Cherepashchuk}, {Chugai}, {Churazov}, {Doroshenko},
  {Eismont}, {Freyberg}, {Gilfanov}, {Haberl}, {Khabibullin}, {Krivonos},
  {Maitra}, {Medvedev}, {Merloni}, {Nandra}, {Nazarov}, {Pavlinsky}, {Ponti},
  {Sanders}, {Sasaki}, {Sazonov}, {Strong}, \& {Wilms}}]{Predehl2020}
{Predehl}, P., {Sunyaev}, R.~A., {Becker}, W., {et~al.} 2020, \nat, 588, 227

\bibitem[{{Ramos-Ceja} {et~al.}(2022){Ramos-Ceja}, {Oguri}, {Miyazaki},
  {Ghirardini}, {Chiu}, {Okabe}, {Liu}, {Schrabback}, {Akino}, {Bahar},
  {Bulbul}, {Clerc}, {Comparat}, {Grandis}, {Klein}, {Lin}, {Merloni},
  {Mitsuishi}, {Miyatake}, {More}, {Nandra}, {Nishizawa}, {Ota}, {Pacaud},
  {Reiprich}, \& {Sanders}}]{Ramos-Ceja2022}
{Ramos-Ceja}, M.~E., {Oguri}, M., {Miyazaki}, S., {et~al.} 2022, \aap, 661, A14

\bibitem[{{Reiprich} {et~al.}(2013){Reiprich}, {Basu}, {Ettori}, {Israel},
  {Lovisari}, {Molendi}, {Pointecouteau}, \& {Roncarelli}}]{Reiprich2013}
{Reiprich}, T.~H., {Basu}, K., {Ettori}, S., {et~al.} 2013, Space Science
  Reviews, 177, 195

\bibitem[{{Remillard} \& {McClintock}(2006)}]{Remillard2006}
{Remillard}, R.~A. \& {McClintock}, J.~E. 2006, \araa, 44, 49

\bibitem[{{Rosati} {et~al.}(2002){Rosati}, {Borgani}, \& {Norman}}]{Rosati2002}
{Rosati}, P., {Borgani}, S., \& {Norman}, C. 2002, \araa, 40, 539

\bibitem[{{Rosen} {et~al.}(2016){Rosen}, {Webb}, {Watson}, {Ballet}, {Barret},
  {Braito}, {Carrera}, {Ceballos}, {Coriat}, {Della Ceca}, {Denkinson},
  {Esquej}, {Farrell}, {Freyberg}, {Gris{\'e}}, {Guillout}, {Heil},
  {Koliopanos}, {Law-Green}, {Lamer}, {Lin}, {Martino}, {Michel}, {Motch},
  {Nebot Gomez-Moran}, {Page}, {Page}, {Page}, {Pakull}, {Pye}, {Read},
  {Rodriguez}, {Sakano}, {Saxton}, {Schwope}, {Scott}, {Sturm}, {Traulsen},
  {Yershov}, \& {Zolotukhin}}]{Rosen2016}
{Rosen}, S.~R., {Webb}, N.~A., {Watson}, M.~G., {et~al.} 2016, \aap, 590, A1

\bibitem[{{Rothschild} {et~al.}(1979){Rothschild}, {Boldt}, {Holt},
  {Serlemitsos}, {Garmire}, {Agrawal}, {Riegler}, {Bowyer}, \&
  {Lampton}}]{1979roths}
{Rothschild}, R., {Boldt}, E., {Holt}, S., {et~al.} 1979, Space Science
  Instrumentation, 4, 269

\bibitem[{Rutledge(1998)}]{rutledge_astronomers_1998}
Rutledge, R.~E. 1998, \pasp, 110, 754, \_eprint: astro-ph/9802256

\bibitem[{{Salvato} {et~al.}(2018){Salvato}, {Buchner}, {Budav{\'a}ri},
  {Dwelly}, {Merloni}, {Brusa}, {Rau}, {Fotopoulou}, \& {Nand
  ra}}]{Salvato2018}
{Salvato}, M., {Buchner}, J., {Budav{\'a}ri}, T., {et~al.} 2018, \mnras, 473,
  4937

\bibitem[{{Salvato} {et~al.}(2022){Salvato}, {Wolf}, {Dwelly}, {Georgakakis},
  {Brusa}, {Merloni}, {Liu}, {Toba}, {Nandra}, {Lamer}, {Buchner}, {Schneider},
  {Freund}, {Rau}, {Schwope}, {Nishizawa}, {Klein}, {Arcodia}, {Comparat},
  {Musiimenta}, {Nagao}, {Brunner}, {Malyali}, {Finoguenov}, {Anderson},
  {Shen}, {Ibarra-Medel}, {Trump}, {Brandt}, {Urry}, {Rivera}, {Krumpe},
  {Urrutia}, {Miyaji}, {Ichikawa}, {Schneider}, {Fresco}, {Boller}, {Haase},
  {Brownstein}, {Lane}, {Bizyaev}, \& {Nitschelm}}]{Salvato2022}
{Salvato}, M., {Wolf}, J., {Dwelly}, T., {et~al.} 2022, \aap, 661, A3

\bibitem[{{Sanders} {et~al.}(2022){Sanders}, {Biffi}, {Br{\"u}ggen}, {Bulbul},
  {Dennerl}, {Dolag}, {Erben}, {Freyberg}, {Gatuzz}, {Ghirardini}, {Hoang},
  {Klein}, {Liu}, {Merloni}, {Pacaud}, {Ramos-Ceja}, {Reiprich}, \&
  {ZuHone}}]{Sanders2022}
{Sanders}, J.~S., {Biffi}, V., {Br{\"u}ggen}, M., {et~al.} 2022, \aap, 661, A36

\bibitem[{{Sanz-Forcada} {et~al.}(2011){Sanz-Forcada}, {Micela}, {Ribas},
  {Pollock}, {Eiroa}, {Velasco}, {Solano}, \&
  {Garc{\'\i}a-{\'A}lvarez}}]{Sanz-Forcada2011}
{Sanz-Forcada}, J., {Micela}, G., {Ribas}, I., {et~al.} 2011, \aap, 532, A6

\bibitem[{{Sarazin}(1986)}]{Sarazin1986}
{Sarazin}, C.~L. 1986, Reviews of Modern Physics, 58, 1

\bibitem[{{Sazonov} {et~al.}(2021){Sazonov}, {Gilfanov}, {Medvedev}, {Yao},
  {Khorunzhev}, {Semena}, {Sunyaev}, {Burenin}, {Lyapin}, {Meshcheryakov},
  {Uskov}, {Zaznobin}, {Postnov}, {Dodin}, {Belinski}, {Cherepashchuk},
  {Eselevich}, {Dodonov}, {Grokhovskaya}, {Kotov}, {Bikmaev}, {Zhuchkov},
  {Gumerov}, {van Velzen}, \& {Kulkarni}}]{Sazonov2021}
{Sazonov}, S., {Gilfanov}, M., {Medvedev}, P., {et~al.} 2021, \mnras, 508, 3820

\bibitem[{{Scargle} {et~al.}(2013){Scargle}, {Norris}, {Jackson}, \&
  {Chiang}}]{Scargle2013}
{Scargle}, J.~D., {Norris}, J.~P., {Jackson}, B., \& {Chiang}, J. 2013, \apj,
  764, 167

\bibitem[{{Schmitt}(1997)}]{Schmitt1997}
{Schmitt}, J.~H.~M.~M. 1997, \aap, 318, 215

\bibitem[{{Schmitt} {et~al.}(2022){Schmitt}, {Czesla}, {Freund}, {Robrade}, \&
  {Schneider}}]{Schmitt2022}
{Schmitt}, J.~H.~M.~M., {Czesla}, S., {Freund}, S., {Robrade}, J., \&
  {Schneider}, P.~C. 2022, \aap, 661, A40

\bibitem[{{Schneider} {et~al.}(2022){Schneider}, {Freund}, {Czesla}, {Robrade},
  {Salvato}, \& {Schmitt}}]{Schneider2022}
{Schneider}, P.~C., {Freund}, S., {Czesla}, S., {et~al.} 2022, \aap, 661, A6

\bibitem[{{Schwope} {et~al.}(2000){Schwope}, {Hasinger}, {Lehmann}, {Schwarz},
  {Brunner}, {Neizvestny}, {Ugryumov}, {Balega}, {Tr{\"u}mper}, \&
  {Voges}}]{Schwope2000}
{Schwope}, A., {Hasinger}, G., {Lehmann}, I., {et~al.} 2000, Astronomische
  Nachrichten, 321, 1

\bibitem[{{Secrest} {et~al.}(2021){Secrest}, {von Hausegger}, {Rameez},
  {Mohayaee}, {Sarkar}, \& {Colin}}]{Secrest2021}
{Secrest}, N.~J., {von Hausegger}, S., {Rameez}, M., {et~al.} 2021, \apjl, 908,
  L51

\bibitem[{{Seppi} {et~al.}(2022){Seppi}, {Comparat}, {Bulbul}, {Nandra},
  {Merloni}, {Clerc}, {Liu}, {Ghirardini}, {Liu}, {Salvato}, {Sanders},
  {Wilms}, {Dwelly}, {Dauser}, {K{\"o}nig}, {Ramos-Ceja}, {Garrel}, \&
  {Reiprich}}]{Seppi2022}
{Seppi}, R., {Comparat}, J., {Bulbul}, E., {et~al.} 2022, \aap, 665, A78

\bibitem[{{Seppi} {et~al.}(2023){Seppi}, {Comparat}, {Nandra}, {Dolag},
  {Biffi}, {Bulbul}, {Liu}, {Ghirardini}, \& {Ider-Chitham}}]{Seppi2023}
{Seppi}, R., {Comparat}, J., {Nandra}, K., {et~al.} 2023, \aap, 671, A57

\bibitem[{{Setti} \& {Woltjer}(1989)}]{Setti1989}
{Setti}, G. \& {Woltjer}, L. 1989, \aap, 224, L21

\bibitem[{{Shapiro} \& {Teukolsky}(1983)}]{Shapiro1983}
{Shapiro}, S.~L. \& {Teukolsky}, S.~A. 1983, {Black holes, white dwarfs and
  neutron stars. The physics of compact objects}

\bibitem[{{Shu} {et~al.}(2019){Shu}, {Koposov}, {Evans}, {Belokurov},
  {McMahon}, {Auger}, \& {Lemon}}]{Shu2019}
{Shu}, Y., {Koposov}, S.~E., {Evans}, N.~W., {et~al.} 2019, \mnras, 489, 4741

\bibitem[{{Str{\"u}der} {et~al.}(2001){Str{\"u}der}, {Briel}, {Dennerl},
  {Hartmann}, {Kendziorra}, {Meidinger}, {Pfeffermann}, {Reppin}, {Aschenbach},
  {Bornemann}, {Br{\"a}uninger}, {Burkert}, {Elender}, {Freyberg}, {Haberl},
  {Hartner}, {Heuschmann}, {Hippmann}, {Kastelic}, {Kemmer}, {Kettenring},
  {Kink}, {Krause}, {M{\"u}ller}, {Oppitz}, {Pietsch}, {Popp}, {Predehl},
  {Read}, {Stephan}, {St{\"o}tter}, {Tr{\"u}mper}, {Holl}, {Kemmer}, {Soltau},
  {St{\"o}tter}, {Weber}, {Weichert}, {von Zanthier}, {Carathanassis}, {Lutz},
  {Richter}, {Solc}, {B{\"o}ttcher}, {Kuster}, {Staubert}, {Abbey}, {Holland},
  {Turner}, {Balasini}, {Bignami}, {La Palombara}, {Villa}, {Buttler},
  {Gianini}, {Lain{\'e}}, {Lumb}, \& {Dhez}}]{Strueder2001}
{Str{\"u}der}, L., {Briel}, U., {Dennerl}, K., {et~al.} 2001, \aap, 365, L18

\bibitem[{{Sunyaev} {et~al.}(2021){Sunyaev}, {Arefiev}, {Babyshkin},
  {Bogomolov}, {Borisov}, {Buntov}, {Brunner}, {Burenin}, {Churazov},
  {Coutinho}, {Eder}, {Eismont}, {Freyberg}, {Gilfanov}, {Gureyev}, {Hasinger},
  {Khabibullin}, {Kolmykov}, {Komovkin}, {Krivonos}, {Lapshov}, {Levin},
  {Lomakin}, {Lutovinov}, {Medvedev}, {Merloni}, {Mernik}, {Mikhailov},
  {Molodtsov}, {Mzhelsky}, {M{\"u}ller}, {Nandra}, {Nazarov}, {Pavlinsky},
  {Poghodin}, {Predehl}, {Robrade}, {Sazonov}, {Scheuerle}, {Shirshakov},
  {Tkachenko}, \& {Voron}}]{Sunyaev2021}
{Sunyaev}, R., {Arefiev}, V., {Babyshkin}, V., {et~al.} 2021, \aap, 656, A132

\bibitem[{{Tamba} {et~al.}(2022){Tamba}, {Odaka}, {Bamba}, {Murakami}, {Mori},
  {Hayashida}, {Terada}, {Mizuno}, \& {Nobukawa}}]{Tamba2022}
{Tamba}, T., {Odaka}, H., {Bamba}, A., {et~al.} 2022, \pasj, 74, 364

\bibitem[{{Tanimura} {et~al.}(2020){Tanimura}, {Aghanim}, {Kolodzig},
  {Douspis}, \& {Malavasi}}]{Tanimura2020}
{Tanimura}, H., {Aghanim}, N., {Kolodzig}, A., {Douspis}, M., \& {Malavasi}, N.
  2020, \aap, 643, L2

\bibitem[{{Treiber} {et~al.}(2021){Treiber}, {Vasilopoulos}, {Bailyn},
  {Haberl}, {Gendreau}, {Ray}, {Maitra}, {Maggi}, {Jaisawal}, {Udalski},
  {Wilms}, {Monageng}, {Buckley}, {K{\"o}nig}, \&
  {Carpano}}]{2021MNRAS.503.6187T}
{Treiber}, H., {Vasilopoulos}, G., {Bailyn}, C.~D., {et~al.} 2021, \mnras, 503,
  6187

\bibitem[{{Treister} \& {Urry}(2005)}]{Treister2005}
{Treister}, E. \& {Urry}, C.~M. 2005, \apj, 630, 115

\bibitem[{{Truemper}(1982)}]{1982truem}
{Truemper}, J. 1982, Advances in Space Research, 2, 241

\bibitem[{{Tub{\'i}n-Arenas} \& {Krumpe}(2023)}]{Tubin2023}
{Tub{\'i}n-Arenas}, D. \& {Krumpe}, M. e.~a. 2023, \aap, in press

\bibitem[{{Ueda} {et~al.}(2014){Ueda}, {Akiyama}, {Hasinger}, {Miyaji}, \&
  {Watson}}]{Ueda2014}
{Ueda}, Y., {Akiyama}, M., {Hasinger}, G., {Miyaji}, T., \& {Watson}, M.~G.
  2014, \apj, 786, 104

\bibitem[{{Vikhlinin} {et~al.}(2009){Vikhlinin}, {Kravtsov}, {Burenin},
  {Ebeling}, {Forman}, {Hornstrup}, {Jones}, {Murray}, {Nagai}, {Quintana}, \&
  {Voevodkin}}]{Vikhlinin2009}
{Vikhlinin}, A., {Kravtsov}, A.~V., {Burenin}, R.~A., {et~al.} 2009, \apj, 692,
  1060

\bibitem[{{Vink}(2012)}]{Vink2012}
{Vink}, J. 2012, \aapr, 20, 49

\bibitem[{{Voges} {et~al.}(1999){Voges}, {Aschenbach}, {Boller},
  {Br{\"a}uninger}, {Briel}, {Burkert}, {Dennerl}, {Englhauser}, {Gruber},
  {Haberl}, {Hartner}, {Hasinger}, {K{\"u}rster}, {Pfeffermann}, {Pietsch},
  {Predehl}, {Rosso}, {Schmitt}, {Tr{\"u}mper}, \& {Zimmermann}}]{Voges1999}
{Voges}, W., {Aschenbach}, B., {Boller}, T., {et~al.} 1999, \aap, 349, 389

\bibitem[{{Voges} {et~al.}(2000){Voges}, {Aschenbach}, {Boller}, {Brauninger},
  {Briel}, {Burkert}, {Dennerl}, {Englhauser}, {Gruber}, {Haberl}, {Hartner},
  {Hasinger}, {Pfeffermann}, {Pietsch}, {Predehl}, {Schmitt}, {Tr{\"u}mper}, \&
  {Zimmermann}}]{Voges2000}
{Voges}, W., {Aschenbach}, B., {Boller}, T., {et~al.} 2000, \iaucirc, 7432, 3

\bibitem[{{Voit}(2005)}]{Voit2005}
{Voit}, G.~M. 2005, Reviews of Modern Physics, 77, 207

\bibitem[{{Warner}(1995)}]{Warner1995}
{Warner}, B. 1995, {Cataclysmic variable stars}, Vol.~28

\bibitem[{{Warwick} {et~al.}(1981){Warwick}, {Marshall}, {Fraser}, {Watson},
  {Lawrence}, {Page}, {Pounds}, {Ricketts}, {Sims}, \& {Smith}}]{Warwick1981}
{Warwick}, R.~S., {Marshall}, N., {Fraser}, G.~W., {et~al.} 1981, \mnras, 197,
  865

\bibitem[{{Webb} {et~al.}(2020){Webb}, {Coriat}, {Traulsen}, {Ballet}, {Motch},
  {Carrera}, {Koliopanos}, {Authier}, {de la Calle}, {Ceballos}, {Colomo},
  {Chuard}, {Freyberg}, {Garcia}, {Kolehmainen}, {Lamer}, {Lin}, {Maggi},
  {Michel}, {Page}, {Page}, {Perea-Calderon}, {Pineau}, {Rodriguez}, {Rosen},
  {Santos Lleo}, {Saxton}, {Schwope}, {Tom{\'a}s}, {Watson}, \&
  {Zakardjian}}]{Webb2020}
{Webb}, N.~A., {Coriat}, M., {Traulsen}, I., {et~al.} 2020, \aap, 641, A136

\bibitem[{{Weinberg} {et~al.}(2013){Weinberg}, {Mortonson}, {Eisenstein},
  {Hirata}, {Riess}, \& {Rozo}}]{Weinberg2013}
{Weinberg}, D.~H., {Mortonson}, M.~J., {Eisenstein}, D.~J., {et~al.} 2013,
  \physrep, 530, 87

\bibitem[{{Wenger} {et~al.}(2000){Wenger}, {Ochsenbein}, {Egret}, {Dubois},
  {Bonnarel}, {Borde}, {Genova}, {Jasniewicz}, {Lalo{\"e}}, {Lesteven}, \&
  {Monier}}]{Wenger2000}
{Wenger}, M., {Ochsenbein}, F., {Egret}, D., {et~al.} 2000, \aaps, 143, 9

\bibitem[{{Whelan} {et~al.}(2022){Whelan}, {Veronica}, {Pacaud}, {Reiprich},
  {Bulbul}, {Ramos-Ceja}, {Sanders}, {Aschersleben}, {Iljenkarevic}, {Migkas},
  {Freyberg}, {Dennerl}, {Kara}, {Liu}, {Ghirardini}, \& {Ota}}]{Whelan2022}
{Whelan}, B., {Veronica}, A., {Pacaud}, F., {et~al.} 2022, \aap, 663, A171

\bibitem[{{Willingale} {et~al.}(2013){Willingale}, {Starling}, {Beardmore},
  {Tanvir}, \& {O'Brien}}]{Willingale2013}
{Willingale}, R., {Starling}, R.~L.~C., {Beardmore}, A.~P., {Tanvir}, N.~R., \&
  {O'Brien}, P.~T. 2013, \mnras, 431, 394

\bibitem[{{Wilms} {et~al.}(2020){Wilms}, {Kreykenbohm}, {Weber}, {Falkner},
  {Dauser}, {Knies}, {Koenig}, {Malyali}, {Rau}, {Merloni}, {Bogensberger},
  {Brunner}, {Buchner}, {Carpano}, {Freyberg}, {Haberl}, {Maitra}, {Salvato},
  {Doroshenko}, {Ducci}, {Ji}, {Schmitt}, \& {Schwope}}]{Wilms2020}
{Wilms}, J., {Kreykenbohm}, I., {Weber}, P., {et~al.} 2020, The Astronomer's
  Telegram, 13416, 1

\bibitem[{Wilms {et~al.}(2020)Wilms, Kreykenbohm, Weber, Falkner, Dauser,
  Knies, Koenig, Malyali, Rau, Merloni, Bogensberger, Brunner, Buchner,
  Carpano, Freyberg, Haberl, Maitra, Salvato, Doroshenko, Ducci, Ji, Schmitt,
  \& Schwope}]{wilms_srgerosita_2020}
Wilms, J., Kreykenbohm, I., Weber, P., {et~al.} 2020, The Astronomer's
  Telegram, 13416, 1

\bibitem[{{Wood} {et~al.}(1984){Wood}, {Meekins}, {Yentis}, {Smathers},
  {McNutt}, {Bleach}, {Byram}, {Chupp}, {Friedman}, \& {Meidav}}]{1984wood}
{Wood}, K.~S., {Meekins}, J.~F., {Yentis}, D.~J., {et~al.} 1984, \apjs, 56, 507

\bibitem[{{Wright} {et~al.}(2011){Wright}, {Drake}, {Mamajek}, \&
  {Henry}}]{Wright2011}
{Wright}, N.~J., {Drake}, J.~J., {Mamajek}, E.~E., \& {Henry}, G.~W. 2011,
  \apj, 743, 48

\bibitem[{{Yeung} {et~al.}(2023){Yeung}, {Freyberg}, {Ponti}, {Dennerl},
  {Sasaki}, \& {Strong}}]{Yeung2023}
{Yeung}, M.~C.~H., {Freyberg}, M.~J., {Ponti}, G., {et~al.} 2023, \aap, 676, A3

\bibitem[{{Zheng} \& {Ponti}(2024)}]{Zheng2023}
{Zheng}, X. \& {Ponti}, G. e.~a. 2024, \aap, in press

\bibitem[{Zonca {et~al.}(2019)Zonca, Singer, Lenz, Reinecke, Rosset, Hivon, \&
  Gorski}]{Zonca2019}
Zonca, A., Singer, L., Lenz, D., {et~al.} 2019, Journal of Open Source
  Software, 4, 1298

\end{thebibliography}
 
% WARNING
%-------------------------------------------------------------------
% Please note that we have included the references to the file aa.dem in
% order to compile it, but we ask you to:
%
% - use BibTeX with the regular commands:
%   \bibliographystyle{aa} % style aa.bst
%   \bibliography{Yourfile} % your references Yourfile.bib
%
% - join the .bib files when you upload your source files
%-------------------------------------------------------------------

\begin{appendix} 

\FloatBarrier\section{The eROSITA Survey PSF}
\label{appendix:psf}

\begin{figure*}
    \centering
    \includegraphics[width=0.9\textwidth]{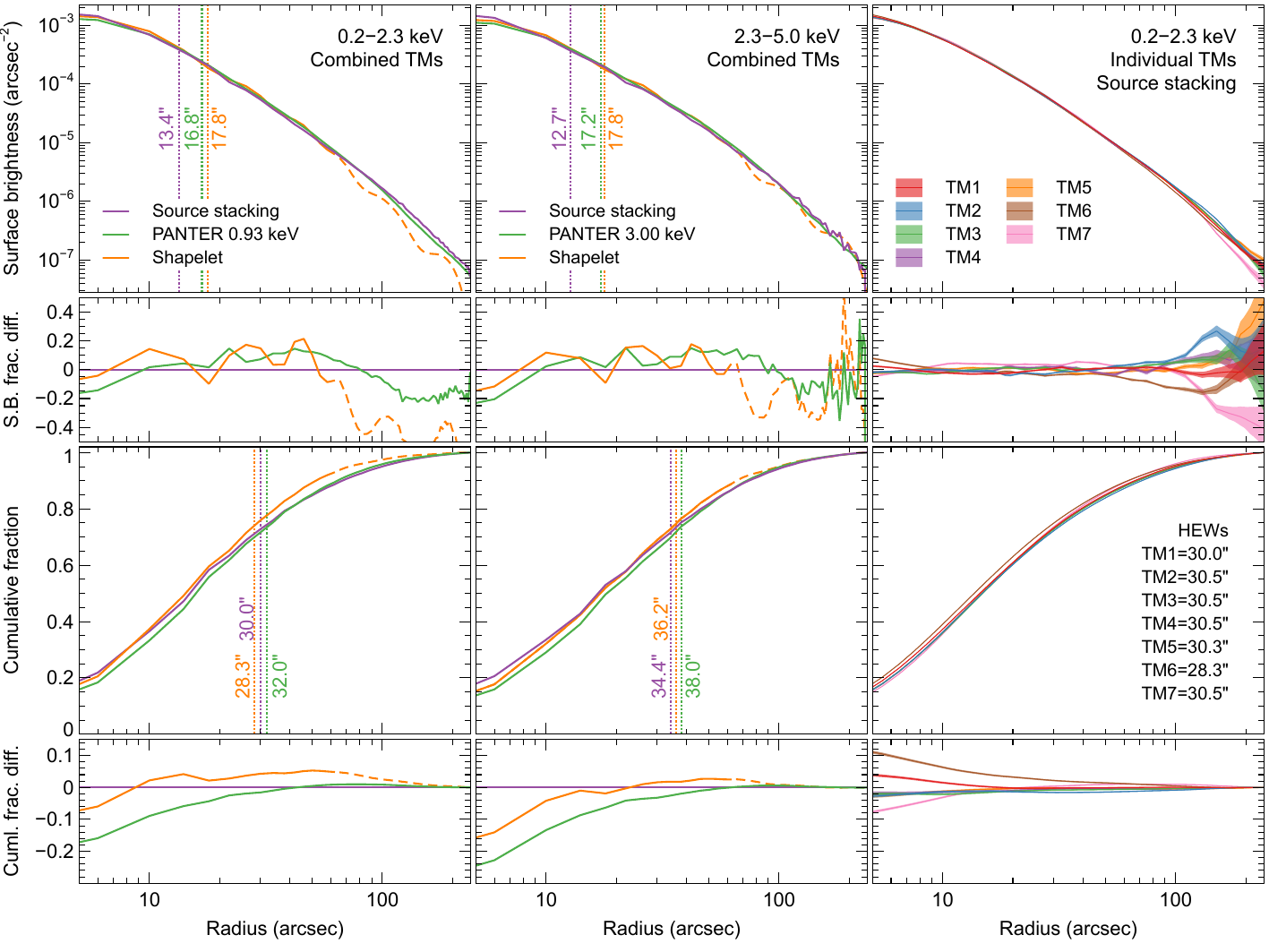}
    \caption{
    Comparison of different survey-averaged eROSITA PSFs.
    Shown are the PSFs in the bands 0.2--2.3 keV (left panels), 2.3--5.0 keV (central panels), or for the individual TMs from stacking in 0.2--2.3 keV (right panels).
    The PSF models shown are the survey-averaged shapelet PSFs, those obtained by stacking sources (see text in Appendix \ref{appendix:psf} for more details) and those from ground-based measurements using PANTER.
    The top row of panels show the PSF surface brightness profiles, normalised within 4 arcmin radius, where the vertical lines plot the FWHM values.
    The second panels down show the fractional difference of the surface brightness of each PSF from the average stacked profile.
    The third panels down show the cumulative signal as a function of radius, plotting the HEWs as vertical lines.
    The lowest panels show the fractional difference of the cumulative profiles to the average stacked profile.
    The shapelet PSFs are those used for fitting the energy band given, the stacked PSF is weighted by the spectra of the input sources, and the PANTER PSF is obtained at the monochromatic energy specified, which is chosen to be representative of the source photons in the band.
    The shapelet PSF is plotted as a dashed line outside a radius of 1 arcmin, the maximum used for fitting in the source detection pipeline.
    In the two leftmost columns, the PSF images were rebinned with 4 arcsec pixels where necessary  for a fair comparison.
    In the rightmost column, we show the best fitting model and uncertainties, rather than stacked, rescaled and rebinned count profiles.
    The FWHMs quoted are sensitive to the pixelisation used and the inner value, and so should be used with care.
    HEWs were computed from minimally-binned images.
    }
\label{fig:psf}
\end{figure*}

The PSF of the eROSITA telescopes can be obtained from the survey itself to verify the PANTER measurements and compare to its shapelet representation used in the source detection pipeline.
We are undertaking a project to obtain the PSF from source stacking and present here preliminary results.
938 sources were taken from the full eROSITA sky, selected from a catalogue of eROSITA point sources with the largest number of counts.
To optimise the signal and to avoid bad pixels we used eRASS:4 for TMs 1, 2, 3 and 6, eRASS:3 for TMs 5 and 7, and eRASS:2 for TM4.
In this analysis we examined the PSF range out to 4 arcmin radius, to be consistent with the ground based measurements, although there is additional flux beyond this radius, estimated to be about 2.7\% from the best fit core PSF (excluding stray light) of \citet{Churazov2023} (their equation A.1).
We excluded the very brightest piled-up sources and those in crowded fields or near large extended sources.
The position of each source was measured by simultaneously fitting 2D models to two X-ray images, one where the source was inside 15 arcmin of the optical axis, and another when it was outside.
We matched against the Gaia G band catalogue to identify those sources potentially affected by optical loading and used the Galactic latitude to flag those potentially affected by scattering halos.
We manually inspected the images and profiles of the sources to look for anomalous profiles, such as evidence of extension, or identify neighbouring contaminating sources.
For sources which could be contaminated in the outskirts by other sources or by non-flat background, we excluded the source, or manually reduced the outer radius used from 4 arcmin to 1, 2 or 3 arcmin as appropriate.
Lightcurves created with \texttt{srctool} were used to quantify pileup, where the inner stacking radius used was chosen as a function of time for a source according to pileup rate thresholds (rates of 15, 44, 100, 155, 250 and $600\,\mathrm{s}^{-1}$ produced inner radii of 20, 30, 40, 60, 90 and 240 arcsec, respectively, based on simulations to estimate the effect of pileup).
Source profiles were stacked in each energy band, and for each combination of inner and outer radius.
A PSF model plus flat background model was simultaneously fitted to the stacked profiles in all the combinations of radial range, allowing the PSF normalisation and background level to vary in each combination of radial range, to produce a model of the PSF as a function of radius.
This process was repeated for each energy band and combinations of TMs.

Figure \ref{fig:psf} shows a comparison between the result of our source stacking analysis, ground-based PANTER measurements \citep{Dennerl2020} and survey-averaged shapelet models of the ground-based data \citep{Brunner2022}.
Also shown are comparisons of the PSF for the different eROSITA TMs.
We note here that the survey-averaged shapelet PSF is never fitted to the data by \texttt{ermldet} in photon imaging mode, but rather, a PSF appropriate for each photon is created.
Within 1 arcmin radius the deviations of the PSF are within 20\%.
Much of the deviation in the shapelet PSF is due to `steps' occurring at intervals of 16 arcsec radius.
At larger radius the shapelet PSF in the soft band is lower than the stacked or ground-based PSF.
This contributes to the shapelet cumulative fraction being higher in the centre, as the profiles are normalised at 4 arcmin radius.
The stacked profiles typically have smaller full-width-half-maximums (FWHMs) than the PANTER and shapelet PSFs, but comparable HEW.
The in-flight eROSITA camera sub-pixel positioning allows a better determination of the PSF core than the 9.6 arcsec pixel would otherwise suggest.

Finally, we tested the uniformity of the PSF across the sky. For this, we divided the sample of 938 point sources described above into four quadrants in ecliptic coordinates (with dividing lines at ecliptic longitude of $180^\circ$ and ecliptic latitude of $0^\circ$) and recompute the HEW of the stacked PSF for TM0 in the 0.2-2.3 keV band. The values are consistent with the all-sky average HEW (30.0\arcsec) within 0.3\arcsec.

\begin{table}[htbp]
    \caption{Point Spread Function Half-Energy Width (HEW) in the 0.2-2.3 keV band, computed by stacking point sources in eRASS1 in four different quadrant across the sky.}
    \label{tab:psf_ecl}
    \centering
    \begin{tabular}{cccc}
       \hline\hline
         Quadrant  &  Ecl. longitude $\lambda$ & Ecl. latitude $\beta$ & HEW  \\
       \hline
       \hline
        QI & $\le 180^\circ$ & $\le 0^\circ$ & 30.0\arcsec \\
        QII & $> 180^\circ$ & $\le 0^\circ$ & 30.1\arcsec \\
        QIII & $\le 180^\circ$ & $> 0^\circ$ & 30.1\arcsec \\
        QIV & $> 180^\circ $& $> 0^\circ$ & 29.7\arcsec \\
       \hline
    \end{tabular}
\end{table}

In the future we plan to improve our stacking procedure. These improvements could include automated detection of contaminating sources and removal of sources with inconsistent profiles.
Future processing versions will improve the accuracy of the boresight calibration as a function of time.
By choosing which sources to include in the stacking as a function of radius to optimise the signal to noise, we would also reduce the effect of the sky background. Finally, we will also verify the obtained PSFs by studying different subsamples of sources.

\subsection{On the azimutal symmetry of PSF and positional uncertainties}
A key assumption for the astrometric calibration and positional uncertainty determination of our source detection algorithm (see Section~\ref{subsection:prelim_astrom}), and of the \texttt{srctool} extraction tool is that the eROSITA PSF can be approximated with a azimutally symmetric function and that the associated positional errors are also symmetric. 

\begin{figure}
    \centering
    \includegraphics[width=0.4\textwidth]{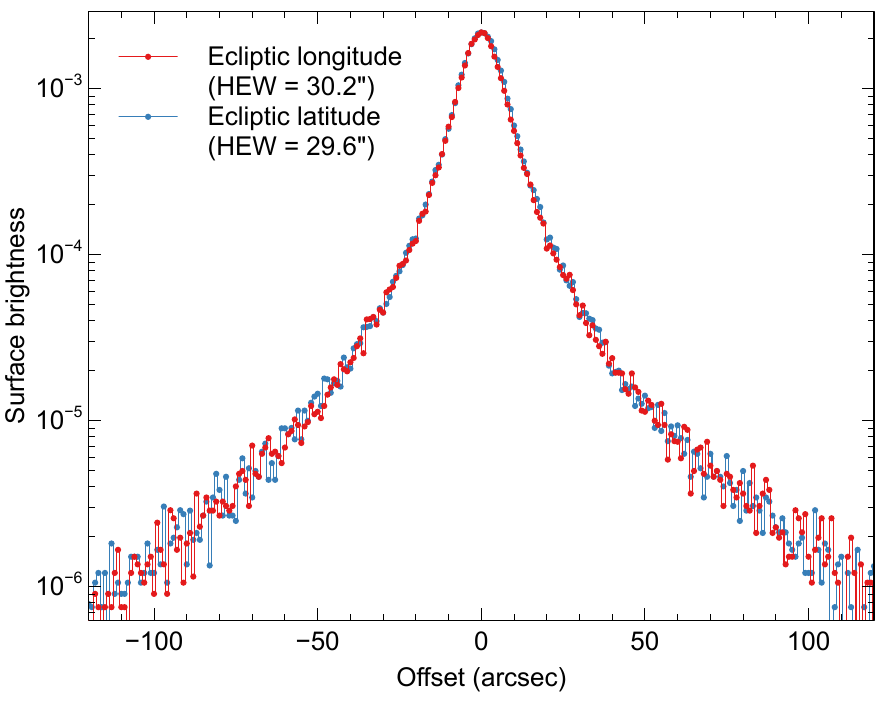} 
    \caption{Stacked PSF profile from the eRASS1 point sources in the 0.2-2.3 keV band along ecliptic longitude (red) and latitude (blue), demonstrating the symmetry of the PSF.  The measured Half-Energy Widths (HEW) are consistent to within 0.6\arcsec.
    }
    \label{fig:ecl_profile_psf}
\end{figure}

Here we validate this assumption in two ways. First of all, we show in Fig.~\ref{fig:ecl_profile_psf} the profile of the average stacked PSF (using the same 938 sources from the all-sky survey as described in the previous section) along ecliptic longitude and latitude. No significant difference in the two profiles can be seen. Secondly, we show in Fig.~\ref{fig:psf_symmetry} the measured positional offset between eRASS1 X-ray source positions and Gaia/unWISE QSOs (see Section~\ref{subsec:astrometric_ext}) in both equatorial and ecliptic coordinates directions. We fit these distributions with a Gaussian function, and obtain for the offset the following values of mean and Standard Deviation for the two cases: $(\Delta\alpha, \Delta\delta)_{\rm mean} = (-0.11\arcsec, 0.08\arcsec)$; $(\Delta\alpha, \Delta\delta)_{\rm SD} = (3.5\arcsec, 3.5\arcsec)$; $(\Delta\lambda, \Delta\beta)_{\rm mean} = (-0.11\arcsec, 0.16\arcsec)$ and $(\Delta\lambda, \Delta\beta)_{\rm SD} = (3.5\arcsec, 3.5\arcsec)$. From this, we conclude that the symmetry approximation is justified, and that the residual offset (of size $<0.2\arcsec$) is small enough when compared to the positional uncertainty to be safely ignored.

\begin{figure*}
    \centering
    \begin{tabular}{cc}
    \includegraphics[width=0.42\textwidth]{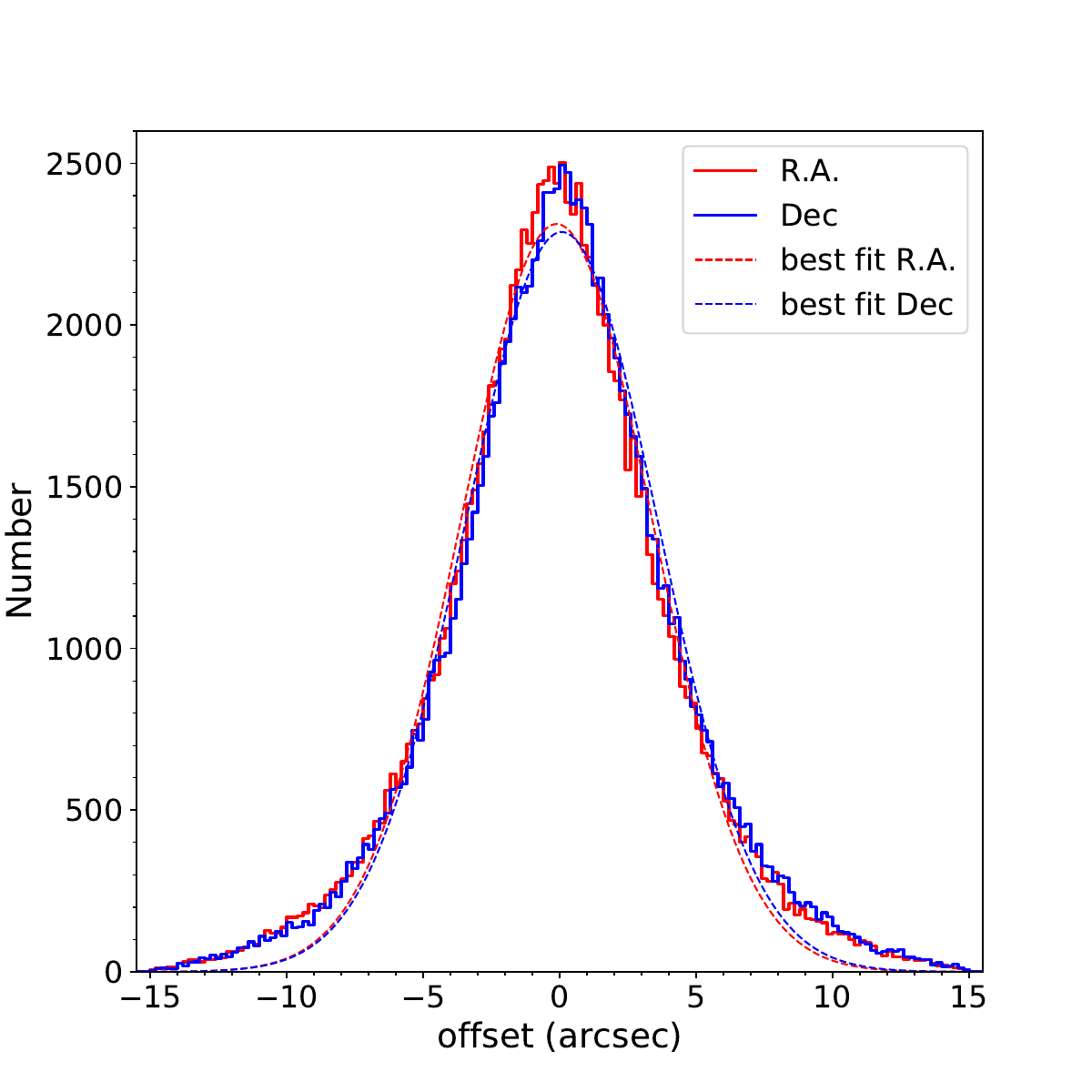} &
    \includegraphics[width=0.42\textwidth]{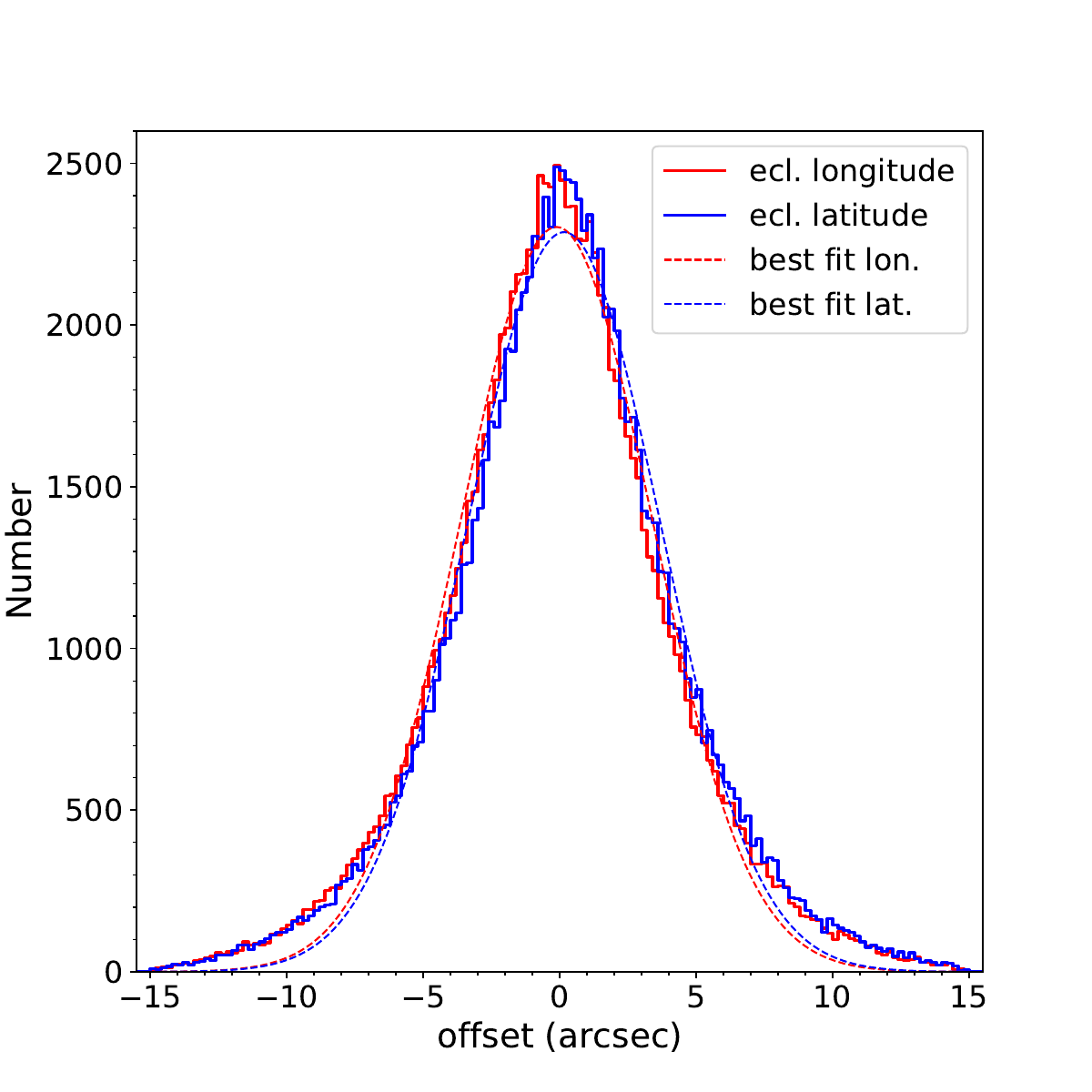}\\ 
    \end{tabular}
    \caption{
    Distribution of measured positional offset between astrometrically corrected eRASS1 sources with \texttt{DET\_LIKE\_0}$>7$ and Gaia/unWISE QSOs (see Section~\ref{subsec:astrometric_ext}) along equatorial (left) and ecliptic (right) coordinates in arcseconds. The dashed lines represent the best fit gaussian function to these distributions, the parameters of which are reported in the text.
    }
    \label{fig:psf_symmetry}
\end{figure*}

\FloatBarrier\section{Testing the energy calibration}
\label{appendix:energy_calibration}

An excellent target for testing the energy calibration is the oxygen--rich SNR 1E\,0102-7219 (in the following abbreviated as 1E0102), the brightest SNR in the SMC. It is characterized by strong emission lines of O, Ne, and Mg, exhibits only little `contaminating' emission from Fe, it is sufficiently compact to utilise the high spectral resolution provided by slitless X--ray gratings (\xmm RGS and \cxo HETG), yet extended enough to minimise any problems with pile--up. This object has been adopted as a standard calibration source by the {\em International Astronomical Consortium for High-Energy Calibration}\/ (IACHEC), which has developed a standard (purely empirical) model specifically designed for calibration \citep{Plucinsky2017}.

In Fig.~\ref{fig:1e0102} we show spectra of 1E0102 taken in dedicated calibration observations on 2019 Nov 7\,--\,8 (60\,--\,61~ks, 16~ks for TM6), and on 2021 Nov 26\,--\,27 (47\,--\,49~ks, 21~ks for TM4), thus covering a time span of more than 2~years. The data were taken with the same onboard processing mode which was used in eRASS. A major difference, however, were the CCD temperatures, which ranged during the first observation between -85.5 and -84.6~C, and during the last observation between -77.9 and -77.0~C. When comparing both observations it should also be considered that, although 1E0102 was observed on--axis, its precise location on the CCDs was different between both observations.

Spectra were extracted with the eSASS tasks {\tt evtool} and {\tt srctool} for each observation and each TM (except for TM5 and TM7, which are affected by the light leak), separately for a source and a background region. A circular source extraction region was used, with a radius of 1~arcmin centered on the FK5 coordinates $\alpha=16.006258\mbox{ deg},\, \delta=-72.032394\mbox{ deg}$, and the background was extracted from a circle with a radius of 5~arc\-min centered on the FK5 coordinates $\alpha=16.169575\mbox{ deg},\, \delta=-71.828828\mbox{ deg}$.

The fits are based on the standard IACHEC model for 1E0102, which consists of 52 narrow Gaussian emission lines, superimposed on an absorbed continuum. The emission lines are organised into 4 groups, corresponding to emission from O\,VII, O\,VIII, Ne\,IX, and Ne\,X\footnote{This model spectrum is available at \url{https://wikis.mit.edu/confluence/display/iachec/\\Thermal+SNR}.}. 

For an assessment of the quality of the energy reconstruction we performed a combined fit of TMs 12346 from both observations (10 spectra in total), using only single pixel events. We applied the standard IACHEC model and treated the normalisations of the  O\,VII, O\,VIII, Ne\,IX, and Ne\,X line complexes as free, but TM independent, parameters (4 free parameters). Only the overall normalisation was adjusted individually for each TM (10 free parameters). We allowed for TM specific shifts of the energy scales by XSPEC `gain fits', with all slopes fixed to 1.0 and the 10 individual offsets as additional free parameters. This resulted in a common fit of 10 spectra with 24 free parameters. The fit yields $\chi^2=4055.2$ for 2290 degrees of freedom, or $\chi^2_r=1.77$ (Fig.\,\ref{fig:1e0102}), and the mean energy shift is $-1.3\mbox{ eV}$ for the first and $+2.2\mbox{ eV}$ for the last observation, with a scatter of $\pm1.0\mbox{ eV}$ and $\pm3.0\mbox{ eV}$.

These long pointed observations of a line--rich SNR represent a benchmark test of the energy calibration. Considering that the calibration requirements for eRASS spectra are more relaxed due to the much shorter exposure times, we conclude that the energy calibration is sufficiently accurate for the sources detected in the eROSITA all--sky survey.

\begin{figure}
    \centering
    \includegraphics[width=0.48\textwidth]{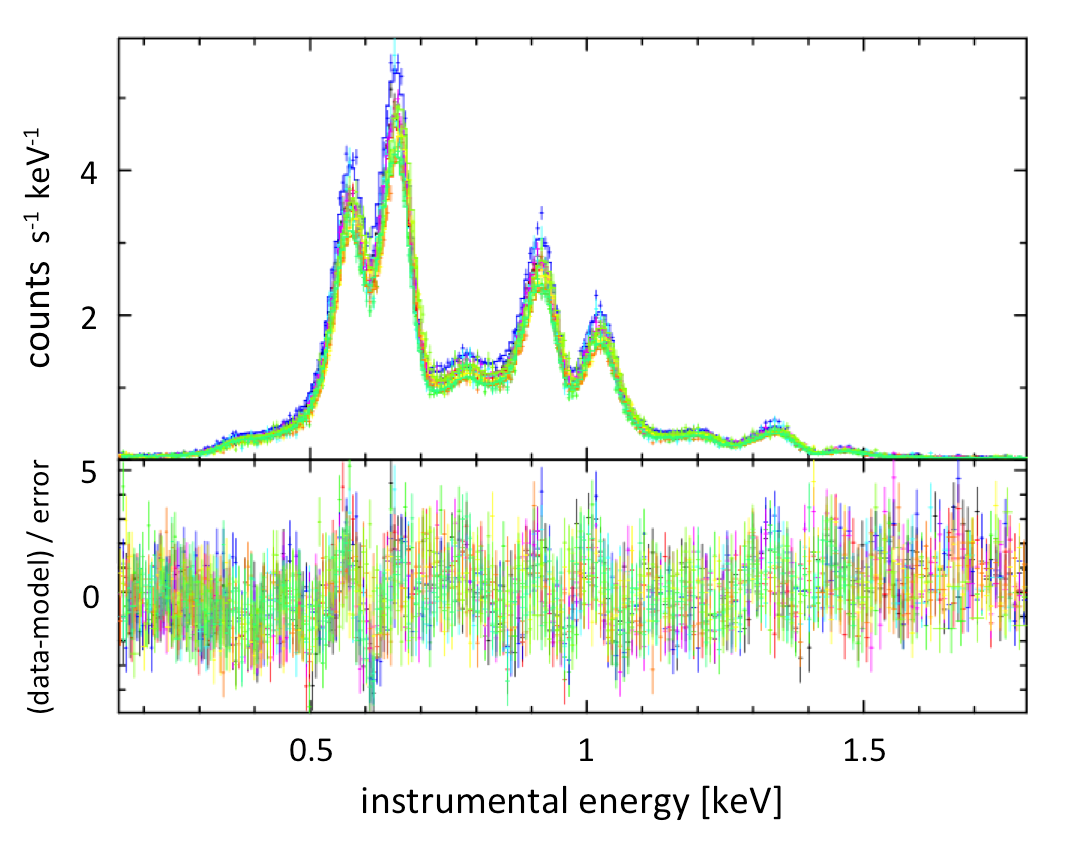}
    \caption{eROSITA spectra of the SNR 1E\,0102-7219, taken with TM12346 in Nov 2019 and 2021.}
\label{fig:1e0102}
\end{figure}

\FloatBarrier\section{The NRTA pipeline}
\label{sec:NRTA}
As outlined in Sect.~\ref{sec:scanning_strategy}, the field of view of eROSITA,
which has a diameter of roughly one degree, scans the sky continuously with a
rotational period of four hours and visits each position in the sky typically
for six consecutive scans. Each position is then revisited roughly six months
later. This cadence provides a unique opportunity to identify and study
transient phenomena in the X-ray sky and offers a compromise between time
resolution and sensitivity. The survey schedule of eROSITA does not allow for
interruptions in order to perform pointed observations, therefore fast
identification of transient events and the communication to other facilities,
also in other wavelength bands, is vital. To this end the NRTA pipeline was
developed. Its purpose is to analyse the science data on ground as soon as it is
ready at MPE and alert the appropriate team of scientists for a given event who
can then decide the correct course of action. Beyond this, the NRTA also aids in
some technical aspects of the maintenance of the eROSITA instrument. 

One of the highest priorities is the identification of bright X-ray transients,
especially if they are not known. To this end, the NRTA runs an implementation
of the Bayesian Blocks algorithm developed by \citet{Scargle2013} on the raw
detector count rates of TMs 1, 2, 3, 4, and 6. The algorithm is disabled for TMs
5 and 7 because their optical light leak causes strong fluctuations in the
count rates. When a bright source passes the field of view a significant excess
of the count rate is expected which can be recognised by the algorithm. This
excess is expected to have a duration of roughly 40\,s. After additional
filtering to exclude such periods of increased count rate caused by artefacts
from the CE, each of these time windows is marked for source detection. An
exemplary detector light curve of TM\,6 for a bright source passing through the
field of view and the resulting Bayesian Blocks is shown in
Fig.~\ref{fig:bayesian_blocks}.

%Since the FITS files created by \texttt{tmsplit} lack the necessary information
%for scientific analysis, the NRTA runs a customized version of the TEL chain
%(see Sect.~\ref{sec:tel_chain}) to calibrate those files. After that, the
%resulting files are split into eroday chunks, since they might cover the overlap
%between two rotations and the NRTA operates on a per eroday basis.

\begin{figure}[htbp] \centering
  \includegraphics[width=\columnwidth]{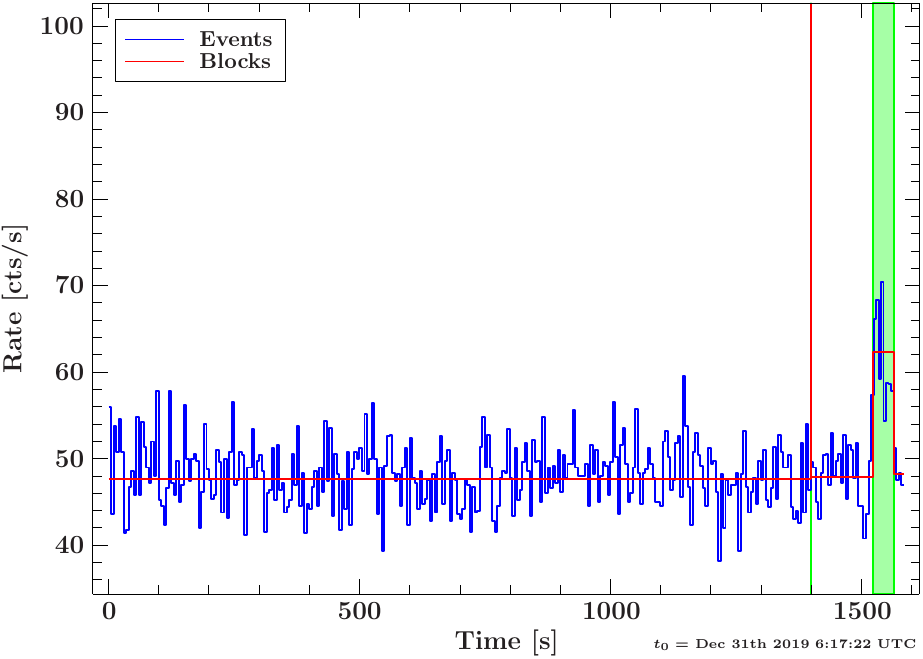}
  \caption{Example detector light curve (blue) of TM 6 of one telemetry file
    with bins 5\,s in length. The segmentation created by the Bayesian Blocks
    algorithm is shown in red and the identified intervals of bright sources in
    the field of view are shaded green. Note that the very short block at $t
    \approx 1400\, \mathrm{s}$ was caused by a remaining corrupt frame. The
    block of elevated count rate towards the end of the light curve was caused
by an unknown transient X-ray source detected on New Year's Eve 2019
\citep{Wilms2020}.} \label{fig:bayesian_blocks} \end{figure}

For all sources, either detected by source detection or ingested as point
sources externally into the pipeline, a variety of products are extracted using
the \texttt{srctool} task from the eSASS, most notably the spectrum and the
light curve for the pass of the source through the field of view. Based on these
data and possibly the information from the source detection, additional custom
parameters tailored for specific science cases are calculated, like count rates
in different energy bands, the signal-to-noise ratio, hardness ratios, and the
minimal distance of the source to the centre of the field of view. To compare
the sources with entries in other external catalogues, the \texttt{Nway}
Bayesian algorithm \citep{Salvato2018} is used to match the positions of the
sources against 54 catalogues covering all wavelengths bands.

The notification to the science teams is based on alerts. An alert for a source
is generated if a given set of criteria, generically called `triggers', which
can be combined using logical \texttt{and}, \texttt{or}, and \texttt{not}
operators, holds true. A variety of these criteria are implemented in a
flexible manner and can be used, for example, to test whether a numerical
parameter for a source falls within a given range, or equals a specific
value, or has a match in a specific catalogue with a specific value. A blacklist
is used to suppress alerts from areas which are known to produce large
amounts of spurious detections, for example from around Sco~X--1 (compare
Fig.~\ref{fig:erass1_spurious}). If a trigger is evaluated as positive for a
given source, an e-mail notification with basic information about the source
is sent to the science team responsible for the trigger. The alert can be
inspected in a web-based front-end which can display more advanced information
about the source (images, light curves, and external surveys). A team of two
scientists is assigned in weekly rotations to inspect the generated alerts and
communicate with the science teams. Some triggers are automatically closed by
the system and stored for later bulk analysis. All results of the NRTA are
stored on a separate archive and available for later inspection. Typically, it takes less than six hours after the data have arrived for
processing until the generation of an alert, which is sufficient given that data
can be stored on-board for up to 24\,hours before being transmitted to the ground
station. 

During eRASS1 the NRTA
analysed the occurrences of 21.4 million possible sources. Roughly one million
of these were found during source detection in regions given externally or
identified as containing a bright source, and the remaining ones were ingested
as known point sources to monitor. In total, roughly 150\,000 alerts were
generated. Out of these, we identified several time critical events, which were
published mostly as Astronomer's Telegrams \citep{rutledge_astronomers_1998}.
The first such event occurred shortly after the begin of the survey on December
31, 2019 and was caused by a bright, and yet otherwise unidentified, X-ray
transient designated SRGt~J123822.3--253206 with a flux of $2 \times
10^{-10}$ erg\,s$^{-1}$\,cm$^{-2}$ in the 0.2--10\,keV
band \citep[see Fig.~\ref{fig:bayesian_blocks},][]{wilms_srgerosita_2020}. Another bright transient,
SRGt~J071522.1--191609, with a flux of $1.3 \times
10^{-11}$ erg\,s$^{-1}$\,cm$^{-2}$, was detected on
April 14, 2020 \citep{gokus_srgt_2020}. On May 22, 2020, the NRTA detected a
flare from the millisecond pulsar PSR~J1023+0038 with a flux of $3.6 \times
10^{-11}$ erg\,s$^{-1}$\,cm$^{-2}$ in the 0.3--10\,keV
band \citep{koenig_srgerosita_2020}. On June 6, 2020, the Be/X-ray binary
RX~J0529.8--6556 in the LMC was found in outburst by the NRTA for multiple
consecutive scans with an average flux of $2.5 \times
10^{-11}$ erg\,s$^{-1}$\,cm$^{-2}$ in the 0.2--8\,keV
band \citep{haberl_srgerosita_2020,2021MNRAS.503.6187T}.

\FloatBarrier\section{Column descriptions}
\label{sec:column_description}

Table~\ref{table:columns} contains a description of the catalogue data model. 
Our single-band and three-band detections result in the 1B and 3B catalogues, which have different columns. Each 1B and 3B source has a unique source ID, namely, \texttt{UID}, or \texttt{DETUID}, which are equivalent. The \texttt{UID} is only unique in the 1B or 3B catalogue, and most of the 1B and 3B sources are identical. The 1B-3B association results (\S~\ref{sec:1B3B}) are saved in additional columns of \texttt{UID\_1B} or \texttt{UID\_Hard}.
The 1B catalog is divided into the Main and the Supp catalogues, which have identical columns. The Hard catalogue is a subsample of the 3B catalogue. 

Most of the source information is calculated in multiple bands and thus a band suffix is printed in the column names. Tables~\ref{table:energy_bands_1B} and \ref{table:energy_bands_3B} provide a dictionary for the energy band definitions in the 1B and 3B catalogues, respectively.
A few columns are from the PSF-fitting of source detection using the \texttt{ermldet} task, e.g., \texttt{DET\_LIKE}, \texttt{ML\_RATE}, \texttt{ML\_CTS}, \texttt{ML\_FLUX}, \texttt{EXT}, and \texttt{EXT\_LIKE}.
In the 3B detection, the band index 1, 2, and 3 indicate 0.2--0.6, 0.6--2.3, and 2.3--5~keV, and band 0 indicates all-band summary values. In the 1B detection, only band 1 (0.2--2.3~keV) is involved, and the all-band summary value (band 0) is identical to that of band 1.
For the 1B catalogues, we performed forced PSF-fitting (\texttt{ermldet}) and forced aperture photometry (\texttt{apetool}) at fixed source positions. They are reported in the bands P1--P9 (Table~\ref{table:energy_bands_1B}). Combining band P1 (0.5--1~keV) and P2 (1--2~keV), we calculated the values in band "S" (0.5--2~keV, see \S~\ref{subsec:flux_limit}).

\begin{table*}[htp]
  \centering
  \small
  \caption{eRASS1 catalogs column desciption. All errors are provided as 68\% confidence intervals (1-$\sigma$).}
  \begin{tabular}{p{0.2\textwidth} p{0.08\textwidth} p{0.64\textwidth}}

   \hline\hline
Column         &Units & Description \\
   \hline
%    \multicolumn{3}{l}{1. Source properties from PSF-fitting detection}\\
%   \hline
\texttt{IAUNAME} & & String containing the official IAU name of the source\\
\texttt{DETUID} & & String unique detection ID \\
\texttt{SKYTILE} & & Sky tile ID \\
\texttt{ID\_SRC} & & Source ID in each sky tile. Use SKYTILE+ID\_SRC to identify the corresponding source products\\
\texttt{UID} & & Integer unique detection ID. It equals CatID$\times 10^{11}$+SKYTILE$\times 10^5$+ID\_SRC, where CatID is 1 for the 1B detected Main and Supp catalogs and 2 for the 3B detected Hard catalog.\\
\texttt{UID\_Hard} & & Hard catalog UID of the source with a strong association, or -UID if the association is weak. 0 means no counterpart found in the Hard catalog. Only in the 1B Main and Supp catalogs\\
\texttt{UID\_1B} & & Main or Supp catalog UID of the source with a strong association, or -UID if the association is weak. 0 means no counterpart found in the 1B catalogs. Only in the Hard catalog\\
\texttt{ID\_CLUSTER} & & Group ID of simultaneously fitted sources\\
\hline
\texttt{RA}             	&deg	& Right ascension (ICRS), corrected\\
\texttt{DEC}            	&deg	& Declination (ICRS), corrected \\
\texttt{RA\_RAW}       	&deg	& Right ascension (ICRS), uncorrected\\
\texttt{DEC\_RAW}      	&deg	& Declination (ICRS), uncorrected\\
\texttt{RA\_LOWERR}        &  arcsec   & 1-$\sigma$ lower error of RA\\
\texttt{RA\_UPERR}         &  arcsec   & 1-$\sigma$ upper error of RA\\
\texttt{DEC\_LOWERR}       &  arcsec   & 1-$\sigma$ lower error of DEC\\
\texttt{DEC\_UPERR}        &  arcsec   & 1-$\sigma$ upper error of DEC\\
\texttt{RADEC\_ERR}     	&arcsec	& Combined positional error, raw output from PSF fitting\\
\texttt{POS\_ERR} & arcsec & 1-$\sigma$ position uncertainty\\
\texttt{LII} & deg & Galactic longitude\\
\texttt{BII} & deg & Galactic latitude\\
\texttt{ELON} & deg & Ecliptic longitude\\
\texttt{ELAT} & deg & Ecliptic latitude\\
\texttt{MJD} & & Modified Julian Date of the observation of the source nearest to the optical axis of the telescope\\
\texttt{MJD\_MIN} & & Modified Julian Date of the first observation of a source\\
\texttt{MJD\_MAX} & & Modified Julian Date of the last observation of a source\\
\hline
\texttt{EXT}            	& arcsec	& Source extent parameter\\
\texttt{EXT\_ERR}       	& arcsec	& 1-$\sigma$ error of EXT\\
\texttt{EXT\_LOWERR}      & arcsec      & 1-$\sigma$ lower error of EXT\\
\texttt{EXT\_UPERR}       & arcsec      & 1-$\sigma$ upper error of EXT\\
\texttt{EXT\_LIKE}      	&	& Extent likelihood\\
\texttt{DET\_LIKE\_{\sl\footnotesize n}}  	&	& Detection likelihood. 1B: n=0; 3B: n=0,1,2,3 \\
\texttt{ML\_CTS\_{\sl\footnotesize n}}     	&cts	& Source net counts. 1B: n=1,P[1-9]; 3B: n=0,1,2,3 \\
\texttt{ML\_CTS\_ERR\_{\sl\footnotesize n}} 	&cts	& 1-$\sigma$ combined counts error. 1B: n=1,P[1-9]; 3B: n=0,1,2,3  \\
\texttt{ML\_CTS\_LOWERR\_{\sl\footnotesize n}} 	&cts	& 1-$\sigma$ lower counts error. 1B: n=1,P[1-9]; 3B: n=1,2,3  \\
\texttt{ML\_CTS\_UPERR\_{\sl\footnotesize n}} 	&cts	& 1-$\sigma$ upper counts error. 1B: n=1,P[1-9]; 3B: n=1,2,3  \\
\texttt{ML\_RATE\_{\sl\footnotesize n}}    	&cts/s	& Source count rate. 1B: n=1,P[1-9]; 3B: n=0,1,2,3  \\
\texttt{ML\_RATE\_ERR\_{\sl\footnotesize n}}	&cts/s	& 1-$\sigma$ combined count rate error. 1B: n=1,P[1-9]; 3B: n=0,1,2,3  \\
\texttt{ML\_RATE\_LOWERR\_{\sl\footnotesize n}}	&cts/s	& 1-$\sigma$ lower count rate error. 1B: n=1,P[1-9]; 3B: n=1,2,3  \\
\texttt{ML\_RATE\_UPERR\_{\sl\footnotesize n}}	&cts/s	& 1-$\sigma$ upper count rate error. 1B: n=1,P[1-9]; 3B: n=1,2,3  \\
\texttt{ML\_FLUX\_{\sl\footnotesize n}}    	&erg/cm$^2$/s	& Source flux. 1B: n=1,P[1-9]; 3B: n=0,1,2,3  \\
\texttt{ML\_FLUX\_ERR\_{\sl\footnotesize n}}	&erg/cm$^2$/s	& 1-$\sigma$ combined flux error. 1B: n=1,P[1-9]; 3B: n=0,1,2,3  \\
\texttt{ML\_FLUX\_LOWERR\_{\sl\footnotesize n}}	&erg/cm$^2$/s	& 1-$\sigma$ lower flux error. 1B: n=1,P[1-9]; 3B: n=1,2,3  \\
\texttt{ML\_FLUX\_UPERR\_{\sl\footnotesize n}}	&erg/cm$^2$/s	& 1-$\sigma$ upper flux error. 1B: n=1,P[1-9]; 3B: n=1,2,3  \\
\texttt{ML\_EXP\_{\sl\footnotesize n}}     	&s	& Vignetted exposure time at the source position. 1B: n=1,P[1-9]; 3B: n=1,2,3  \\
\texttt{ML\_BKG\_{\sl\footnotesize n}}     	&cts/arcmin$^2$	& Background at the source position. 1B: n=1,P[1-9]; 3B: n=0,1,2,3   \\
\texttt{ML\_EEF\_{\sl\footnotesize n}}     	    &   &Encloded energy fraction. 1B: n=1,P[1-9]; 3B: n=1,2,3  \\
\texttt{APE\_CTS\_{\sl\footnotesize n}}    	&cts	& Total counts extracted within the aperture. 1B: n=1,P[1-9],S; 3B: n=1,2,3 \\
\texttt{APE\_EXP\_{\sl\footnotesize n}}    	&s	& Exposure map value at the given position. 1B: n=1,P[1-9],S; 3B: n=1,2,3 \\
\texttt{APE\_BKG\_{\sl\footnotesize n}}    	&cts	& Background counts extracted within the aperture, excluding nearby sources using the source map. 1B: n=1,P[1-9],S; 3B: n=1,2,3 \\
\texttt{APE\_RADIUS\_{\sl\footnotesize n}} 	&pixels	& Extraction radius in pixels (4"). 1B: n=1,P[1-9]; 3B: n=1,2,3 \\
\texttt{APE\_POIS\_{\sl\footnotesize n}}   	&	& Poisson probability that the extracted counts (APE\_CTS) are a background fluctuation. 1B: n=1,P[1-9],S; 3B: n=1,2,3 \\
\hline
\texttt{FLAG\_SP\_SNR} & & Source may lie within an overdense region near a supernova remnant\\
\texttt{FLAG\_SP\_BPS} & & Source may lie within an overdense region near a bright point source\\
\texttt{FLAG\_SP\_SCL} &  & Source may lie within an overdense region near a stellar cluster\\
\texttt{FLAG\_SP\_LGA} & & Source may lie within an overdense region near a local large galaxy\\
\texttt{FLAG\_SP\_GC\_CONS} & &  Source may lie within an overdense region near a galaxy cluster  \\
\texttt{FLAG\_NO\_RADEC\_ERR} & &  Source contained no \texttt{RA\_DEC\_ERR} in the pre-processed version of the catalogue \\
\texttt{FLAG\_NO\_CTS\_ERR} & &  Source contained no \texttt{ML\_CTS\_0\_ERR} in the pre-processed version of the catalogue \\
\texttt{FLAG\_NO\_EXT\_ERR} & & Source contained no \texttt{EXT\_ERR} in the pre-processed version of the catalogue \\
\texttt{FLAG\_OPT} & & Source matched within 15\arcsec with a bright optical star, likely contaminated by optical loading \\

%FLAG\_SP\_SNR & & Move Table 5 here?\\
%FLAG\_SP\_BPS & & \\
%FLAG\_SP\_SCL & & \\
%FLAG\_SP\_LGA & & \\
%FLAG\_SP\_GC\_CONS & & \\
%FLAG\_NO\_RADEC\_ERR & & \\
%FLAG\_NO\_EXT\_ERR & & \\
%FLAG\_NO\_CTS\_ERR & & \\
%\hline
%ID\_CLUSTER & & \\
%OWNER: & & \\
%LII & & Galactic longitude of the source (non-corrected). \\
%BII & & Galactic latitude of the source (non-corrected)\\
%DIST\_NN : & & \\
%SRCDENS & & \\
%TSTART & & determined based on the TSTOP and TSTART keywords in the event files, done somewhere in catprep (what was earliest observation contributing to this source, in spacecraft clock), and carried over to source lists. It is not the time when the source first entered the field of view, but is when the first event fell in the sky tile that contributed to the source.\\
% GL: With the introduction of the source specific MJD_MIN, MJD_MAX columns 
%these columns are probably NO_RADEC_ERR, NO_CTS_ERR obsolete 
%TSTOP & & \\
%MASKFRAC & & Fraction of good pixels in ML\_RADIUS\\
%HR\_1  & & \\
%HR\_1\_ERR  & & \\
%HR\_2  & & \\
%HR\_2\_ERR  & & \\
%HR\_3  & & \\
%HR\_3\_ERR  & & \\
%VIGNET\_{\sl\footnotesize n}  & & \\
%ELON & & Ecliptic longitude of the source (non-corrected)\\
%ELAT & & Ecliptic latitude of the source (non-corrected)\\
%ELON\_CORR & & Ecliptic longitude of the source, corrected.\\
%ELAT\_CORR & & Ecliptic latitude of the source, corrected.\\
\hline	
  \end{tabular}
%  \tablefoot{
%    The hard catalogue has almost identical columns, except that the PSF-fitting output columns (the ones marked with $_\ast$) have four sets of data with an energy-band suffix (0,1,2,3) in the column name rather than only one set of data (0, 1).}
      \label{table:columns}
\end{table*}

\begin{table}
  \small
  \caption{Dictionary of energy band suffixes in the eRASS1 1B (Main and Supplementary) catalogs}
  \centering
  %\begin{tabular}{p{0.09\textwidth} p{0.16\textwidth}}
  \begin{tabular}{cc}
   \hline\hline
Band       & Energy range (keV) \\  
\hline	
0, 1 & 0.2--2.3 \\
P1 & 0.2--0.5   \\
P2 & 0.5--1.0   \\
P3 & 1.0--2.0   \\
P4 & 2.0--5.0   \\
P5 & 5.0--8.0   \\
P6 & 4.0--10.0  \\
P7 & 5.1--6.1   \\
P8 & 6.2--7.1   \\
P9 & 7.2--8.2   \\
S & 0.5--2      \\
\hline
  \end{tabular}
      \label{table:energy_bands_1B}
\end{table}

\begin{table}
  \small
  \caption{Dictionary of energy band suffixes in the eRASS1 Hard catalog}
  \centering
  %\begin{tabular}{p{0.09\textwidth} p{0.16\textwidth}}
  \begin{tabular}{cc}
   \hline\hline
Band       & Energy range (keV) \\  
\hline	
0  & 0.2--5.0 \\
1  & 0.2--0.6 \\
2  & 0.6--2.3 \\
3  & 2.3--5.0 \\
%P1 & 0.2--0.5 \\
%P2 & 0.5--1.0 \\
%P3 & 1.0--2.0 \\
%P4 & 2.0--5.0 \\
%P5 & 5.0--8.0 \\
%P6 & 4.0--10.0 \\
%P7 & 5.1--6.1 \\
%P8 & 6.7--7.1 \\
%P9 & 7.2--8.2 \\
\hline
  \end{tabular}
      \label{table:energy_bands_3B}
\end{table}

%\end{document}

\FloatBarrier\section{Software and calibration versions used in this work}
\label{appendix:esass}
The eROSITA data presented in this work were processed in the time period Nov. 2021 to  Jan. 2022. The software and calibration versions used (pipeline configuration 010) are therefore different from those of the earlier eROSITA early data release (EDR, pipeline configuration 001), as well as from later work based on proprietary eROSITA data (pipeline configuration 020). Here we provide a summary of the main differences between the EDR, DR1 and later software (Tab.~\ref{tab:esass_versions}) and calibration (Tab.~\ref{table:caldb}) versions. Further details on eSASS software versions are available on the eROSITA DR1 website\footnote{\url{https://erosita.mpe.mpg.de/dr1/eSASS4DR1/}}.

\begin{table*}[htp]
  \centering 
  \small
  \caption{EDR/DR1 eSASS task version changes}
  \begin{tabular}{clcccl}
  \hline\hline
  pipeline & eSASS task & \multicolumn{3}{c}{task versions} & Important functional improvements \\
  chain  & & 001 (EDR) & 010 (DR1) & 020 & \\
  \hline
TEL & {\tt evprep}  & 3.11-3.11.2   & 3.46.1-3.46.3 & 3.46.3-3.49 & \underline{010}: using high cadence science housekeeping; \\
    &               &               &               &             & flagging of image artefacts; improved handling of CCD \\
    &               &               &               &             & frames split across telemetry frames \\
    &               &               &               &             & \underline{020}: improved flagging of image artefacts \\
    & {\tt ftfindhotpix}& 1.7       & 1.22          & 1.22.1      & \underline{010}: corrected flagging of neighbouring pixels; time \\
    &               &               &               &             & filtering of bad pixels added \\
    & {\tt pattern} & 2.10          & 4.3.1-4.3.3   & 4.3.5-4.3.7 & \underline{010}: improved treatment of secondary event threshold and \\
    &               &               &               &             & of bad pixels; PAT\_INF assigned to all events of a pattern; \\
    &               &               &               &             & stronger TM-specific noise suppression of doubles \& triples \\
    &               &               &               &             & \underline{020}: further minor improvements of PAT\_INF \\
    & {\tt energy}  & 2.10          & 3.2.1         & 3.2.2-3.2.5 & \underline{010} CCD temperature and long-term degradation con- \\
    &               &               &               &             & sidered; keyword PAT\_MIN\_EV is evaluated; workaround \\
    &               &               &               &             & for missing time values in housekeeping data \\
    &               &               &               &             & \underline{020}: error in subpixel position computation fixed \\
    & {\tt attprep} & 1.6- 1.8      & 1.10          & 1.10        & \underline{010}: quaternions renormalised to increase accuracy \\ 
\hline
%EXP & {\tt evtool}  & 2.10.1        &  2.29.2      &2.29.2-2.29.4& \underline{010}:          \\ 
%EXP & {\tt flaregti}& 1.20          &  1.26        & 1.26        &            \\
DET & {\tt expmap}  &[1.33.1-1.33.3]& 2.11.3        & 2.11.3-2.13 & \underline{010}: improved accuracy of projection; corrected flagging \\
    &               &               &               &             & of pixels next to bad pixels; powerlaw-weighted energy \\
    &               &               &               &             & dependence of vignetting (index specified on command line) \\
    & {\tt erbackmap}& [1.17]       & 1.26          & 1.26        & \underline{010}: number of kernel scales changed from 8 to 16 \\
%    & {\tt erbox}   & [1.15]        & 1.18          & 1.18        &            \\
    & {\tt ermldet} & [1.35]        & 1.56          & 1.56-1.58   & \underline{010}: upper and lower errors of source parameters added \\
    &               &               &               &             & \underline{020}: increased maximum number of fit iterations and \\
    &               &               &               &             & minimum step width \\
%    & {\tt apetool} & [1.9.6]       & 1.23          & 1.23-1.24   &            \\  
%    &{\tt ersensmap}& [1.8]         & 1.10          & 1.10-1.11   &            \\
 \hline   
SOU & {\tt srctool}& [1.61-1.63]    & 1.81/[1.76.1] & 1.81-1.85   & \underline{010}: computation of BACKRATIO improved; asymmetric \\
    &              &                &               &             & errors and additional columns added to light-curve tables; \\
\hline
  \end{tabular}
\tablefoot{Only tasks with functional improvements are listed. Task versions for pipeline configuration 020 are as of 2023-01-15. The EDR dataset only provided calibrated event lists; task versions available in the EDR user eSASS package are listed in square brackets. Task {\tt srctool} has different but functionally equivalent version numbers in the DR1 pipeline and in the 
corresponding DR1 user eSASS (user eSASS version marked in square brackets).\label{tab:esass_versions}}
\end{table*}

\begin{table*}[htp]
  \centering
  \small
  \caption{EDR/DR1 calibration version changes}
  \begin{tabular}{lcccl}
  \hline\hline
  calibration & \multicolumn{3}{c}{versions} & Comments \\
  component     & 001 (EDR) & 010 (DR1) &  020 &        \\
  \hline
  {\tt gyro\_boresight\_191208v0<x>} & -- & 10 & 10 & boresight correction\\
  {\tt gyro\_boresight\_200611v0<x>} & -- & -- & 11 & boresight correction\\
  {\tt tm<x>\_badcamt\_190701v0<x>} & -/-/1/-/-/1/- & 2/3/3/8/4/6/3 & 2/3/3/8/4/6/3 & for TM1/2/3/4/5/6/7\\
  {\tt tm2\_badpix\_<date><version>} & -- & 210117v01 & 210117v01 & new bad pixels which appeared during flight\\
  {\tt tm3\_badpix\_<date><version>} & -- & 201127v01 & 201127v01 & --- \glqq\ --- \\
  {\tt tm4\_badpix\_<date><version>} & -- & 210223v10 & 210223v10 & new bad pixels due to micrometeoroid impact\\
  {\tt tm4\_badpix\_<date><version>} & -- & 210616v02 & 210616v02 & --- \glqq\ ---\\
  {\tt tm4\_badpix\_<date><version>} & -- & 210805v03 & 210805v03 & --- \glqq\ ---\\
  {\tt tm5\_badpix\_<date><version>} & -- & 210310v13 & 210310v13 & --- \glqq\ ---\\
  {\tt tm5\_badpix\_<date><version>} & -- & 210616v02 & 210616v02 & --- \glqq\ ---\\
  {\tt tm6\_badpix\_<date><version>} & -- & 191023v01 & 191023v01 & new bad pixels which appeared during flight\\
  {\tt tm7\_badpix\_<date><version>} & -- & 210129v05 & 210129v05 & new bad pixels due to micrometeoroid impact\\
  {\tt tm7\_badpix\_<date><version>} & -- & 210207v02 & 210207v02 & --- \glqq\ ---\\
  {\tt tm7\_badpix\_<date><version>} & -- & 210210v02 & 210210v02 & --- \glqq\ ---\\
  {\tt tm7\_badpix\_<date><version>} & -- & 210211v08 & 210211v08 & --- \glqq\ ---\\
  {\tt tm7\_badpix\_<date><version>} & -- & 210616v02 & 210616v02 & --- \glqq\ ---\\
  {\tt tm<x>\_corr\_algos\_190701v0<x>} & -- & -/-/-/-/2/-/2 & -/-/-/-/2/-/2 & for TM1/2/3/4/5/6/7 \\
  {\tt tm<x>\_energy\_190714v0<x>} & 3 & 6 & 6 & for TM1/2/3/4/5/6/7 \\
  {\tt tm<x>\_timeoff\_190701v0<x>} & 1/1/1/1/1/1/1 & 1/2/2/2/2/2/2 & 1/2/2/2/2/2/2 & for TM1/2/3/4/5/6/7\\
  \hline
  \end{tabular}
    \tablefoot{
    Listed are the versions numbers which are used in the respective processings. That does not necessarily mean, that they are used for eRASS1 processing, because their application is only necessary at a later date. 
    
    Different versions are separated by "/". In case a calibration file does not exist for a TM it is marked by "-".
      \label{table:caldb}}
\end{table*}

\FloatBarrier\section{Essential Dictionary and List of Acronyms}
\label{sec:dictionary}
Table~\ref{tab:erass1_dictionary} presents a list of useful terms and acronyms found in this paper.

\begin{table*}
    \centering
    \caption{Essential dictionary and list of acronyms}
    \label{tab:erass1_dictionary}
    %\begin{tabular}{p{2.6cm}|p{13.5cm}}
    \begin{tabular}{ll}
        \hline\hline
        Term or Acronym & Definition \\
        \hline
        1B & Catalogue created by DET chain run on a single broad band (0.2--2.3 keV)\\
        3B & Catalogue created by DET chain run simultaneously on three bands (0.2--0.6; 0.6--2.3; 2.3--5~keV)\\
        AGN & Active Galactic Nuclei\\
        ARF&Ancillary Response Function\\
        CalPV & Calibration and Performance Verification\\
        CCD & Charge-Coupled Device\\
        CE& Camera Electronics\\
        CGM & Circum-Galactic Medium\\
        CME & Coronal Mass Ejection\\
        CXB & Cosmic X-ray Background\\
        DR1 & Data Release 1\\
        ECF & Energy Conversion Factor\\
        EDR & Early Data Release\\
        eFEDS & eROSITA Final Equatorial Depth Survey\\
        eRASS($n$) & ($n$th) eROSITA All-Sky Survey\\
        eRASS:$n$ & Cumulative dataset from the first $n$ eROSITA All-Sky Surveys\\
        eROSITA & extended ROentgen Survey with an Imaging Telescope Array\\
        eROSITA-DE & German eROSITA Consortium\\
        Eroday & Time-interval of 4 hours, equivalent to a scan\\
        eSASS & eROSITA Science Analysis Software System\\
        FAP & False Alarm Probability\\
        FOV & Field Of view\\
        FWC & Filter Wheel Closed\\
        GTI & Good Time Interval\\
        HiPS & Hierarchical Progressive Survey\\
        HEW & Half Energy Width\\
        ITC & Interface and Thermal Controller\\
        L2& Second Lagrange point of the Earth-Sun system\\
        LHB & Local Hot Bubble\\
        LMC & Large Magellanic Cloud\\
        MIP & Minimum Ionizing Particles\\
        MW & Milky Way\\
        NRTA & Near Real Time Analysis\\
        PS & Point Source\\
        PSF & Point Spread Function\\
        QSO & Quasi Stellar Objects\\
        RASS & ROSAT All-Sky Survey\\
        RMF & Redistribution Matrix File\\
        Scan & One revolution of \SRG around its axis during all-sky survey operations\\
        SEP & South Ecliptic Pole\\
        SEU & Single Event Upset\\
        S/N & Signal to Noise\\
        SMBH & SuperMassive Black Hole\\
        SMC & Small Magellanic Cloud\\
        SNR & SuperNova Remnants\\
        \SRG & {\it Spektrum Roentgen Gamma} Observatory\\
        SWCX & Solar Wind Change Exchange\\
        SZ & Sunyaev-Zedovich\\
        TM($n$) & ($n$th) Telescope Module\\
        UTC & Coordinated Universal Time\\
        Visit & Single pass (approximately 40s long) of eROSITA over a source during a scan\\
        ZEA & Zenith Equal Area (projection)\\
    \hline    
    \end{tabular}
\end{table*}

\end{appendix}
\end{document}